\documentclass{aa}
\usepackage{graphicx}
\usepackage[varg]{txfonts}
\usepackage{natbib}
\bibpunct{(}{)}{;}{a}{}{,} 
\begin{document}

\title{Aperture synthesis imaging of the carbon AGB 
star R~Sculptoris
\thanks{Based on observations made with the VLT Interferometry (VLTI)
at Paranal Observatory under programme IDs 090.D-0136, 093.D-0015, 096.D-0720.}}
\subtitle{Detection of a complex structure and a dominating spot on the 
stellar disk}
\author{
M.~Wittkowski\inst{1}\and
K.-H.~Hofmann\inst{2}\and
S.~H\"ofner\inst{3}\and
J.~B.~Le~Bouquin\inst{4}\and
W.~Nowotny\inst{5}\and
C.~Paladini\inst{6}\and
J.~Young\inst{7}\and
J.-P.~Berger\inst{4}\and
M.~Brunner\inst{5}\and
I.~de~Gregorio-Monsalvo\inst{8,9}\and
K.~Eriksson\inst{3}\and
J.~Hron\inst{5}\and
E.~M.~L.~Humphreys\inst{1}\and
M.~Lindqvist\inst{10}\and
M.~Maercker\inst{10}\and
S.~Mohamed\inst{11,12,13}\and
H.~Olofsson\inst{10}\and
S.~Ramstedt\inst{3}\and
G.~Weigelt\inst{2}
}
\institute{
European Southern Observatory, Karl-Schwarzschild-Str. 2,
85748 Garching bei M\"unchen, Germany,
\email{mwittkow@eso.org}
\and
Max-Planck-Institut f\"ur Radioastronomie, Auf dem H\"ugel 69, 53121 Bonn,
Germany
\and
Department of Physics and Astronomy, Uppsala University,
Box 516, 75120 Uppsala, Sweden
\and
Univ. Grenoble Alpes, CNRS, IPAG, F-38000 Grenoble, France
\and
Department of Astrophysics, University of Vienna, T\"urkenschanzstraße 17,
1180 Vienna, Austria
\and
Institut d'Astronomie et d'Astrophysique,
Universit{\'e} Libre de Bruxelles, CP.226,  Boulevard du Triomphe,
1050 Brussels, Belgium
\and
Astrophysics Group, Cavendish Laboratory, JJ Thomson Avenue,
Cambridge CB3 0HE, UK
\and
Joint ALMA Office, Alonso de C{\'o}rdova 3107, Vitacura, Casilla 19001,  Santiago 19, Chile
\and
European Southern Observatory, Alonso de C{\'o}rdova 3107, Vitacura, Santiago, Chile
\and
Department of Earth and Space Sciences, Chalmers University of Technology,
Onsala Space Observatory, 43992 Onsala, Sweden
\and
South African Astronomical Observatory, PO Box 9, Observatory 7935,
South Africa
\and
Astronomy Department, University of Cape Town, 7701, Rondebosch, South Africa
\and
National Institute for Theoretical Physics, Private Bag X1, Matieland, 7602, South Africa
}
\date{Received \dots; accepted \dots}
\abstract{}
{We present near-infrared interferometry of the carbon-rich 
asymptotic giant branch (AGB) star R Sculptoris (R~Scl).
}
{We employ medium spectral resolution $K$-band interferometry 
obtained with the instrument AMBER at the Very Large Telescope
Interferometer (VLTI) and $H$-band low spectral resolution 
interferometric imaging observations obtained with the VLTI 
instrument PIONIER. We compare our data to a recent grid of 
dynamic atmosphere and wind models.
We compare derived fundamental parameters to stellar evolution 
models.}
{The visibility data indicate a broadly circular resolved stellar
disk with a complex substructure.
The observed AMBER squared visibility values show drops at the 
positions of CO and CN bands,
indicating that these lines form in extended layers above
the photosphere. The AMBER visibility values
are best fit by a model without a wind.
The PIONIER data are consistent with the same model.
We obtain a Rosseland
angular diameter of 8.9$\pm$0.3\,mas, corresponding to a Rosseland
radius of 355$\pm$55\,$R_\sun$, an effective temperature of
2640$\pm$80\,K, and a luminosity of $\log L/L_\sun$\,=\,3.74$\pm$0.18. 
These parameters match evolutionary tracks of initial mass
1.5$\pm$0.5\,$M_\sun$ and current mass 1.3$\pm$0.7\,$M_\sun$.
The reconstructed PIONIER images exhibit a complex structure within 
the stellar disk including a dominant bright spot located at the 
western part of the stellar disk.
The spot has an $H$-band peak intensity of 40\% to 60\% above 
the average intensity of the limb-darkening-corrected stellar disk.
The contrast between the minimum and maximum intensity on the stellar
disk is about 1:2.5.}
{
Our observations are broadly consistent with predictions by 
dynamic atmosphere and wind models, although models with wind
appear to have a circumstellar envelope that is too extended 
compared to our observations.
The detected complex structure within
the stellar disk is most likely caused by giant convection 
cells, resulting in large-scale shock fronts, and their effects on
clumpy molecule and dust formation seen against the photosphere
at distances of 2--3 stellar radii.
}
\maketitle
\section{Introduction}
Low- to intermediate-mass stars evolve into red giant
and subsequently into asymptotic giant branch (AGB) stars.
Mass loss becomes increasingly important during
the AGB phase, both for the evolution of the star, and for the return of
material to the interstellar medium and thus the chemical enrichment
of galaxies.
Indeed, the evolution of the mass-loss rate throughout the AGB limits 
the time the star spends on the AGB, and hence the number of thermal 
pulses. This very strongly affects the yields from AGB stars and the 
total amount of dust produced by AGB stars.
Depending on whether or not carbon has been dredged up from the core into 
the atmosphere, 
AGB star atmospheres appear to have an oxygen-rich (C/O<1) or 
a carbon-rich (C/O>1) chemistry.
A canonical model of the mass-loss process has been developed for
the case of a carbon-rich chemistry, where atmospheric carbon dust has a
sufficiently large opacity to be radiatively accelerated and driven out of
the gravitational potential of the star and where the dust drags along the gas
\citep[e.g.,][]{Fleischer1992,wachter2002,Mattsson2010}. 
Recently, \citet{Eriksson2014}
presented an extensive grid of state-of-the-art dynamic atmosphere and 
wind models for carbon-rich AGB stars of different stellar parameters and 
resulting mass-loss rates.
These models are mostly tested by comparison to low-resolution 
spectra and photometry,
and generally show a satisfactory agreement.
However, questions on the details of the mass-loss process remain.
For example, \citet{Sacuto2011}, \citet{Rau2015}, and \citet{Rau2017}
reported that models with dust-driven winds show discrepancies with observations
of several semi-regular (SR) carbon-rich AGB stars, although
the stars are known to be surrounded by circumstellar material.

Some carbon-rich AGB stars are known to exhibit an extreme clumpiness
of their circumstellar environment. For example, near-infrared images
of IRC\,+10\ 216 showed several clumps within the circumstellar environment
that changed dramatically on timescales of several years
\citep{Weigelt1998,Osterbart2000,Weigelt2002,Kim2015,Stewart2016}.
Recent Atacama Large Millimeter Array (ALMA) observations 
suggest the presence of a binary-induced spiral
structure in the innermost region of the circumstellar environment
of IRC\,+10\ 216 \citep{Cernicharo2015,Decin2015,Quintana2016}. 
\citet{Paladini2012} detected an asymmetry in the circumstellar envelope 
(CSE) of a carbon Mira, R~For,
that they interpreted as caused by a dust clump or a substellar companion.
\citet{vanBelle2013} reported for their sample of carbon stars, observed
with the Palomar Testbed Interferometer, a general tendency for detection
of statistically significant departures from sphericity with increasing
interferometric signal-to-noise. They discussed that most -- and potentially
all -- carbon stars may be non-spherical and that this may be caused
by surface inhomogeneities and a rotation-mass-loss connection.

\citet{Maercker2012,Maercker2016} 
obtained ALMA images of the 
$^{12}$CO $J$\,=\,1–0, 2–1, and 3–2 line emission
of the carbon-rich AGB star R~Sculptoris (R~Scl). 
These images surprisingly 
revealed a spiral structure within and connected 
to a previously known detached CO shell \citep{Olofsson1996},
indicating the presence of a previously undetected binary companion,
which is shaping the CSE with its gravitational interaction.
Maercker et al. interpreted their observations
by a shell entirely filled with gas, a variable mass-loss 
history after a thermal pulse 1800 years ago, and a binary 
companion. The star
R~Scl is thus an interesting target to study mass-loss
evolution during the AGB phase. The ALMA data and other previous observations 
give us a good view of the CSE and the 
longer-term evolution of the wind, 
but lack detailed information on the very close atmospheric environment,
where the extension of the atmosphere and the dust formation take place,
and thus on the current state of R~Scl.

Here, we present near-infrared $K$-band spectro-interferometry
and $H$-band interferometric imaging
of R~Scl with the goals of (1) further constraining 
and testing available dynamic atmosphere and wind models for
carbon-rich AGB stars,
(2) revealing the detailed morphology 
of the stellar atmosphere and innermost mass-loss region,
and (3) constraining fundamental stellar properties of R~Scl.

\section{Observations and data reduction}
\label{sec:obs}
We obtained interferometric observations of R~Scl with the VLTI/AMBER
instrument between October and December 2012 and with the VLTI/PIONIER
instrument between August and September 2014 as well as between
November and December 2015. The details of the AMBER and PIONIER
observations and their data reduction are described in 
Sects.~\ref{sec:obs_amber} and \ref{sec:obs_pionier}, respectively.
In particular, the PIONIER observations obtained in 2014 allowed 
us to reconstruct images of R~Scl at three spectral channels
within the near-infrared $H$-band.
\subsection{Lightcurve and properties of R~Scl}
\label{sec:lightcurve}
\begin{figure}
\resizebox{\hsize}{!}{\includegraphics{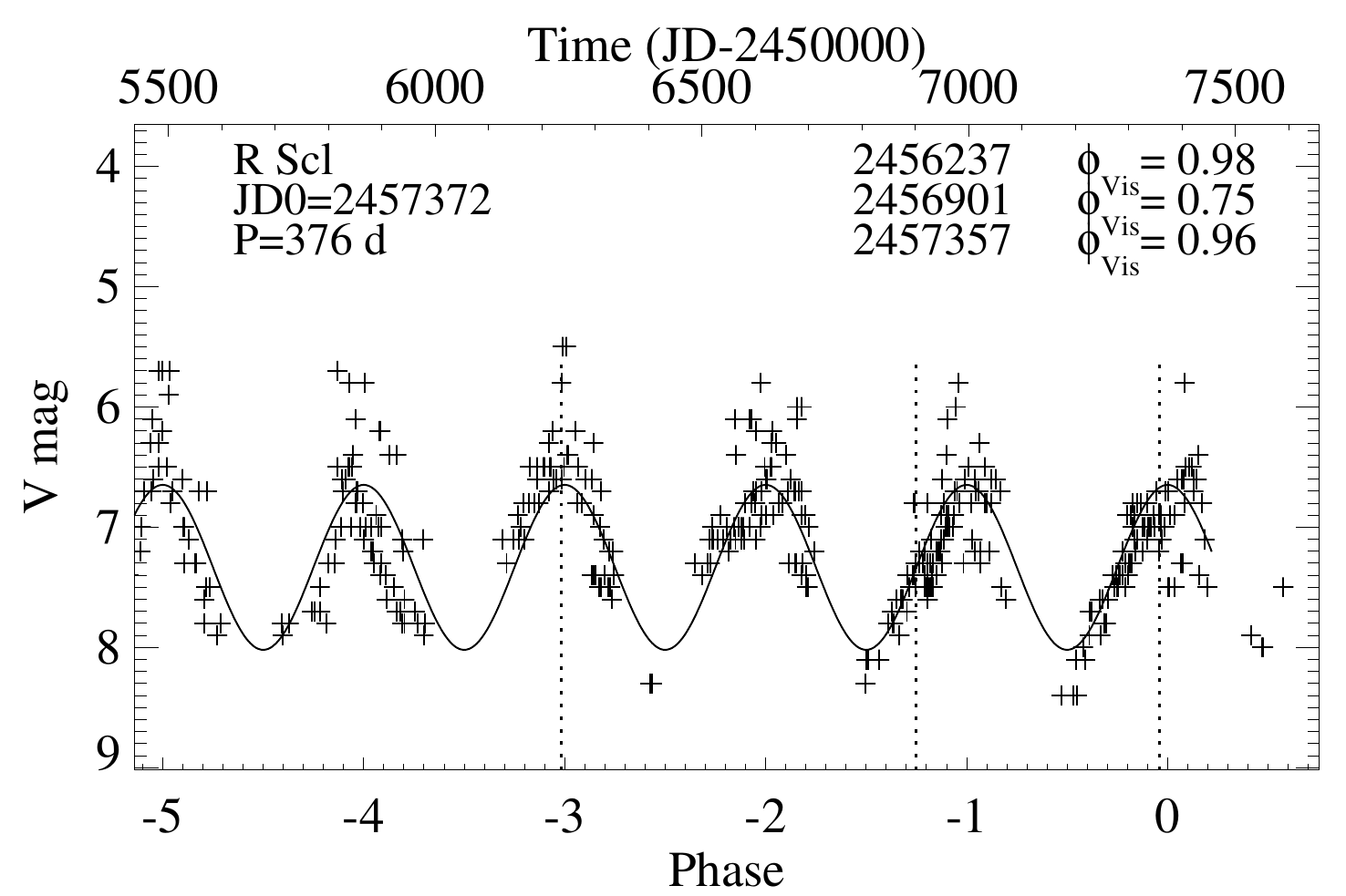}}
\caption{Visual lightcurve of R~Scl based on the AAVSO and
AFOEV databases. The solid line represents a sine fit to the 
ten most recent cycles. The dashed vertical lines denote the
mean epochs of our 2012 AMBER and 2014 \& 2015 PIONIER observations.}
\label{fig:lightcurve}
\end{figure}
\begin{table*}
\centering
\caption{Log of our AMBER observations.\label{tab:obs_amber}}
\begin{tabular}{rrrrrrl}
\hline\hline
No. & Date          & JD\tablefootmark{a} & Baseline & Proj. baseline length & Proj. baseline angle & Calibrators \\
    &               &                     &          & (m)                   & ($\deg$ N of E)      & before, after\\\hline
1   & 2012-10-13    & 56214.29            & A1/B2/D0 & 10.56/31.05/31.03     & 41.3/-58.6/-30.0     &HR~109, $\chi$~Phe\\ 
2   & 2012-10-13    & 56214.34            & A1/B2/D0 &  9.93/26.83/28.60     & 27.7/-62.4/-42.1     &$\chi$~Phe, $\iota$Eri\\
3   & 2012-10-17    & 56218.15            & A1/B2/D0 & 11.01/35.77/33.89     & 70.4/-38.3/-20.4     &HR~109, $\chi$~Phe\\
4   & 2012-10-17    & 56218.19            & A1/B2/D0 & 11.31/35.46/33.57     & 62.6/-46.3/-27.7     &HR~109, $\chi$~Phe\\
5   & 2012-10-17    & 56218.23            & A1/B2/D0 & 11.14/34.16/32.80     & 53.9/-52.6/-33.6     &$\chi$~Phe, $\iota$Eri\\
6   & 2012-10-17    & 56218.29            & A1/B2/D0 & 10.43/30.92/30.61     & 38.8/-59.4/-39.7     &$\chi$~Phe, $\iota$Eri\\
7   & 2012-10-18    & 56219.12            & A1/B2/C1 & 10.65/11.30/15.57     & -74.1/16.2/59.4      &HR~109, HR~109\\
8   & 2012-11-17    & 56249.16            & D0/H0/I1 & 40.76/59.70/78.77     & -19.1/82.5/67/0      &HR~109, $\chi$~Phe\\
9   & 2012-12-08    & 56270.12            & D0/G1/I1 & 42.74/70.59/76.57     & -51.5/29.5/62.9      &$\chi$~Phe, $\iota$Eri\\
10  & 2012-12-08    & 56270.15            & D0/G1/I1 & 39.45/69.44/70.77     & -55.0/20.5/53.2      &$\chi$~Phe, $\chi$~Phe\\
11  & 2012-12-08    & 56270.18            & D0/G1/I1 & 36.78/68.77/66.61     & -56.7/14.3/45.8      &$\chi$~Phe, $\iota$Eri\\
12  & 2012-12-03    & 56265.20            & A1/G1/K0 & 66.97/76.46/90.26     & 35.7/-42.0/-88.5     &-, $\chi$~Phe\\
\hline
\end{tabular}
\tablefoot{
\tablefootmark{a}{JD-2450000.}}
\end{table*}
The source R~Scl is classified as a carbon-rich SR pulsating 
AGB star with a period of 370 days \citep{Samus2009}. 
\citet{Maercker2016} estimated a present-day mass-loss rate of 
up to $10^{-6}$M$_\odot$/yr and a present-day outflow velocity
of about 10 km/s.

The distance to R~Scl based on the empirical
period-luminosity ($P$-$L$) relationships by \citet{Knapp2003}
results in 370$^{-70}_{+100}$\,pc for SR variables and
340$^{-70}_{+100}$\,pc for Mira variables.
The $P$-$L$ relationships for Miras by \citet{Groenewegen1996} and
\citet{Whitelock2008} give similar distances of 360\,pc and 370\,pc,
respectively.
\citet{Sacuto2011} derived a distance of 350$\pm$50\,pc
by fitting the SED to hydrostatic atmosphere models.
The original Hipparcos parallax was uncertain with 2.11\,$\pm$\,1.52\,mas,
corresponding to a distance of 474$^{-200}_{+1200}$\,pc.
The revised Hipparcos parallax \citep{vanLeeuwen2007} of 3.76\,$\pm$\,0.75\,mas
corresponds to a distance of 266$^{-45}_{+66}$\,pc.
Both Hipparcos parallaxes might be biased because R~Scl has a
large angular diameter and exhibits surface features.
\citet{Maercker2012} and \citet{Vlemmings2013} used a distance of 290\,pc
in the tradition of \citet{Olofsson2010} and \citet{Schoeier2005}.
\citet{Maercker2016} corrected the distance to the $P$-$L$ distance of
370\,pc. In total, this shows that the distance to R~Scl is
uncertain. In the following we use the $P$-$L$ distance
$d_\mathrm{PL}=370^{-70}_{+100}$\,pc.

Figure~\ref{fig:lightcurve} shows the
recent visual lightcurve of R~Scl based on data obtained from the
AAVSO (American Association of Variable Star Observers) and 
AFOEV (Association Francaise des Observateurs d'Etoiles Variables)
databases. The solid line represents a fit of a sine
curve to the ten most recent cycles, and results in a period of 376 days
and a Julian Date of last maximum of 2457372. 
We indicated in Fig.~\ref{fig:lightcurve} the
mean epochs of our observations, that is,  AMBER at a mean JD 2456237
and PIONIER at mean JDs 2456901 and 2457357,
corresponding to a maximum phase of 0.98 for our AMBER observations
and pre-maximum \& maximum phases of 0.75 \& 0.96 for our PIONIER
observations.
The visual lightcurve 
shows an observed amplitude of about 1.5--2\,mag and a regular sinusoidal 
pulsation. This lightcurve
may be more typical for a Mira variable than for a SR variable. 
In fact, carbon-rich stars generally have a lower visual amplitude
than oxygen-rich stars of the same bolometric amplitude,
because molecules in the carbon-rich case have a lower temperature 
sensitivity and lower opacity in the visual compared to oxygen-rich
stars. As a result, carbon stars are often classified as
SR variables while oxygen-rich stars of the same
bolometric lightcurve are classified as Miras \citep[e.g.,][]{Kerschbaum1992}.
We placed R~Scl on the diagram of the period-luminosity ($P$-$L$) 
sequences by \citet[][Fig. 10]{Wood2015}. 
This diagram is for the Large Magellanic Cloud (LMC), but 
\citet{Whitelock2008} did not find significantly different $P$-$L$ 
relations for different metallicities. We corrected the 
apparent $J$ and $K$ magnitudes from the 2\,MASS catalog
\citep{Cutri2003} to the distance of the LMC,
and obtained a $W_\mathrm{JK}$ value \citep[cf.][]{Wood2015} 
of 9.1$\pm$0.4. Together with the pulsation period, R~Scl 
is located on the sequence that corresponds to fundamental
mode pulsation.
Here, the long
pulsation period of 376 days excludes overtone modes,
which have significantly lower pulsation periods below about
250\,days. Fundamental mode pulsation is again more typical for a Mira 
variable than for a SR variable, although \citet{Wood1999} 
noted that some SRs pulsate in fundamental mode.

We integrated available spectro-photometry. We used $B$ and $V$ 
magnitudes from the SIMBAD (Set of Identifications, Measurements,
and Bibliography for Astronomical Data)
database \citep{Wenger2000}, 
the mean $J$, $H$, $K$, $L$ magnitudes 
from \citet{Whitelock2006},
the $M$ magnitude from \citet{Lebertre1992}, and the IRAS 12\,$\mu$m,
25\,$\mu$m, 60\,$\mu$m, 100\,$\mu$m fluxes from \citet{Beichmann1985},
adopting $A_V=0.07$ from \citet{Whitelock2008} and using the reddening
law from \citet{Schlegel1998}. We obtained
a bolometric flux of $f_\mathrm{bol}=1.28 10^{-9}$\,W/m$^2$, with an 
assumed error of 5\%. 
\subsection{AMBER observations and data reduction}
\label{sec:obs_amber}
\begin{figure}
  \includegraphics[width=\hsize]{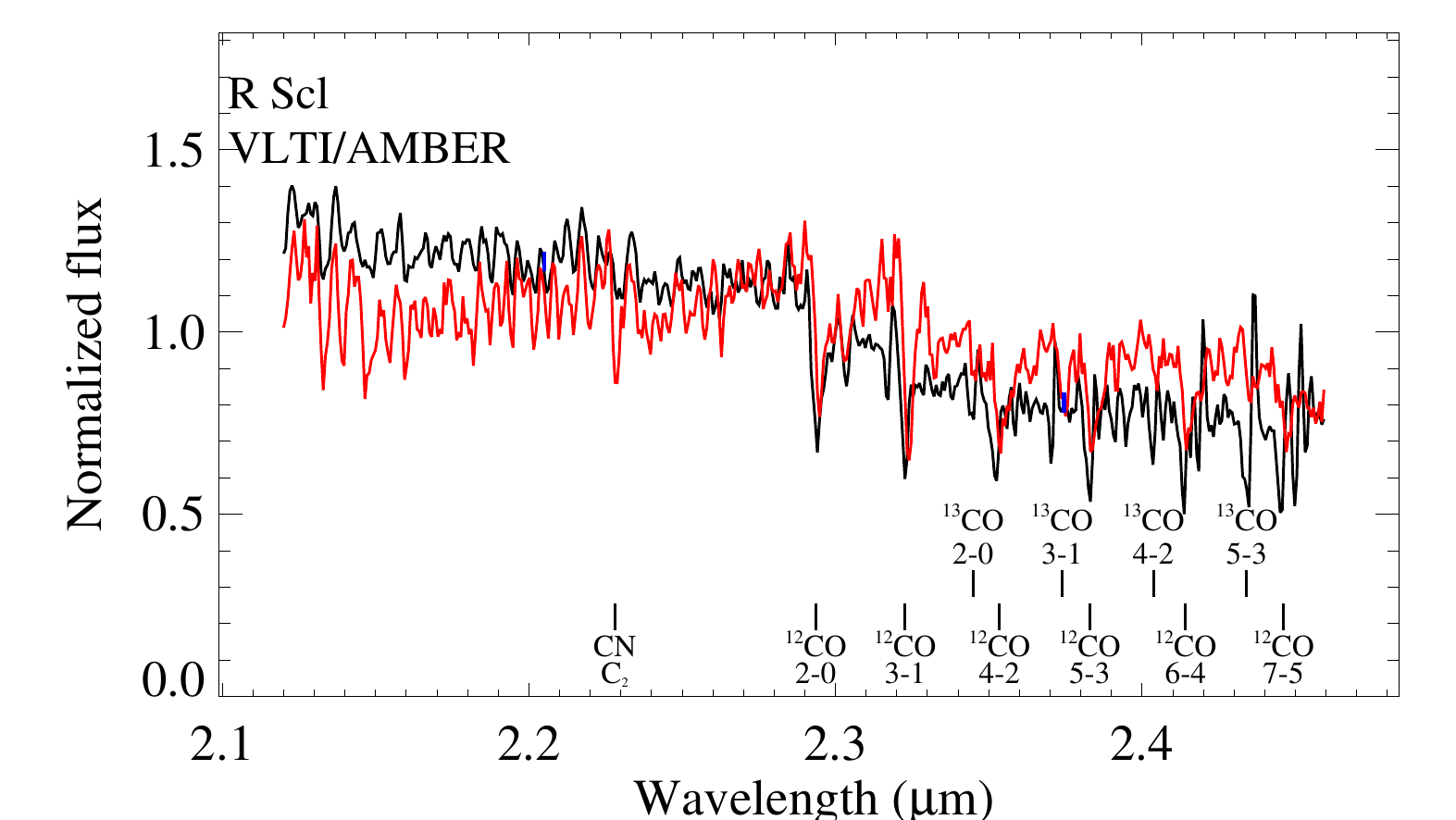}
\caption{AMBER flux spectrum (black)
compared to the prediction by the best-fit dynamic model (red).
The blue bars denote the mean errors for the first and second halves 
of the wavelength range. Instrumental and telluric signatures
are calibrated and removed.}
\label{fig:amber_flux}
\end{figure}
\begin{figure}
  \includegraphics[width=0.9\hsize]{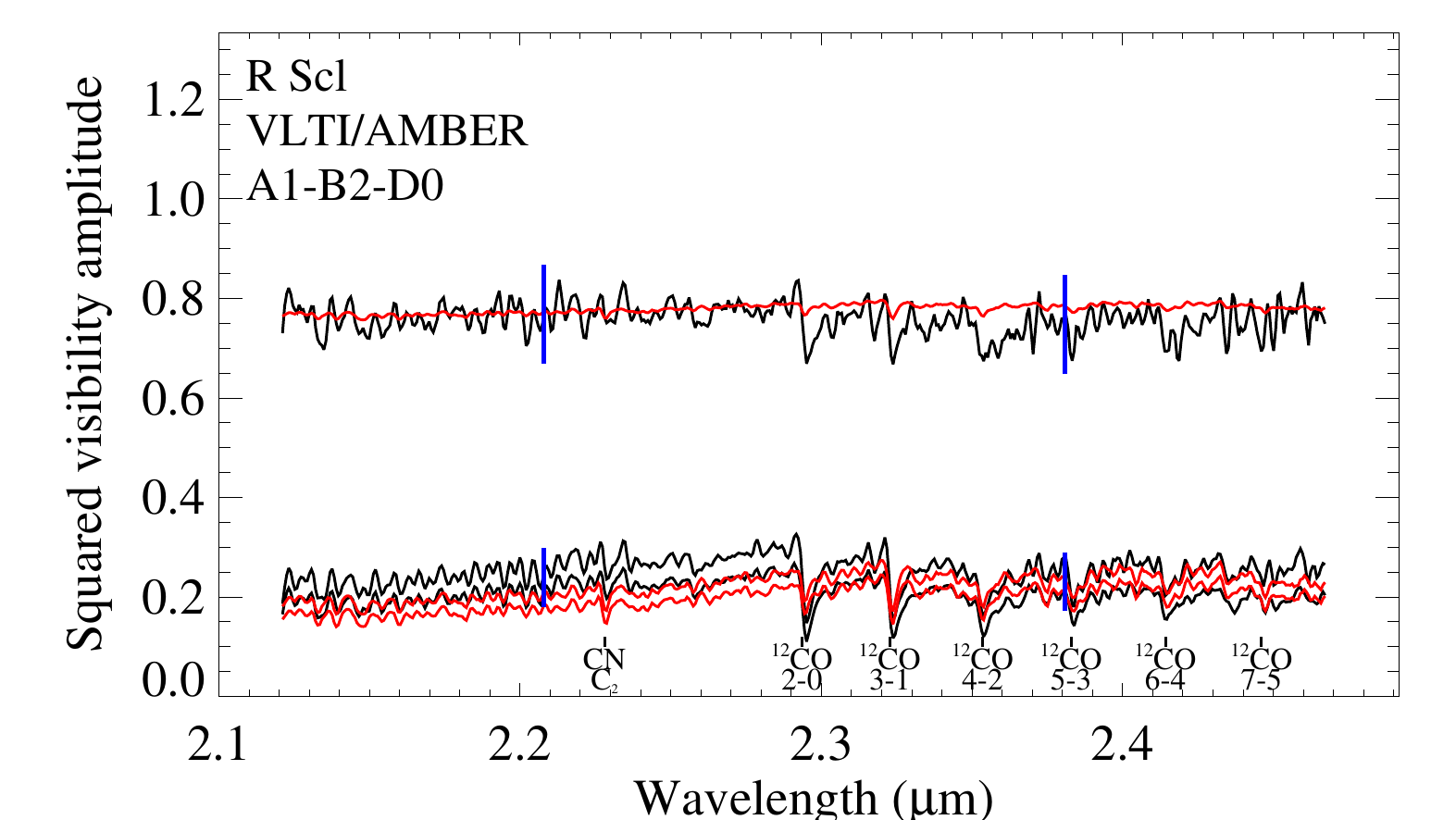}
  \includegraphics[width=0.9\hsize]{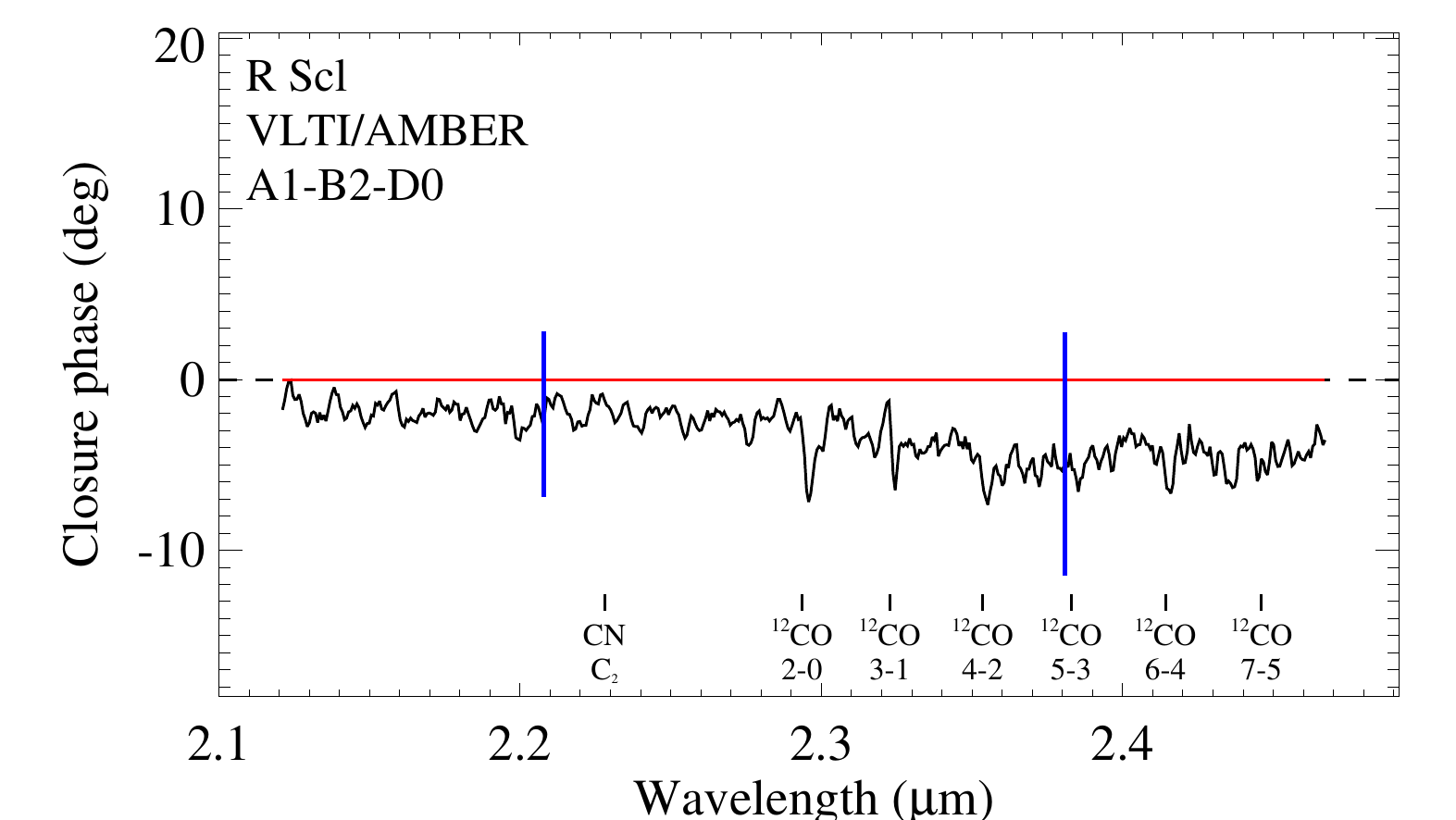}

  \includegraphics[width=0.9\hsize]{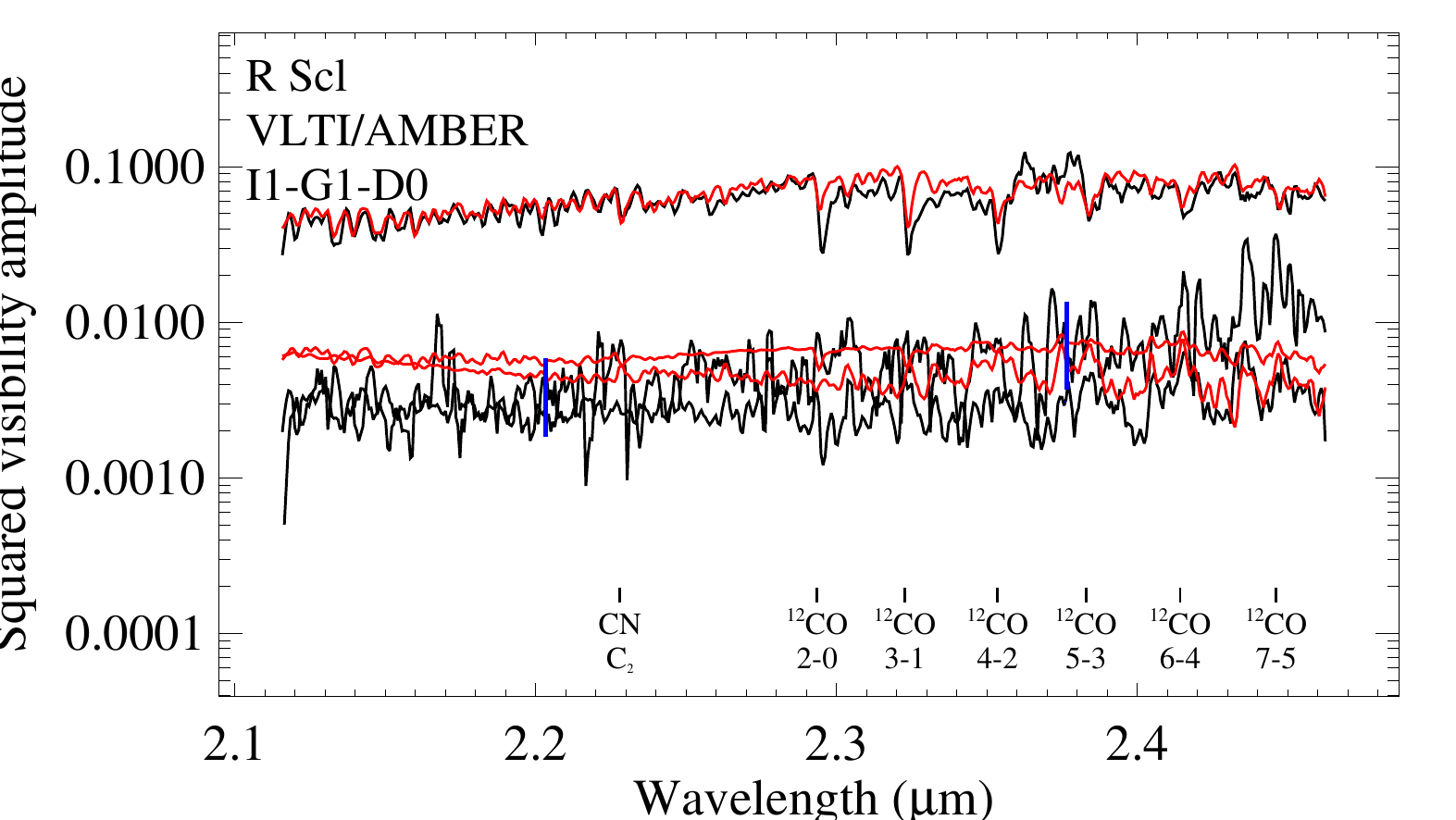}
  \includegraphics[width=0.9\hsize]{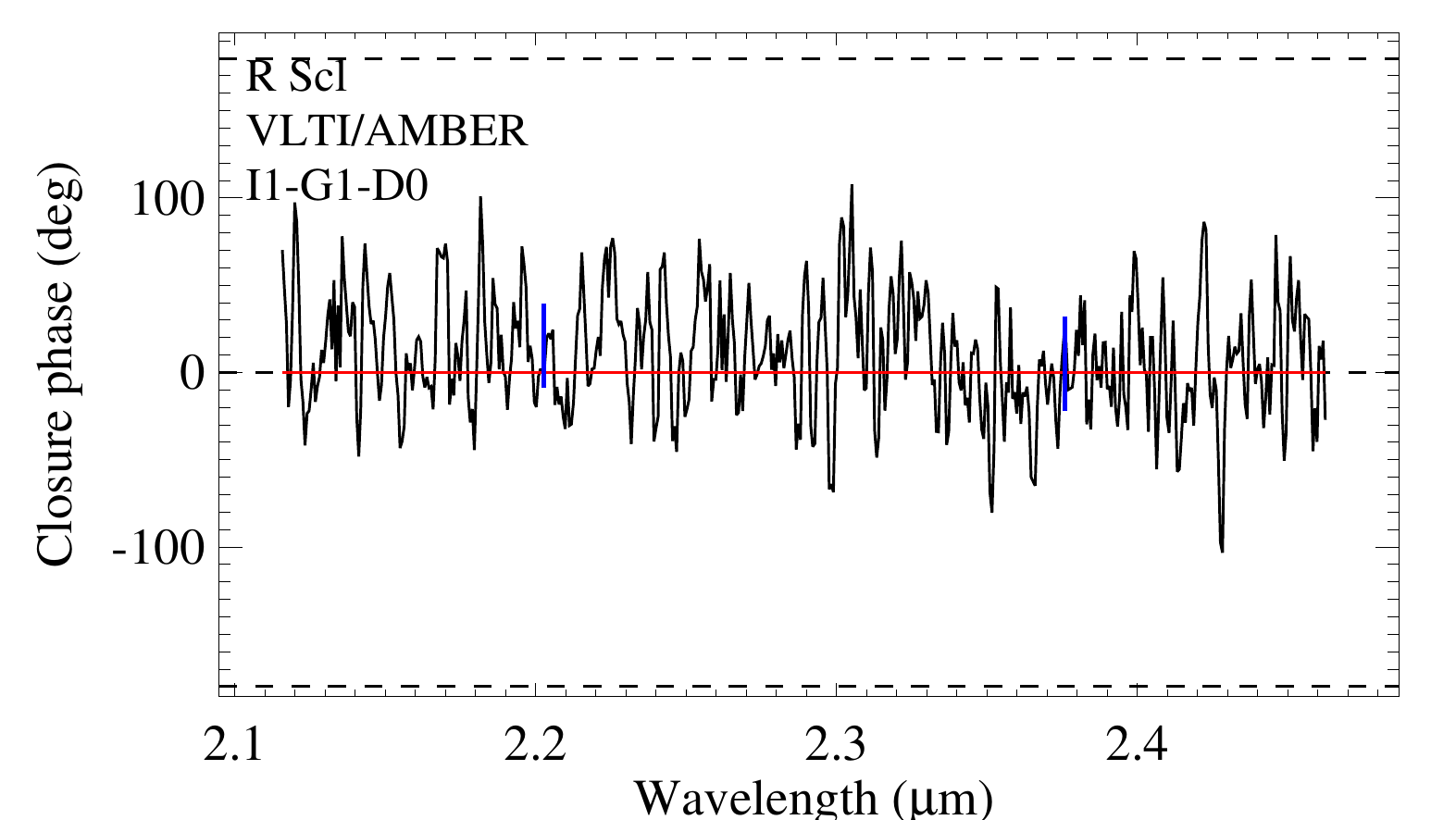}
\caption{AMBER results (black)
compared to the prediction by the best-fit dynamic model (red).
The two top panels show the squared visibilities
and closure phases versus wavelength for the example of a 
compact baseline configuration (data set no. 5
in Table~\protect\ref{tab:obs_amber}), followed by
an example of an extended configuration (data set no. 9).
The blue bars denote the
mean errors for the first and second halves of the wavelength range.
The three lines correspond to the three baselines of the
baseline triplet.
}
\label{fig:amber_visspectra}
\end{figure}
\begin{figure}
  \includegraphics[width=\hsize]{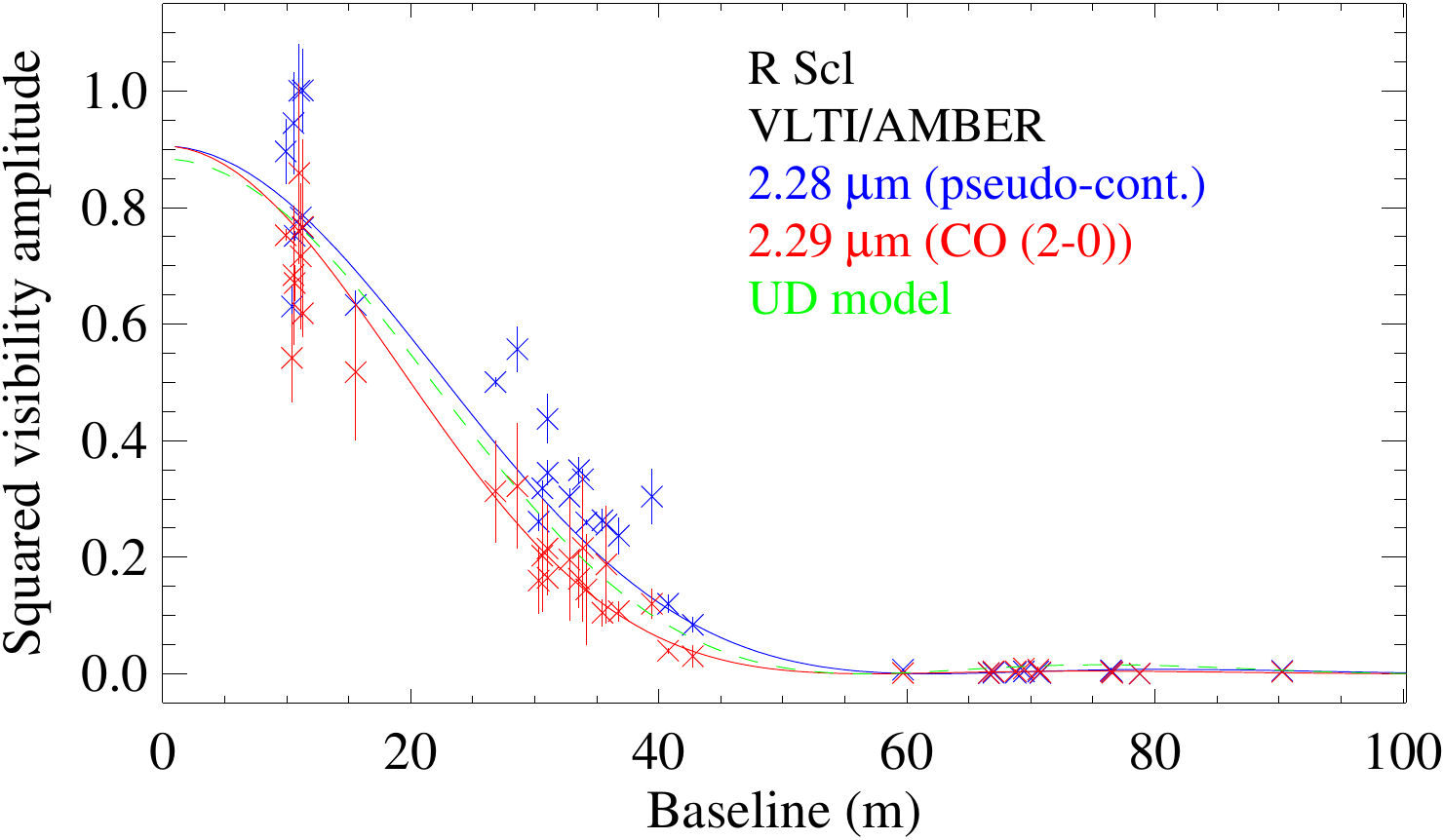}
  \includegraphics[width=\hsize]{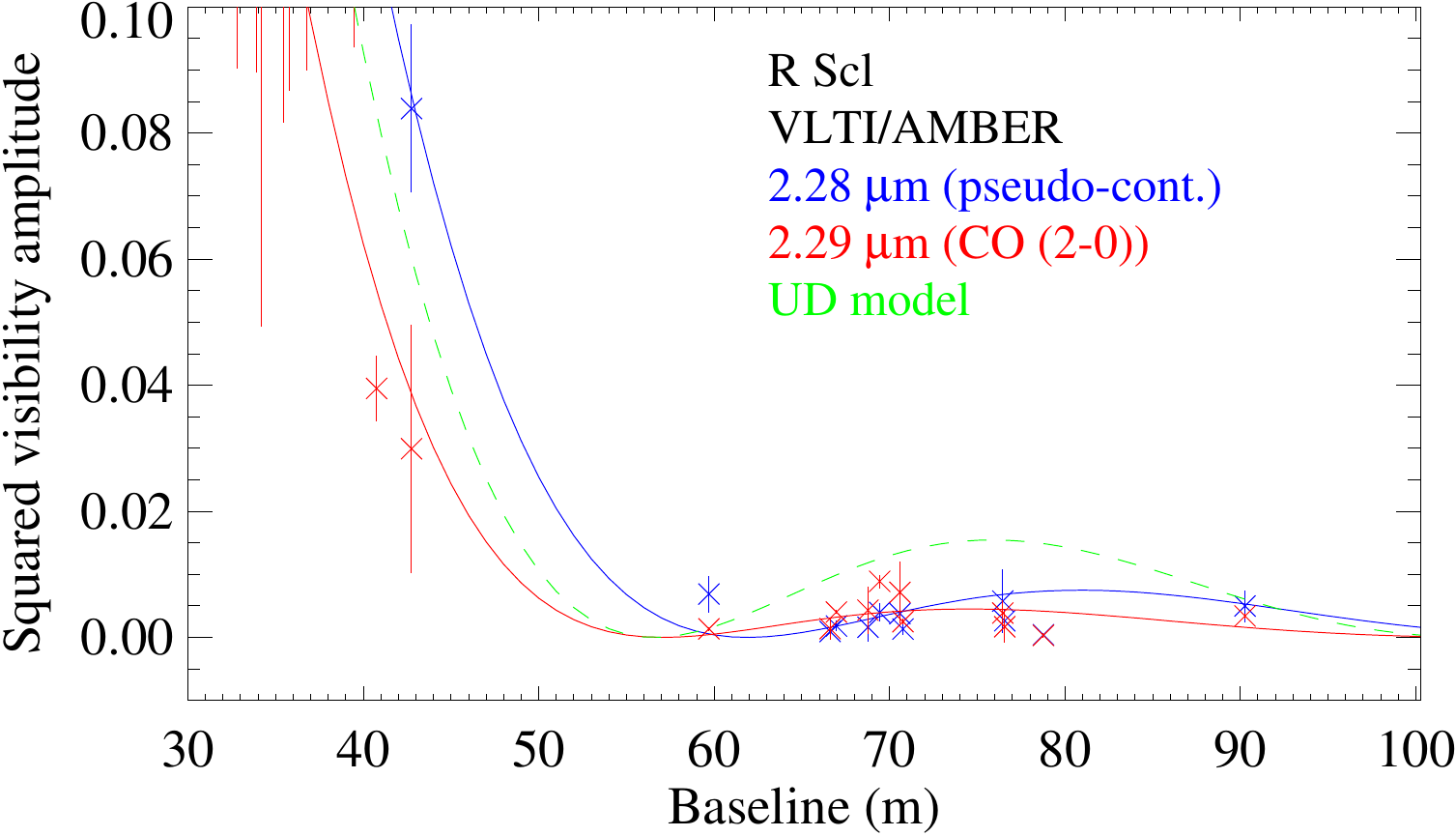}
\caption{AMBER visibility results as a function of baseline length
for the examples of bandpasses in the CO (2-0) line at 2.29\,$\mu$m
and the nearby pseudo-continuum at 2.28\,$\mu$m. The points with
error bars denote the observations. The solid lines denote
the synthetic visibility based on the best-fit model atmosphere.
The fit was derived from the full data set.
The green dotted line
shows the result based on the uniform disk (UD) fit.
The upper panel shows the full visibility range. The lower
panel shows an enlargement of the small visibility values
at long baselines.}
\label{fig:amber_visspfr}
\end{figure}
We obtained service mode observations of R~Scl with the AMBER instrument \citep{Petrov2007}
in the $K$-band 2.3\,$\mu$m medium resolution mode employing external 
FINITO fringe tracking between October and December 2012.
Table~\ref{tab:obs_amber} shows the details of our AMBER observations, 
including the dates, the Julian Day, the baseline configurations
with the projected baseline lengths and angles, and the interferometric
calibrators used for each observation. Observations of R~Scl were 
interleaved with observations of calibrators in sequences of 
a calibrator observation taken before the science target observation, the
science target observation, and another calibrator observation taken 
afterward.
Calibrators were selected 
with the ESO Calvin tool, and their angular $K$-band diameters were adopted 
from \citet{Lafrasse2010} (HR~109: 3.06\,$\pm$\,0.22\,mas;
$\chi$~Phe: 2.69\,$\pm$\,0.03\,mas; $\iota$Eri: 2.12\,$\pm$\,0.02\,mas).
Each of the observations typically consists of five files of 170 scans each, 
accompanied by a dark and a sky file.
It is essential for a good calibration of the interferometric transfer 
function that the performance of the FINITO fringe tracking is comparable 
between corresponding calibrator and science target observations. 
To this purpose, we inspected the FINITO lock ratios and the FINITO phase 
rms values as reported in the fits headers of each file 
\citep[cf.][]{LeBouquin2009b,Merand2012}. In a few cases, 
we deselected individual files for which these values deviated significantly
from the mean of the calibrator and science files.
In a few cases, we had to discard a complete data set because the FINITO phase
rms values were systematically very different for all files between science 
target observation and calibrator observations.

We obtained averaged visibility and closure phase values from the raw data
using the latest version, 3.0.9, of the {\tt amdlib}
data reduction package \citep{Tatulli2007,Chelli2009}. The absolute wavelength
calibration and the calibration of the interferometric transfer function
were performed using IDL (Interactive Data Language) scripts. For the absolute 
wavelength calibration, we correlated the AMBER flux spectra with a
reference spectrum that included the AMBER transmission curve, the telluric
spectrum estimated with ATRAN \citep{Lord1992}, and the expected stellar
spectrum using the spectrum of the K giant BS~4371 from
\citet{Lancon2000}, which has a similar spectral type as our calibrators. 
Moreover, we performed a relative flux calibration
using the calibrator stars and the BS~4371 spectrum to remove the 
instrumental and telluric signatures.
For calibrating the interferometric transfer
function, we used an average transfer function of the calibrator observations
obtained before and after that of the science target. 
The error of the final visibility spectra includes the statistical noise
of the raw data and the systematic error of the transfer function calculated
as the standard deviation between the two calibrator observations.
The resulting calibrated flux and visibility data are shown in 
Figs.~\ref{fig:amber_flux}--\ref{fig:amber_visspfr} and are discussed
in Sect.~\ref{sec:results_amber}
together with the atmosphere model results.
\subsection{PIONIER observations and data reduction}
\label{sec:obs_pionier}
\begin{table}
\centering
\caption{Log of our PIONIER observations.\label{tab:obs_pionier}}
\begin{tabular}{lrrlr}
\hline\hline
Target          & Date          & JD\tablefootmark{a} & Baseline    & \# obs.\tablefootmark{b}\\\hline 
R~Scl           & 2014-08-25    & 56895               & A1/G1/J3/K0 & 10       \\      
R~Scl           & 2014-08-26    & 56896               & A1/G1/J3/K0 & 7        \\      
R~Scl           & 2014-09-02    & 56903               & D0/G1/H0/I1 & 15       \\      
R~Scl           & 2014-09-06    & 56907               & A1/B2/C1/D0 & 11       \\      
R~Scl           & 2014-09-07    & 56908               & A1/B2/C1/D0 & 14       \\      
R~Scl           & 2015-11-28    & 57355               & A0/G1/J2/J3 & 1        \\      
R~Scl           & 2015-12-02    & 57359               & A0/B2/C1/D0 & 1        \\      
\hline
\end{tabular}
\tablefoot{
\tablefootmark{a}{JD-2450000.}\\
\tablefootmark{b}{Each observation consists typically of five consecutive measurements. 
Each of these measurements contains six squared visibility values and four closure phases
at three (2014) or six (2015) spectral channels.}
}
\end{table} 
\begin{figure}
  \includegraphics[width=0.95\hsize]{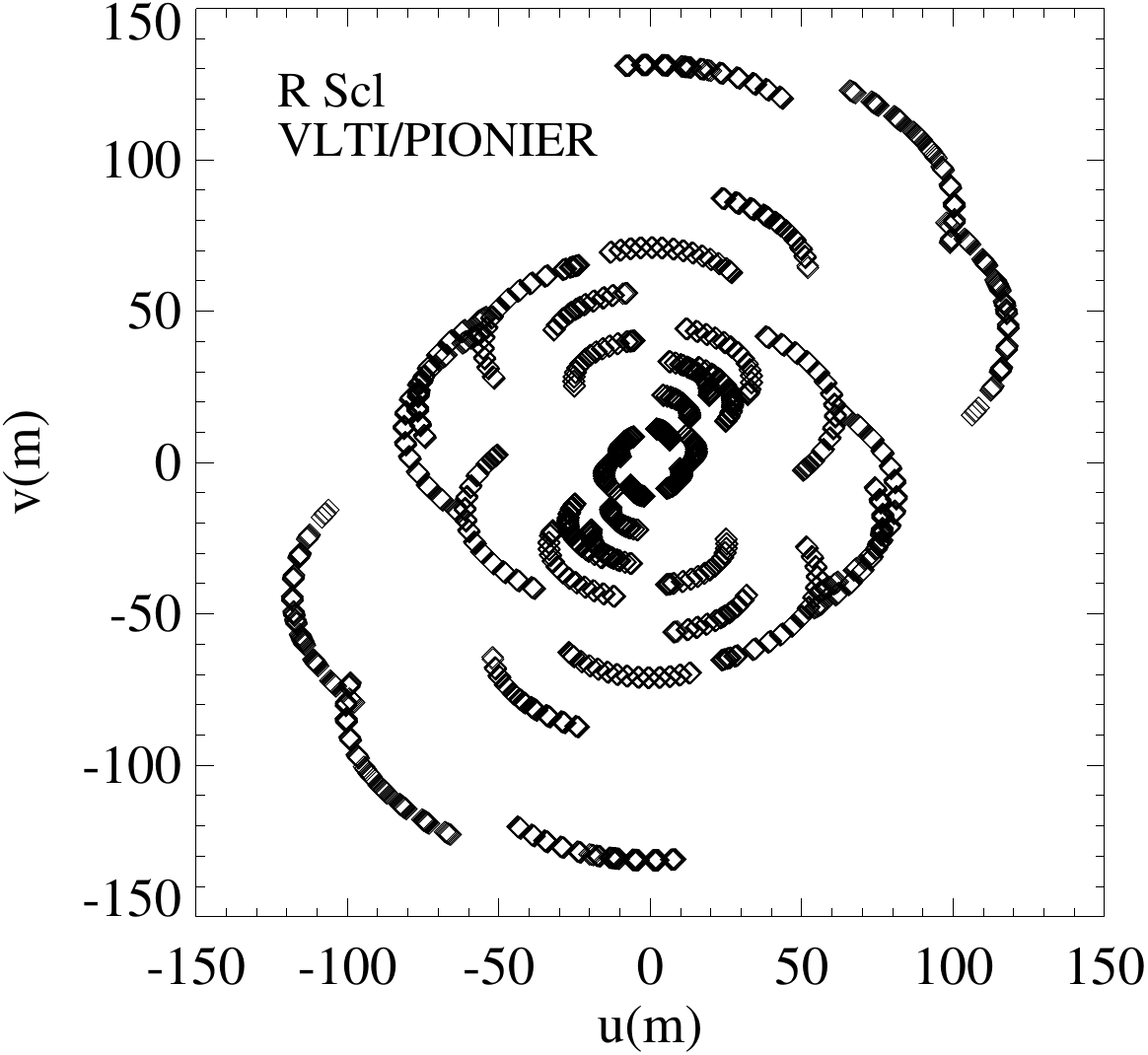}
  \includegraphics[width=0.95\hsize]{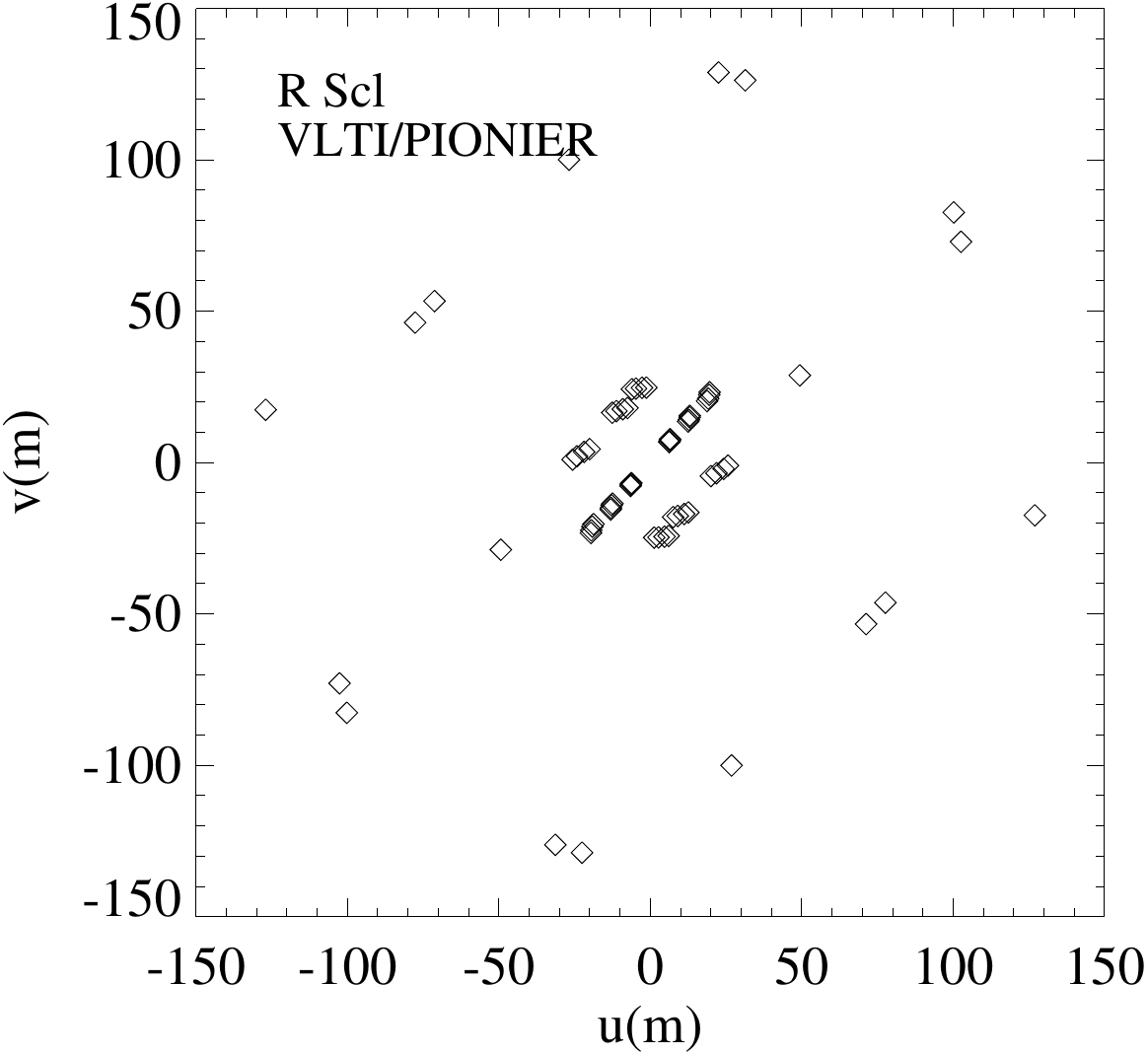}
\caption{The $uv$ coverage obtained for our PIONIER observations of R~Scl
in 2014 (top) and 2015 (bottom).}
\label{fig:pionier_uv}
\end{figure}
We obtained interferometry of R~Scl with the PIONIER instrument 
\citep{LeBouquin2011} and the four auxiliary telescopes (ATs) of the 
Very Large Telescope Interferometer (VLTI)
between 25 August and 7 September 2014 in designated visitor mode,
and on 29 November and 3 December 2015 in service mode. The 2014 observations 
were taken with the original PICNIC camera \citep{LeBouquin2011} and data 
were dispersed over three spectral channels with central wavelengths 
1.59\,$\mu$m, 1.68\,$\mu$m, 1.76\,$\mu$m, and channel widths 
$\sim$\,0.09\,$\mu$m.  In 2014, we used all three available baseline 
configurations, where the ATs were positioned on stations A1/B2/C1/D0 
(``small'', ground baseline lengths between 11.3\,m and 35.8\,m), 
D0/G1/H0/I1 (``medium'', 40.8--82.4\,m), and A1/G1/J3/K0 
(``large'', 56.6--140.0\,m).
The 2015 observations were taken after the upgrade to the RAPID camera 
\citep{Guieu2014} and data were dispersed over six spectral channels 
with central wavelengths 1.53\,$\mu$m, 1.58\,$\mu$m, 1.62\,$\mu$m,
1.68\,$\mu$m, 1.72\,$\mu$m, 1.76\,$\mu$m, and channel widths 
$\sim$\,0.05\,$\mu$m. In 2015, we obtained data on only two of the three 
baseline configurations, where the ATs were positioned on stations 
A0/B2/C1/D0 (``small'', 11.3--33.9\,m) and A0/G1/J2/J3 
(``large'', 58.2--132.4\,m), and we obtained only one observation per
configuration.
Observations of the science
targets were interleaved with observations of interferometric 
calibrators. We selected calibrators from the ESO calibrator
selector CalVin, which is based on the catalog by \citet{Lafrasse2010}.
The calibrators include 
HD~6629 (spectral type K4\,III, 
$\Theta_\mathrm{UD}^\mathrm{H}=$1.36\,$\pm$\,0.01\,mas,
used during 2014-08-25, 2014-08-26, 2014-09-02, 2014-09-06, 2014-09-07),
HR~400 (M0\,III, 2.45\,$\pm$\,0.01\,mas, 2014-08-25),
$\xi$~Scl (K1\,III, 1.32\,$\pm$\,0.02\,mas, 2014-08-25),
HD~8887 (K0\,III, 0.52\,$\pm$\,0.02\,mas, 2014-08-26, 2014-09-02, 
2014-09-06, 2014-09-07, 2015-11-28, 2015-12-02),
HD~9961 (K2\,III, 0.87\,$\pm$\,0.01\,mas, 2014-08-26, 2014-09-02, 
2014-09-06, 2014-09-07),
HD~8294 (K2\,III, 0.70\,$\pm$\,0.02\,mas, 2015-11-28, 2015-12-02), and
HR~453 (K0\,III, 0.90\,$\pm$\,0.06\,mas, 2015-12-02).
Table~\ref{tab:obs_pionier} lists the details of our R~Scl PIONIER 
observations.
Figure~\ref{fig:pionier_uv} shows the $uv$ coverage that we obtained
in 2014 and 2015.

\begin{figure*}
  \includegraphics[width=0.49\hsize]{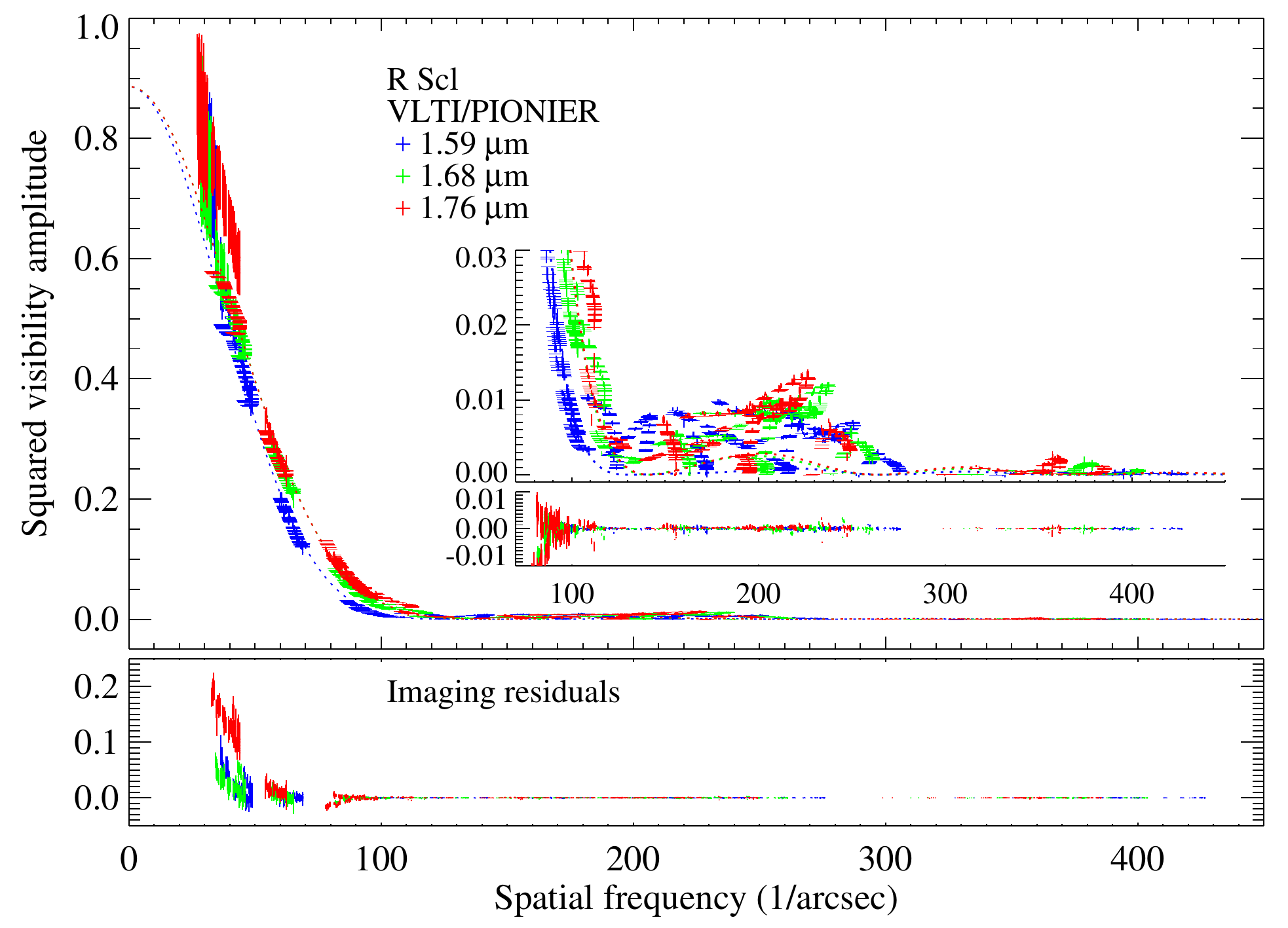}
  \includegraphics[width=0.49\hsize]{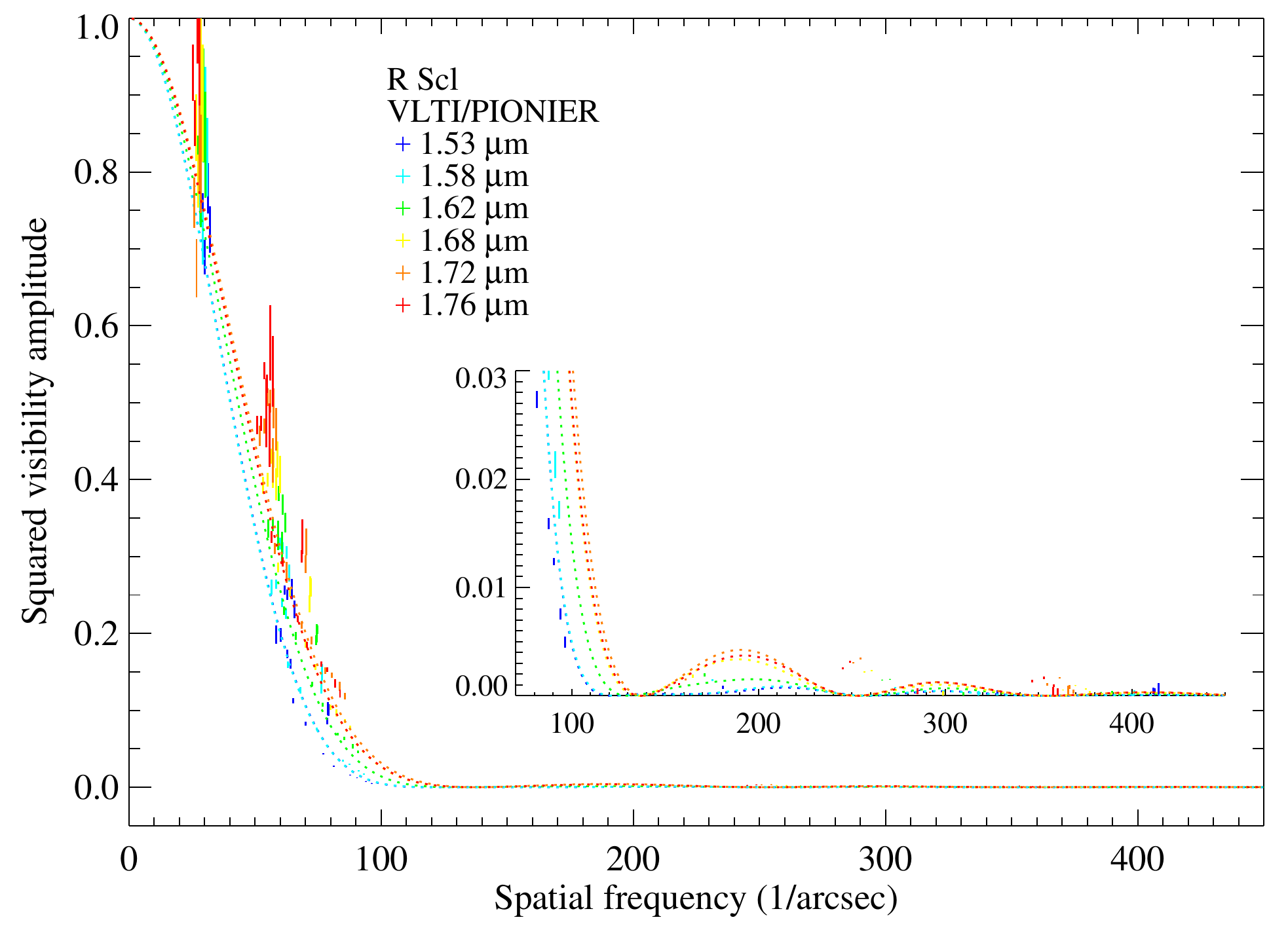}

  \includegraphics[width=0.49\hsize]{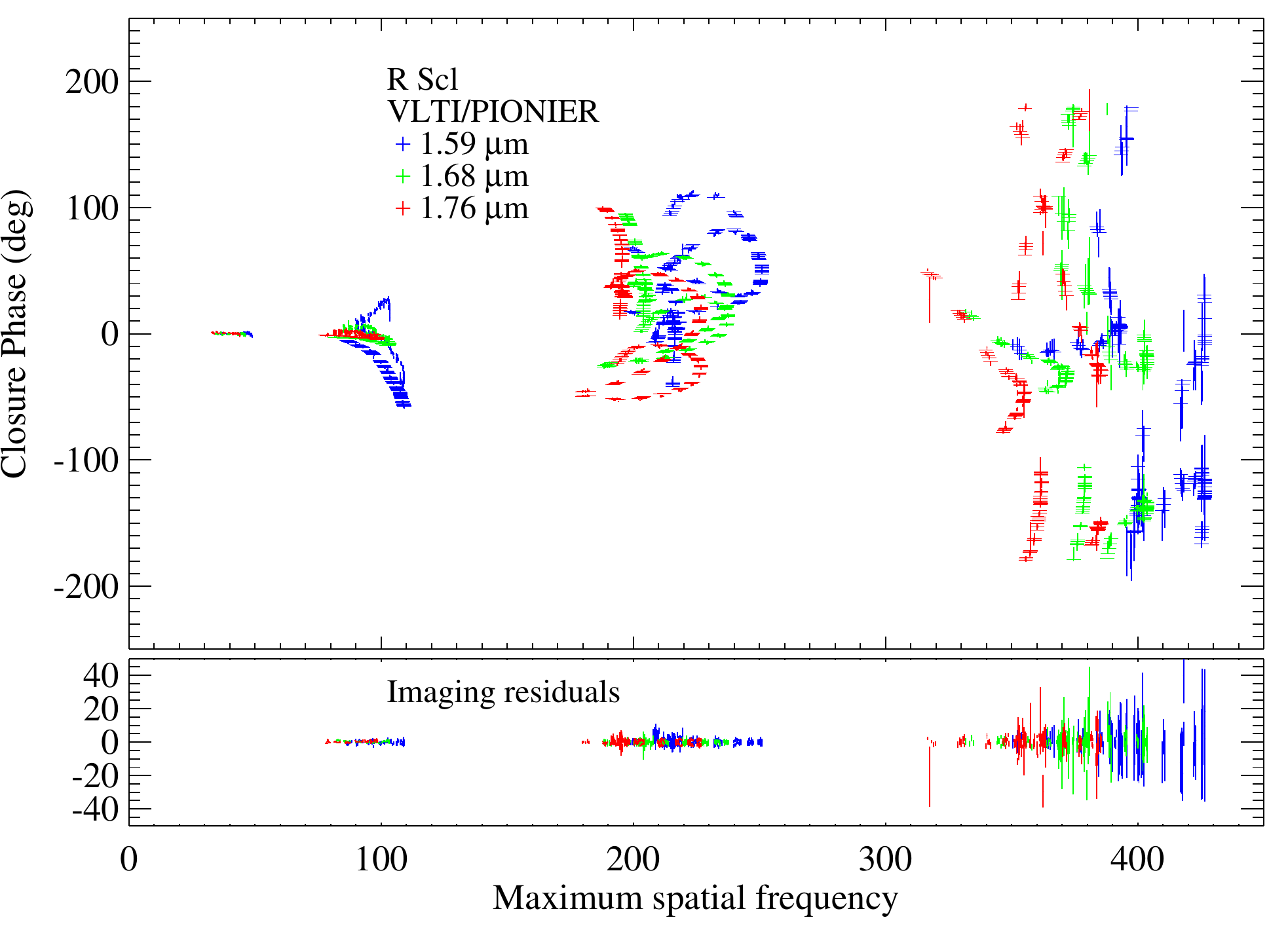}
  \includegraphics[width=0.49\hsize]{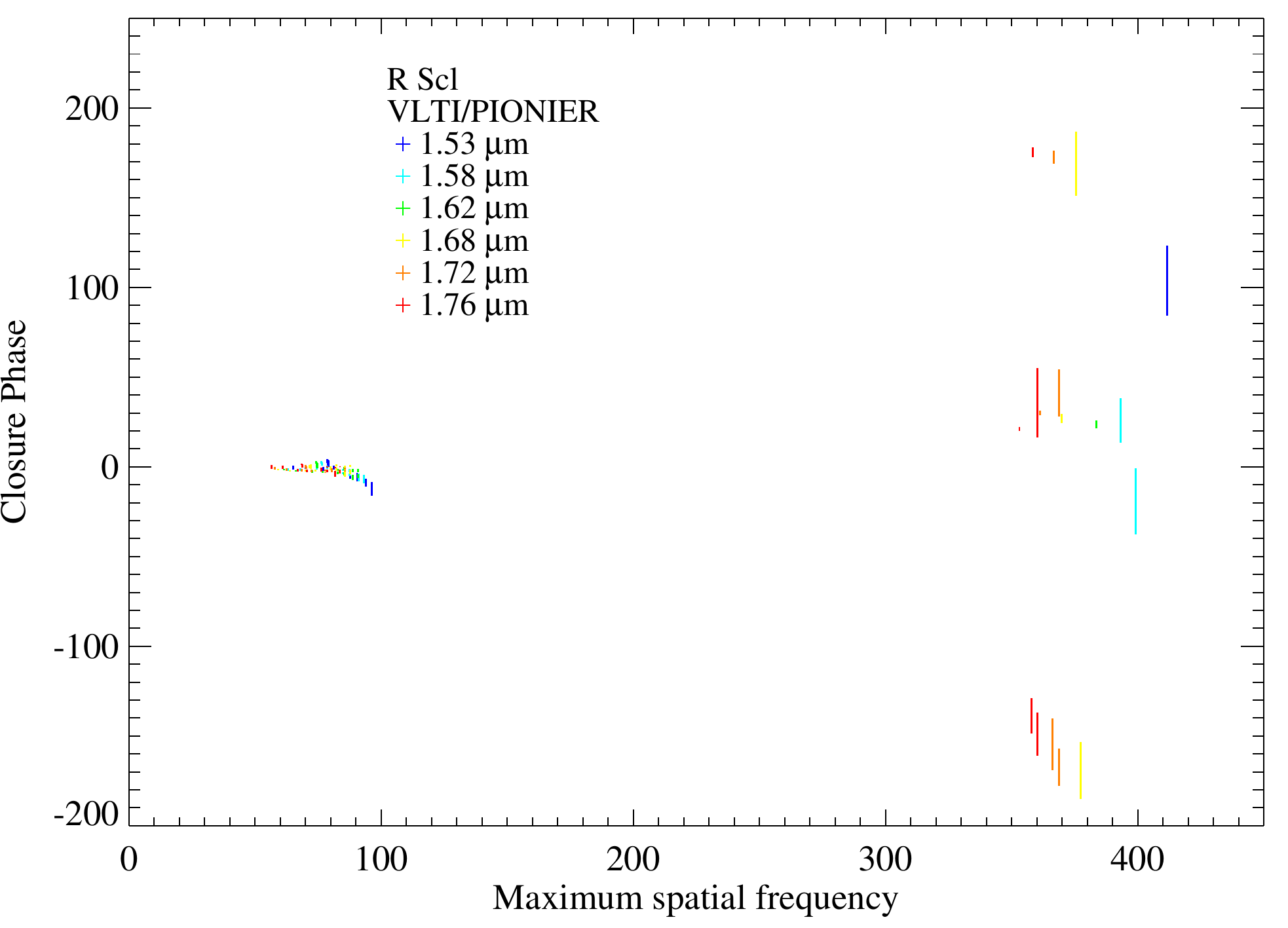}
\caption{PIONIER results of R~Scl as a function of spatial
frequency including the squared visibility values (top) and the
closure phases (bottom). The left panels shows the results obtained
in 2014, and the right panels the results obtained in 2015.
The vertical bars indicate the symmetric error bars. The central
positions of the measured values are omitted for the sake
of clarity.
The different colors denote the different
spectral channels.
The dashed lines indicate the best-fit model atmospheres.
For the 2014 data, the synthetic values based on the reconstructed
images are shown by horizontal bars. Here, the lower small panels
provide the residuals between observations and reconstructed images.
}
\label{fig:pionier2014_visspfr}
\end{figure*}

We reduced and calibrated the data with the {\tt pndrs} package 
\citep{LeBouquin2011}. Each of the listed observations
typically consists of five consecutive measurements. Each of these results
in six squared visibility amplitudes and four closure phases at 
three (2014) or six (2015) spectral channels. Not all of these
points passed the quality checks that are included in the
{\tt pndrs} package. In total, we secured 4959/198 valid squared visibility
amplitudes and 3103/113 valid closure phase values in 2014/2015 on R~Scl.
The resulting calibrated visibility values are shown in 
Fig.~\ref{fig:pionier2014_visspfr} and are discussed 
below in Sect.~\ref{sec:results_pionier}
together with the atmosphere model results. The quantity and quality
of the 2014 data allowed us to reconstruct images of R~Scl, which
are described in Sect.~\ref{sec:imaging_pionier}. The 2015
data are too few to reconstruct images.

In addition to R~Scl, we secured data on the resolved 
K5/M0 giant $\upsilon$~Cet as a ``check star'' in order to assess the 
validity and accuracy of the visibility and closure phase data, in 
particular for the low fringe contrasts, as well as the consistency 
between different nights and between the different epochs in 2014 and 2015.
The details of these observations are described in 
appendix~\ref{sec:checkstar}. These observations confirmed a high
accuracy of the low visibility values close to and beyond
the first minimum of the visibility function and of the closure
phase data. These data allowed us to confirm the 
strength of the model-predicted limb-darkening effect with high
significance. However, the high visibility data at short baselines
lay systematically above the model prediction, which is a known
effect of PIONIER calibrations and most likely caused by different 
magnitudes or airmass between science target and calibrators
(see appendix~\ref{sec:checkstar} for more details). 
In the following, we took this known
effect into account for the modeling and imaging processes of R~Scl by 
excluding the short baseline visibility data (baselines shorter than 
$B=11.5$\,m), and by accepting that high synthetic visibility values
based on the best-fit model prediction or the reconstructed image
may lie below the measured values.
\section{Atmosphere modeling}
\label{sec:models}
We employed the atmosphere model grid for carbon-rich AGB
stars constructed by \citet{Mattsson2010} and \citet{Eriksson2014}.
In particular, we used the data by \citet{Eriksson2014},
which correspond to a sub-sample of the \citet{Mattsson2010} grid
in terms of stellar parameters, focusing on typical values
of solar-metallicity carbon-rich AGB stars. The models assume 
spherical symmetry and cover dynamic models
with effective temperatures between 2600\,K to 3200\,K,
luminosities
$\log L/L_\sun$ between 3.55 and 4.00, masses between
0.75\,$M_\sun$ and 2.0\,$M_\sun$, and C/O abundance ratios of
1.35, 1.69, and 2.38. The corresponding pulsation periods lie
between 221 days and 525 days. The effects of stellar
pulsation are simulated by a variable inner boundary
(so-called piston models), characterized by
velocity 
amplitudes $\Delta u_p$ of 2\,km/s, 4\,km/s, and 6\,km/s.
Finally, a parameter $f_L$ describes whether the
original luminosity amplitude 
\citep[as described in][]{Hoefner2003} was used,
or twice the luminosity amplitude of the \citet{Mattsson2010} grid, 
following \citet{Nowotny2010}, to better describe observed photometric 
variations. 
We note that the stellar parameters mentioned above are those
of the hydrostatic starting model.
Models were calculated over different cycles. 
A typical model run for a given set of stellar and pulsation
parameters covers several hundred pulsation phases.
Of the total of 540 models, some of the models turn out to produce
a wind (229), and some do not produce a wind (311).
\citet{Eriksson2014} performed detailed
radiative transfer calculations to obtain low
resolution spectra in the wavelength range 0.35--25\,$\mu$m
and a number of filter magnitudes. For this calculation, they 
used the opacity generation code {\tt COMA} \citep{Aringer2009} 
to compute continuous atomic, molecular, and amorphous carbon (amC) 
dust opacities for all atmospheric layers and each wavelength point, 
assuming local thermodynamic equilibrium (LTE) and assuming 
the small particle limit for the dust opacities.

\begin{table*}
\label{tab:cons_models}
\centering
\caption{Parameters of the considered dynamic models.}
\begin{tabular}{c|llllllllll|rrr}
\hline\hline
No. & $T_\mathrm{eff}$ & $L$ & $M$         & $\log g$ & $C/O$ & $\Delta u_p$ & $f_L$ & $\overline{\dot{M}}$ & $\overline{u_\mathrm{inf}}$ & $\overline{P}$ & \multicolumn{3}{c}{$\chi^2_\nu$} \\
    & (K)              & $_\sun$           & ($M_\sun$) &  cgs     &       & (km/s)       &       & ($M_\sun$/yr)       & (km/s)                      & (d)            & A   & P\,1   & P\,2  \\\hline
1   & 2600             & 5000             & 0.75        & -0.79    & 1.35  & 6            & 1     & 3.9e-07             & 2.3                         & 294            & 1.33   &  28.8     & 53.5 \\
2a  & 2600             & 5000             & 1.0         & -0.66    & 1.69  & 4            & 1     & 7.4e-07             & 6.5                         & 294            & 1.61   & 112.0     & 94.2 \\
2b  & 2600             & 5000             & 1.0         & -0.66    & 1.69  & 4            & 1     & 7.4e-07             & 6.5                         & 294            & 1.44   &  90.1     & 80.4 \\ 
2c  & 2600             & 5000             & 1.0         & -0.66    & 1.69  & 4            & 1     & 7.4e-07             & 6.5                         & 294            & 1.45   &  86.7     & 77.2 \\
3a  & 2800             & 7000             & 1.0         & -0.69    & 1.35  & 4            & 1     &         /           &      /                      & 390            & 1.28   &  31.8     & 43.4  \\
3b  & 2800             & 7000             & 1.0         & -0.69    & 1.35  & 4            & 1     &         /           &     /                       & 390            & 1.28   &  32.1     & 42.4  \\
\hline
\end{tabular}
\tablefoot{The last three columns indicate the $\chi^2_\nu$ of the 
best-fit snapshot of each model for the AMBER (A) and PIONIER (P1 and P2) epochs.}
\end{table*}

From this model grid, we pre-selected models that 
are broadly consistent with the known parameters of R~Scl,
the available broad-band photometry (cf.~Sect.~\ref{sec:lightcurve}),
and the available Infrared Space Observatory (ISO) spectra from 
the Short Wavelength Spectrometer (SWS) \citep{Sloan2003}. The parameters 
of the considered models are listed in Table~\ref{tab:cons_models}. 
We employed three models, and for two of them we selected snapshots
at two or three different cycles.
Every entry in Table~\ref{tab:cons_models} contains
40--60 snapshots at continuous phases. 
In addition to the stellar parameters described above, 
Table~\ref{tab:cons_models} also lists the corresponding surface gravity, 
the average mass-loss rate, outflow velocity, and period.
The first two of the considered models in Table~\ref{tab:cons_models}
correspond to effective temperatures of 2600\,K 
with masses 0.75\,$M_\sun$ and 1.0\,$M_\sun$.
They have mass-loss rates of $3.9\times10^{-7}$\,$M_\sun$/yr
and $7.4\times10^{-7}$\,$M_\sun$/yr, respectively, and periods
of 294\,d. The third model
corresponds to an effective temperature of 2800\,K, a 
mass of 1.0\,$M_\sun$ and a period of 390\,d, 
and does not produce a mass loss. 
Other models of the grid with
a higher mass or a higher effective temperature lead to 
luminosities or periods that are not consistent with the observed
values for R Scl. However, we cannot fully exclude that another
model at certain phases may also explain the data.

Using {\tt COMA}, we tabulated for every snapshot
intensity profiles between 1.4\,$\mu$m and 2.5\,$\mu$m
at intervals of 0.0001\,$\mu$m at typically 200-300 radial points
between the center and the outer boundary, which
depends on the extension of the particular model.
The wavelength intervals are chosen such that we can average
about ten wavelengths per AMBER spectral channel in its medium
spectral resolution mode.
We computed synthetic squared visibility amplitudes using
the Hankel transform and averaging over the bandpass of every
AMBER or PIONIER spectral channel, as described in detail
by, for example, \citet{Wittkowski2004}. Our primary fit parameter
is the angular diameter that corresponds to the arbitrary 
outermost model layer where the tabulated intensity profiles
stop.
We use the model structure to relate this arbitrary outermost
model layer to the model layer where the Rosseland optical depth 
is 2/3, and provide a best-fit Rosseland angular diameter
$\Theta_\mathrm{Ross}$ corresponding
to the Rosseland radius $R (\tau_\mathrm{Ross} = 2/3)$.
We also 
integrated the intensities over the stellar disk to calculate the
synthetic flux at the AMBER and PIONIER spectral channels.
\section{Results}
\label{sec:results}
\subsection{Diameter fit results}
\begin{table*}
\centering
\caption{Diameter fit results.\protect\label{tab:fitresults}}
\begin{tabular}{lrrr|rrrr|rrrr}
\hline\hline
Target         & JD             &  Phase   &  Instrument &  Model    &  $\Theta_\mathrm{Ross}$ &  $A$            & $\chi^2_\nu$ & Phase  & R$_\mathrm{Ross}$ & $\log L/L_\sun$ & $T_\mathrm{eff}$ \\
               & -2450000       &  obs.    &             &           &  mas                    &                 &              &        & ($R_\sun$)       &                  & (K)              \\\hline
R~Scl          & 6237           &  0.98    &  AMBER      &  3/72326  &  9.25 $\pm$ 0.30        & 0.95 $\pm$ 0.02 & 1.28         & 0.98   & 363               & 3.92             & 2900             \\
R~Scl          & 6901           &  0.75    &  PIONIER    &  3/72376  &  8.79 $\pm$ 0.15        & 0.96 $\pm$ 0.04 & 31.8         & 0.44   & 303               & 3.76             & 2895             \\
R~Scl          & 7357           &  0.96    &  PIONIER    &  3/76266 &  9.17 $\pm$ 0.40        & 1.00 $\pm$ 0.05 & 42.4         & 0.50   & 301               & 3.75             & 2892             \\
\hline
\end{tabular}
\tablefoot{The model designation refers to the numbers in 
Table~\protect\ref{tab:cons_models} and includes the number of the 
best-fit snapshot of the model.}
\end{table*}
As a first step, we calculated best-fit uniform-disk (UD) diameters
for our AMBER data and the two epochs of PIONIER data, and obtained
values of 10.5\,mas, 10.7\,mas, and 11.4\,mas, with reduced
$\chi^2_\nu$ values of 2.0, 191, and 106, respectively. 
We estimate the errors to be about 0.5\,mas. These values are 
consistent with the previous estimates of 10.1\,mas 
by \citet{Cruzalebes2013} based on previous AMBER data and 
of 10.2\,mas by \citet{Sacuto2011} based on MIDI data.
The $\chi^2_\nu$ values of the PIONIER fits are high because 
of the known systematic calibration problems at high visibilities 
as discussed in Sect.~\ref{sec:obs_pionier} and 
appendix~\ref{sec:checkstar}.

Then we compared our data to the dynamic atmosphere and
wind models as introduced in Sect.~\ref{sec:models}.
We used a visibility model that takes into 
account a possible over-resolved component due 
to large-scale circumstellar dust emission of the form:
\begin{equation*}
V= A * V_\mathrm{model} (\mathrm{Model}, \Theta_\mathrm{Ross}),
\end{equation*}
where $A$ is the flux fraction of the synthetic stellar 
contribution, depending on the chosen model from 
Table~\ref{tab:cons_models} and $\Theta_\mathrm{Ross}$, and
$(1-A)$ the flux fraction of an over-resolved dust component.

We added to Table~\ref{tab:cons_models} the reduced $\chi^2_\nu$ values of 
the best-fit snapshot of each model as compared to our AMBER data and our 
two epochs of PIONIER data. Table~\ref{tab:fitresults} lists the details 
of the best-fit snapshots, including the 
best-fit values of $\Theta_\mathrm{Ross}$ and $A$, as well as the
phase, luminosity, and derived effective temperature of the 
best-fit snapshot.
All dynamic models provide a better fit to both the AMBER and PIONIER 
data than the simple UD fit.

The best-fit model overall is model 3 from Table~\ref{tab:cons_models} 
with $T_\mathrm{eff}=$2800\,K, $L=$7000\,$L_\sun$, $M=$1\,$M_\sun$, that 
-- however -- does not produce a wind. The maximum outer boundary for this
model is situated a bit above two stellar radii, and the condensation degree 
for carbon is only a few 10$^{-4}$.
The weighted mean of the best-fit 
$\Theta_\mathrm{Ross}$ values of the three epochs is 8.9$\pm$0.3\,mas, 
and the flux fraction of an over-resolved dust component is 
0.05$\pm$0.02 for AMBER in the $K$-band, and 0.02$\pm$0.03 for 
PIONIER in the $H$-band. Together with the 
bolometric flux and the distance
estimates from Sect.~\ref{sec:lightcurve}, this value of 
$\Theta_\mathrm{Ross}$ corresponds to $T_\mathrm{eff}$ of 2640$\pm$80\,K,
$\log L/L_\sun$ of 3.74$\pm$0.18, and R$_\mathrm{Ross}$ of 
355$\pm$55\,$R_\sun$. These values are consistent with the theoretical
values of the best-fit snapshots, except for $T_\mathrm{eff}$, which 
is lower. The period of model 3 is closer to the period of
R~Scl than that of models 1 and 2.
We note that model 3 is the same model that was identified by
\citet{Sacuto2011} to best fit their MIDI data of R~Scl.

Model 1 provides the next-best $\chi^2_\nu$ values. 
It has values
$T_\mathrm{eff}=$2600\,K, $L=$5000\,$L_\sun$, $M=$0.75\,$M_\sun$,
and produces a mass-loss rate of $3.9\times10^{-7}$\,$M_\sun$/year with 
an outflow velocity of 2.3\,km/s. 
However, the best-fit $\Theta_\mathrm{Ross}$
values based on this model lie between 5.6\,mas and 6.1\,mas,
which is neither consistent
with the UD fit nor with the previous measurements by \citet{Sacuto2011}
and \citet{Cruzalebes2013}, nor with our imaging result shown below. 
This angular diameter would correspond to
$T_\mathrm{eff}$ as high as 3200$\pm$120\,K and R$_\mathrm{Ross}$ as
low as 230$\pm$40,$R_\sun$, which are also not consistent with 
independent estimates.
A closer inspection of the synthetic intensity profiles of this model
showed that the Rosseland angular diameter is too small,
but the CSE of this model is more extended, so that 
overall the 10\% and 30\% intensity diameters correspond to those of the 
best-fit snapshot of model~3. In other words, this 
model provides formally a similarly good fit to the data, but with a 
Rosseland radius that is too small compared to independent
estimates. This is compensated for by a CSE that
is too extended compared to our observations.
As a result, we discarded this model. In general, the considered
models that produce mass-loss (models 1 and 2) appear to have a 
CSE that is too extended compared to our
observations. This may be related to the fact that it is difficult 
to produce carbon-rich wind models corresponding to the low end of 
the observed range of mass loss rates 
\citep[see, e.g.,][for a discussion of this problem]{Mattsson2010,Eriksson2014}.

\begin{figure}
  \includegraphics[width=\hsize]{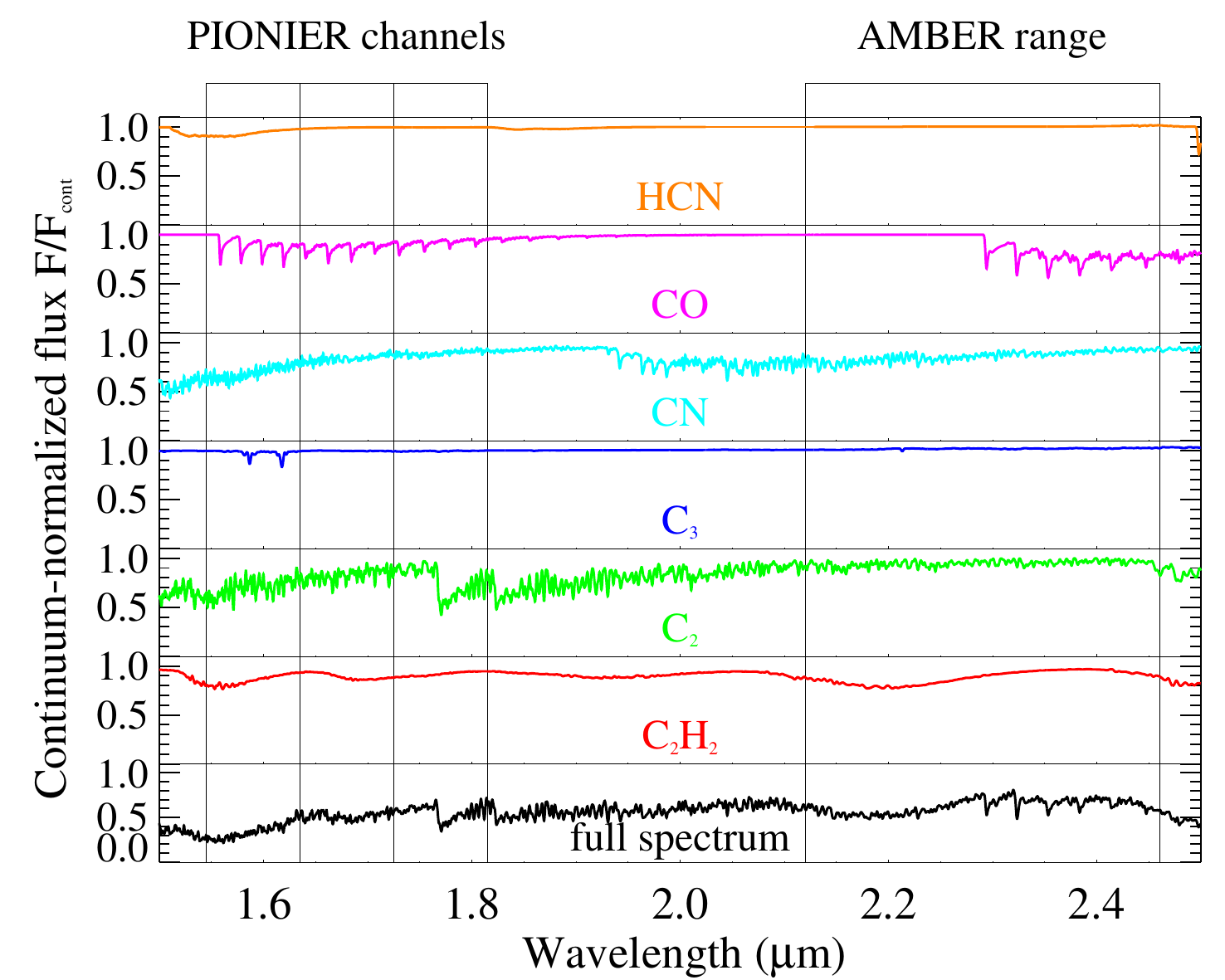}
\caption{Synthetic continuum-normalized spectra using {\tt COMA} \citep{Aringer2009}
for the most important molecular species and for the full spectrum.
The example spectra of this figure are based on the best-fit 
model snapshot to our first PIONIER epoch.
The resolution of the spectra is 1500, except for C$_3$ and 
C$_2$H$_2$, which are treated in opacity sampling.
Also indicated are the three spectral channels of our 
PIONIER observations, as well as the range
of the AMBER MR 2.3\,$\mu$m mode.}
\label{fig:mol_contrib}
\end{figure}
Spectra of carbon AGB stars do not show any genuine 
continuum within the near-IR observing bands, most importantly due to
molecular contributions by CO, CN, C$_2$, C$_2$H$_2$, HCN, 
and C$_3$ \citep[cf.,e.g.,][]{Gautschy-Loidl2004,Paladini2009}.
All spectral channels of our PIONIER and AMBER observations are thus
affected by atomic and molecular features.
As an illustration, Fig.~\ref{fig:mol_contrib} shows an example of
a synthetic continuum-normalized spectrum, together with the 
contributions by the individual molecular opacities listed above. 
This example is based on the best-fit model to our first PIONIER epoch.
We used {\tt COMA} \citep{Aringer2009} to calculate 
the spectra, using all available atomic and molecular opacities for the 
full spectrum,
and selecting one molecular opacity at a time for the 
individual molecular contributions. We show the combined 
wavelength range of our PIONIER and AMBER observations,
covering 1.5--2.5\,$\mu$m at the resolution of the AMBER
medium spectral resolution mode of 1500.
The molecules C$_3$ and 
C$_2$H$_2$ are treated in the opacity sampling method.
We indicate the three spectral channels of our PIONIER observations, 
as well as the range of the used AMBER MR 2.3\,$\mu$m mode.
\subsection{AMBER visibility results and model comparison}
\label{sec:results_amber}
Figure~\ref{fig:amber_flux} shows the observed AMBER flux 
spectrum of R~Scl
derived from an average of all AMBER data, together with 
the prediction by the best-fit dynamic model to the AMBER data. Observed and modeled flux spectra are generally consistent.
The spectrum shows an overall flat distribution between 
2.12\,$\mu$m and 2.29\,$\mu$m with a number of
features due to atomic and molecular 
lines. The most prominent feature within this wavelength range
is an absorption line near 2.228\,$\mu$m. We identified
this feature to be mostly a CN band with some contribution
from C$_2$, using {\tt COMA} runs with various variations of 
atomic and molecular abundances.
This feature is relatively weak in the observed
flux spectrum but more prominent in the visibility spectra,
indicating that it forms at extended layers above the pseudo-continuum.
Observed and modeled flux spectra between 2.29\,$\mu$m 
and 2.46\,$\mu$m clearly exhibit the $^{12}$CO (2-0, 3-1, 4-2, 5-3, 6-4)
and $^{13}$CO (2-0, 3-1, 4-2, 5-3) bandheads and show
a generally lower flux level due to CO absorption.
``Emission''-like features next to CO bandheads are residuals
of the telluric correction.

Figure~\ref{fig:amber_visspectra} shows the squared visibility
amplitudes and closure phases as a function of wavelength
for two examples, data set no. 5 from Table~\ref{tab:obs_amber}
obtained with a compact baseline configuration and data set no. 9
obtained with an extended baseline configuration.
The synthetic values based on the best-fit dynamic model
are shown as well. The shown errors are
dominated by the systematic uncertainty of the global height
of the interferometric transfer function. The wavelength-differential
errors are small compared to the systematic error, as evidenced
by the small pixel-to-pixel variations.
The observed visibility spectra are generally flat
as a function of wavelength
and consistent with the dynamic model. The observed squared visibility
values show drops at the positions of the $^{12}$CO bandheads between
2.3\,$\mu$m and 2.5\,$\mu$m as well as at the
position of the CN band near 2.228\,$\mu$m.
This is best visible at squared visibility levels of 0.1--0.2
and indicates that these lines are formed at extended layers above
the pseudo-continuum. This behavior is comparable to oxygen-rich
SR AGB stars \citep[e.g.,][]{Marti-Vidal2011}.
The $^{13}$CO bandheads are less prominent in the visibility spectra
than in the flux spectrum, which may indicate that $^{13}$CO is less
extended than $^{12}$CO.
Wavelength-differential closure phase features
at the positions of the CO bandheads indicate photocenter
displacements between the CO-forming regions and the nearby
pseudo-continuum.

Figure~\ref{fig:amber_visspfr} shows the
AMBER visibility results of all data sets as a function of baseline length
for the examples of bandpasses in the CO (2-0) line at 2.29\,$\mu$m
and the nearby pseudo-continuum at 2.28\,$\mu$m together with the
model prediction in these bandpasses. It illustrates that
R~Scl is more extended in the CO (2-0) bandpass compared to the
nearby pseudo continuum, and that this behavior is consistent
with the model prediction. The difference between the two 
bandpasses might be slightly larger in the observations than 
in the model prediction. The UD fit lies in between the 
model prediction at these bandpasses and -- as a wavelength independent
model -- cannot reproduce the visibility differences. 

\subsection{PIONIER visibility results and model comparison}
\label{sec:results_pionier}
Figure~\ref{fig:pionier2014_visspfr} shows all our PIONIER squared 
visibility (top) and closure phase (bottom) data as a function of 
spatial frequency. PIONIER data obtained
in 2014 are shown in the left panels, and those obtained in 2015
in the right panels. We indicate the synthetic visibility
based on the best-fit snapshots by dashed lines. For the 2014 data,
we also show the visibility values based on the reconstructed
image as discussed below in Sect.~\ref{sec:imaging_pionier}, as well
as their residuals relative to the measured values.

The observed visibility values show a characteristic curve of 
a resolved stellar disk. The squared visibility amplitudes
in the first lobe up to the first minimum at a spatial
frequency of $\sim$130\,cycles/arcsec are consistent with the synthetic 
curve based on a spherical model, while the squared visibility
amplitudes at higher spatial frequencies cannot be well described by
the spherical model. Consistently, the closure phase values are consistent
with zero up to maximum spatial frequencies of about 
90\,cycles/arcsec, 
shortly before the minimum of the squared visibility, 
then start to deviate from zero values between about 90-110 cycles, 
and show a very complex behavior at higher observed maximum spatial 
frequencies between 180\,cycles/arcsec and 440\,cycles/arcsec. 
This behavior of the squared visibility amplitudes and 
closure phases indicate a broadly circular resolved stellar
disk with a complex substructure. 

The squared visibility amplitudes at short spatial frequencies
up to about 50/cycles lie systematically above the model
predictions. We have seen the same behavior for the well-known
check star $\upsilon$\,Cet (appendix~\ref{sec:checkstar}) and attribute 
it to known systematic calibration effects owing to different magnitudes
or airmass between the science target and the calibrator
(cf. Sect. \ref{sec:obs_pionier}).

The squared visibility amplitudes in the first lobe indicate that
the source is slightly more extended at the 1.59\,$\mu$m
spectral channel compared to the 1.68\,$\mu$m and 1.76\,$\mu$m 
spectral channels in the 2014 data, as well as at the bluer
spectral channels at 1.53\,$\mu$m and 1.58\,$\mu$m compared
to the redder spectral channels at 1.68\,$\mu$m to 1.76\,$\mu$m
in the 2015 data. This result is also predicted by the 
best-fit model snapshots. A larger extension at central wavelengths
of 1.53--1.59\,$\mu$m might be related to the ''1.53\,$\mu$m feature'',
which was discussed by  \citet{Gautschy-Loidl2004} and attributed 
to C$_2$H$_2$ and HCN (cf. also Fig.~\ref{fig:mol_contrib}).

The closure phases starting from maximum spatial frequencies of 
about 100\,cycles/arcsec and more clearly around 200\,cycles/arcsec show 
closure phase values decreasingly different from zero across the three 2014
spectral channels from 1.59\,$\mu$m to 1.76\,$\mu$m, indicating 
differences in the substructure of the stellar disk
at the three spectral channels.
\subsection{PIONIER imaging}
\label{sec:imaging_pionier}
\begin{figure*}
  \includegraphics[width=0.33\hsize]{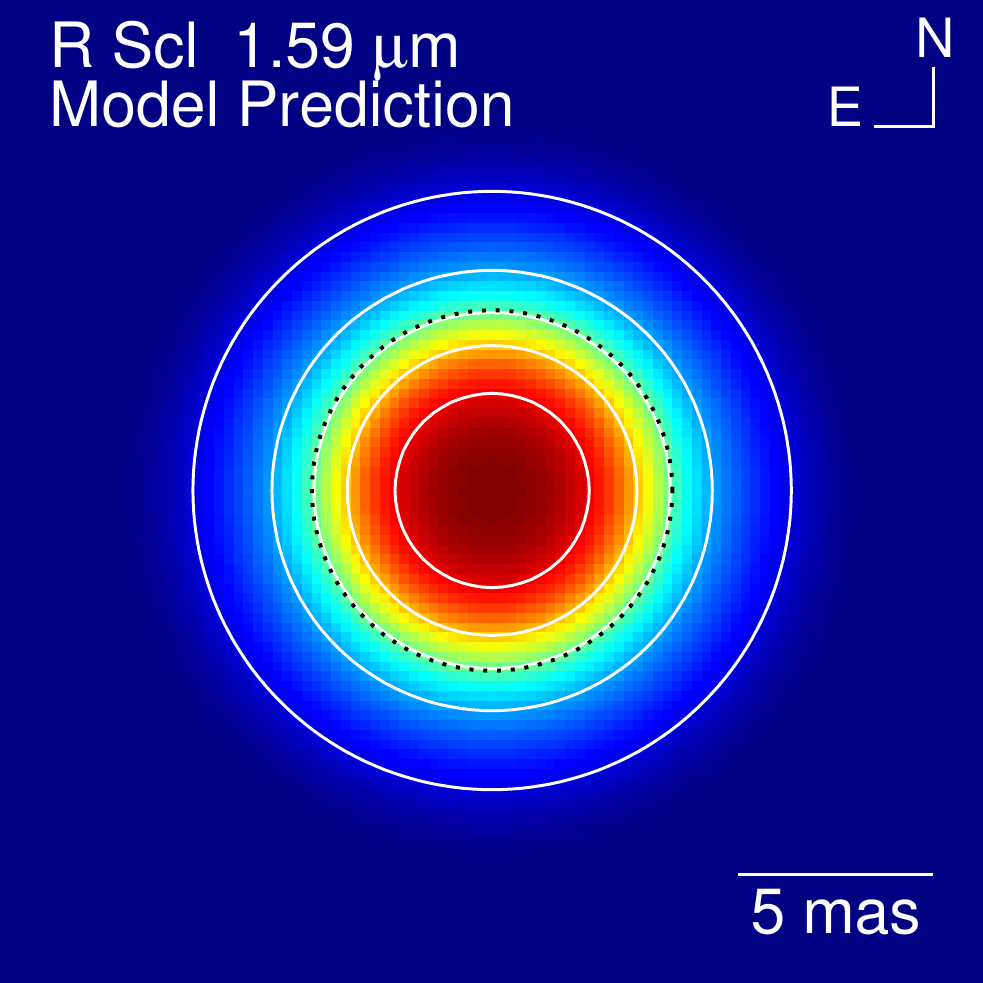}
  \includegraphics[width=0.33\hsize]{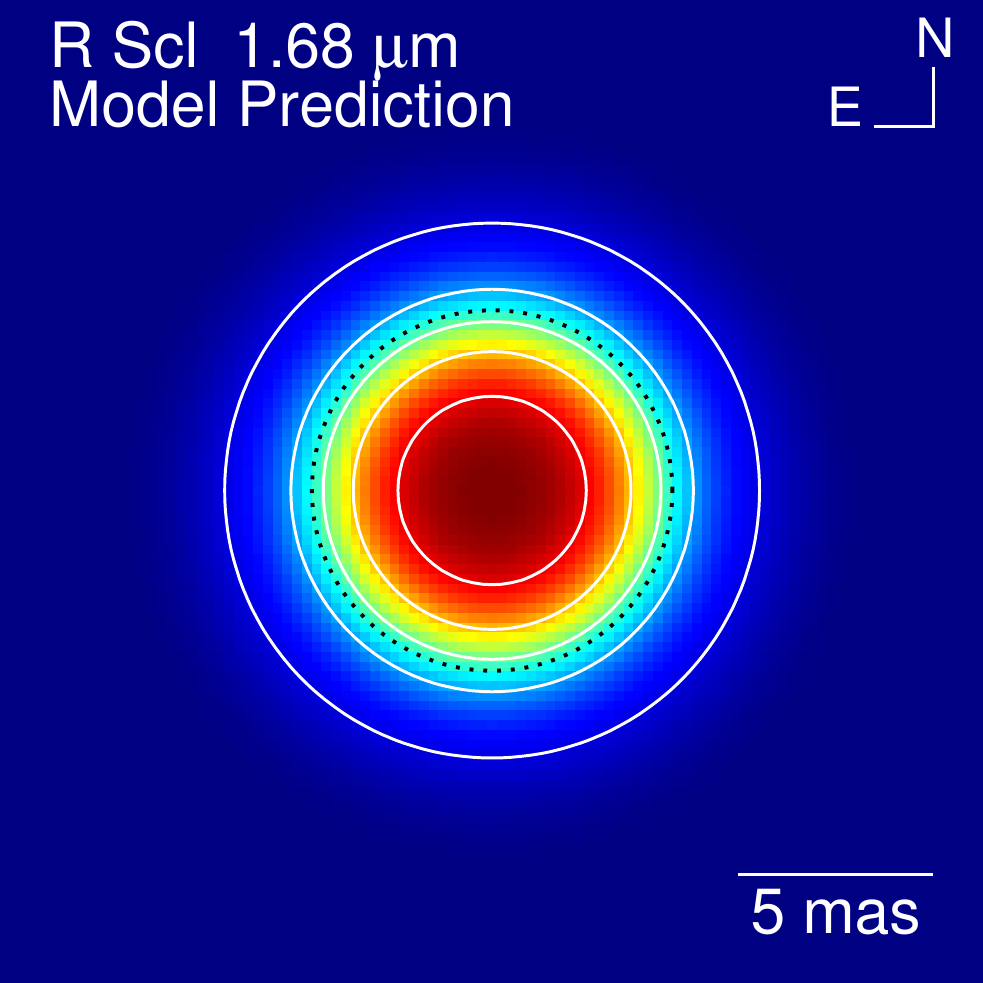}
  \includegraphics[width=0.33\hsize]{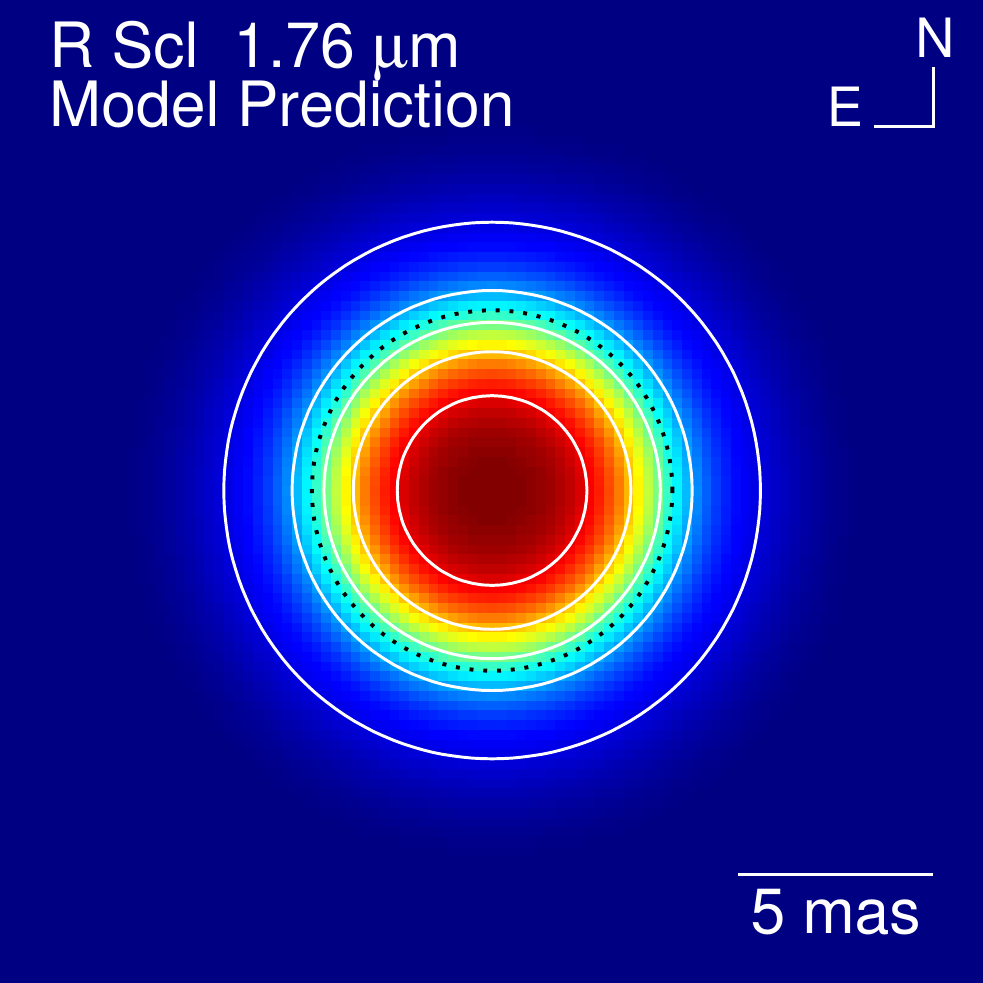}
\caption{Image representations of the intensity profiles of the
best-fit dynamic model snapshot. The three images represent the three
PIONIER spectral channels with central wavelengths 1.59\,$\mu$m (left),
1.68\,$\mu$m (middle), and 1.76\,$\mu$m (right). All images are based
on the same Rosseland angular diameter. The images are  
convolved with a theoretical point spread function (PSF) using a Gaussian
with FWHM $\lambda_\mathrm{central}/B_\mathrm{max}$.
The Rosseland angular diameter is indicated by the dashed black circle.
Contours are drawn at levels of 0.9, 0.7, 0.5, 0.3, 0.1.}
\label{fig:pionier_image_model}
\end{figure*}
\begin{figure*}
  \includegraphics[width=0.33\hsize]{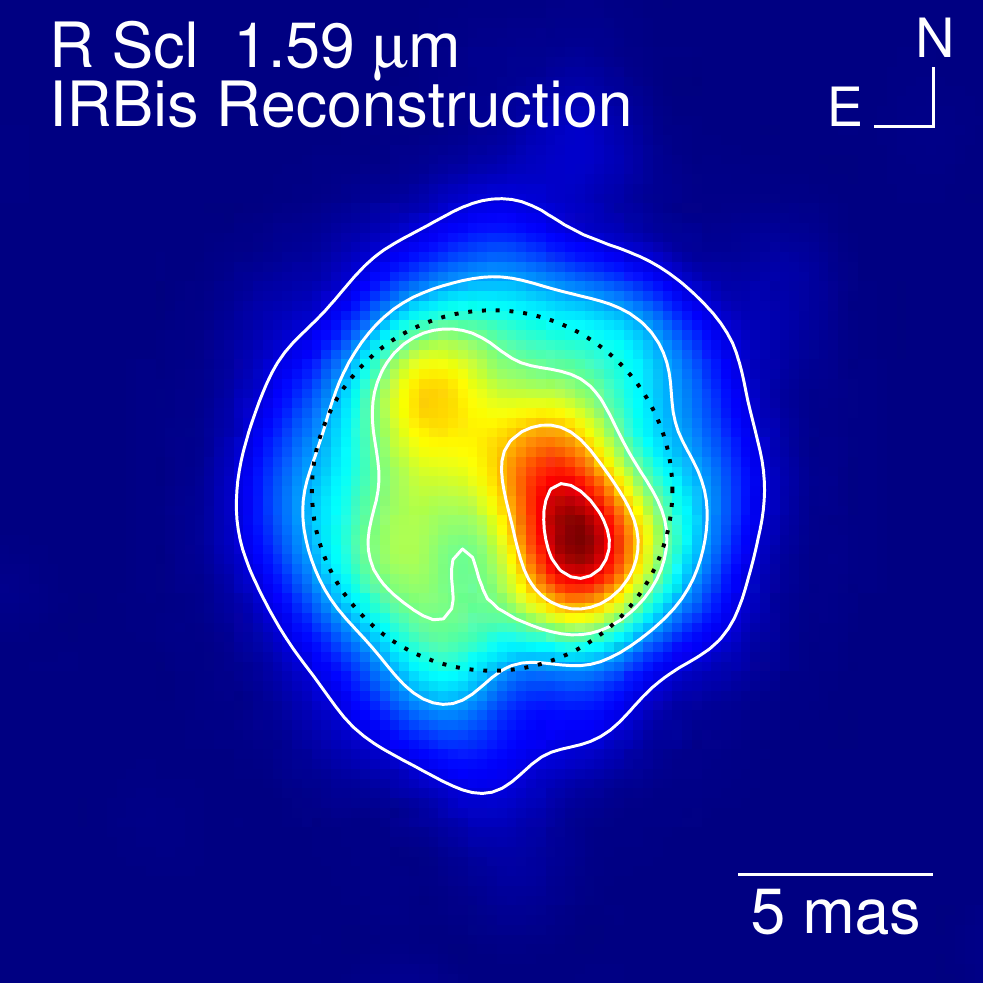}
  \includegraphics[width=0.33\hsize]{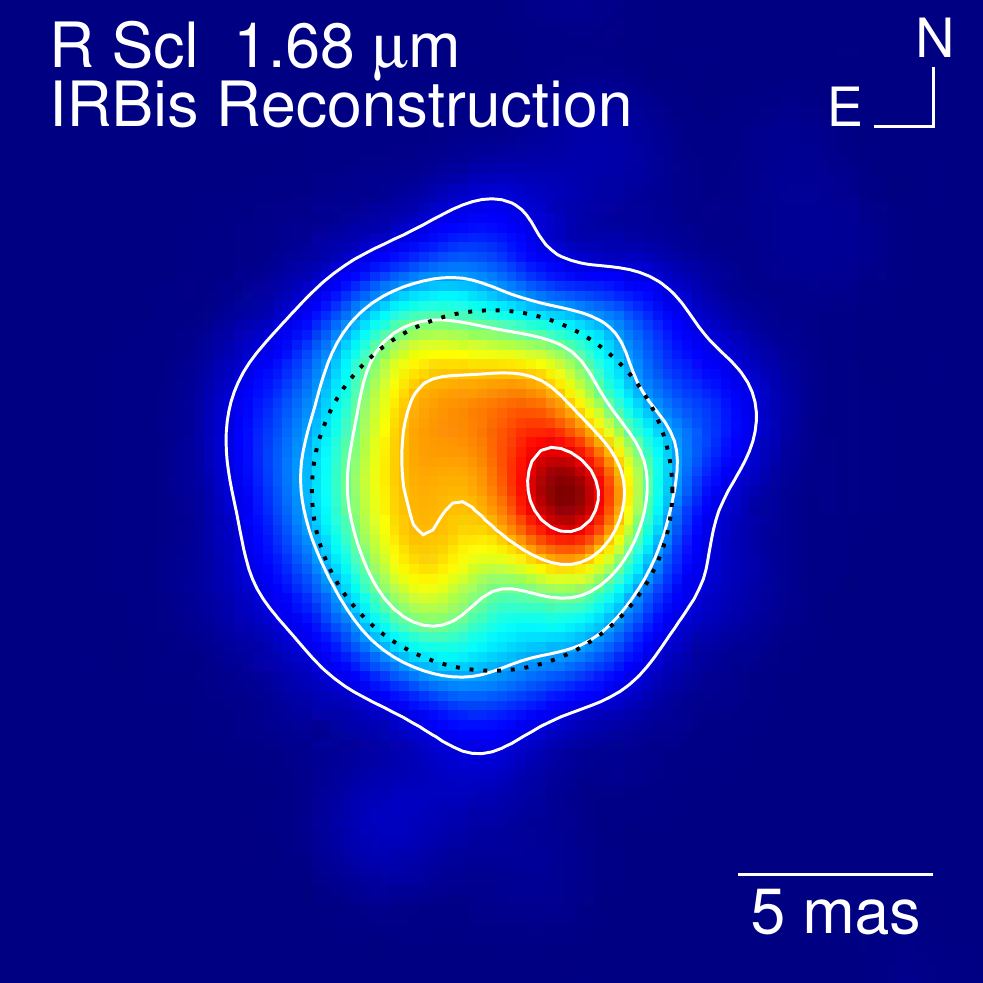}
  \includegraphics[width=0.33\hsize]{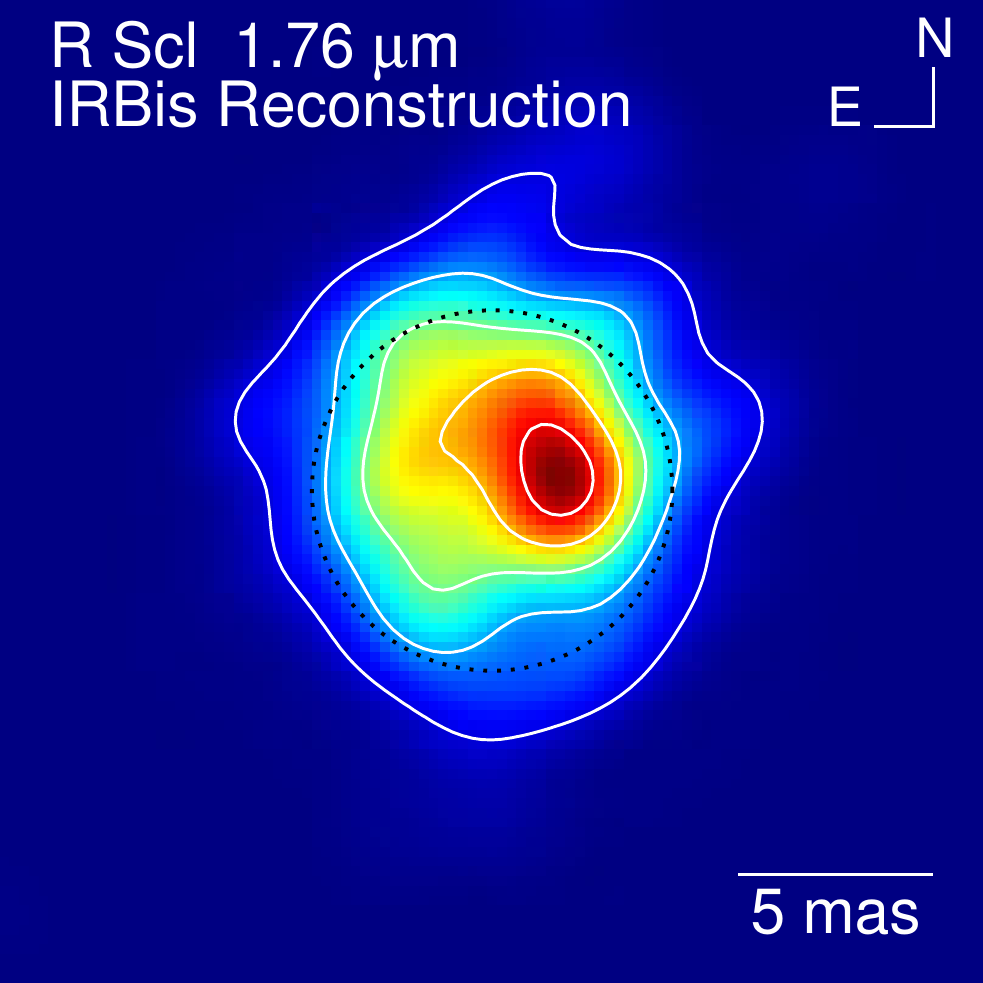}
\caption{Image reconstructions of R~Scl based on our 2014
PIONIER data and the {\tt IRBis} image reconstruction package
\protect\citep{Hofmann2014}. The three images represent the three
PIONIER spectral channels with central wavelengths 1.59\,$\mu$m (left),
1.68\,$\mu$m (middle), and 1.76\,$\mu$m (right). The images are 
convolved with a theoretical point spread function (PSF) using a Gaussian
with FWHM $\lambda_\mathrm{central}/B_\mathrm{max}$.
Our estimate of
the Rosseland angular diameter is indicated by the dashed black circle.
Contours are drawn at levels of 0.9, 0.7, 0.5, 0.3, 0.1.}
\label{fig:pionier_image_IRB}
\end{figure*}

\begin{figure*}
  \includegraphics[width=0.33\hsize]{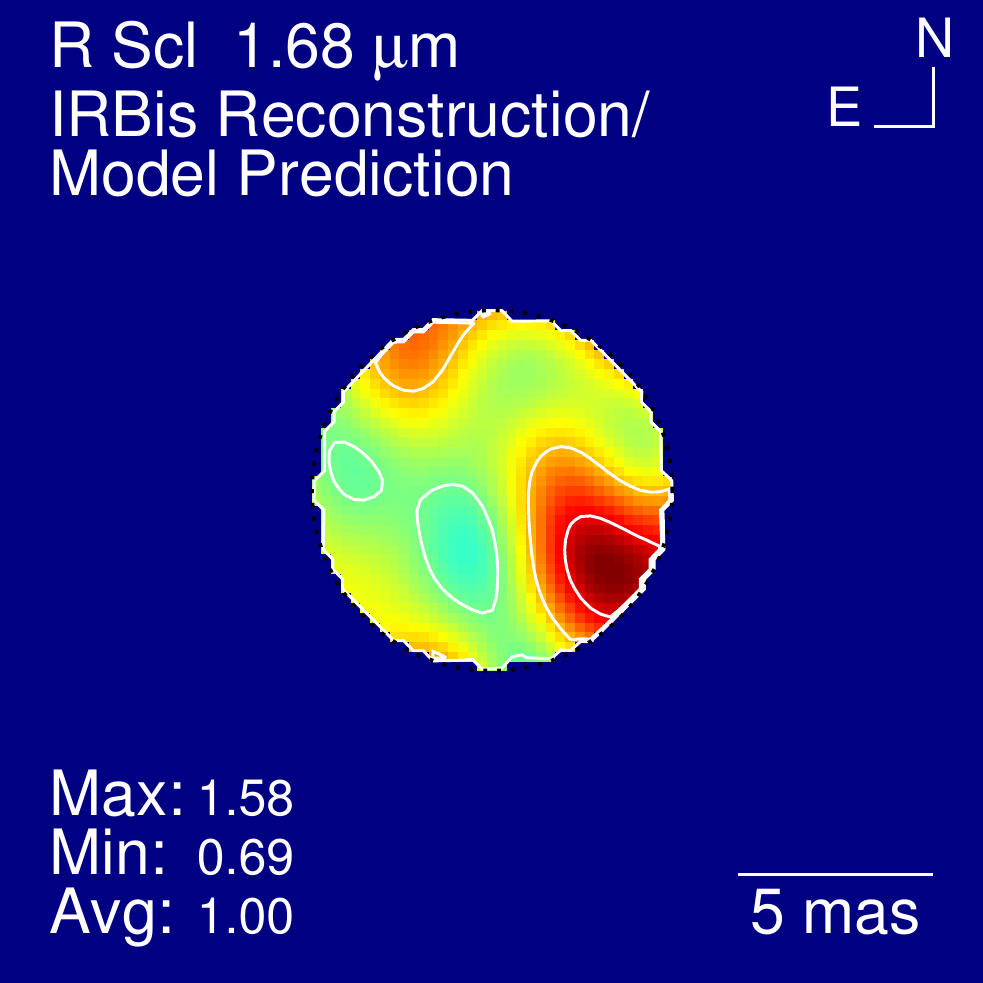}
  \includegraphics[width=0.33\hsize]{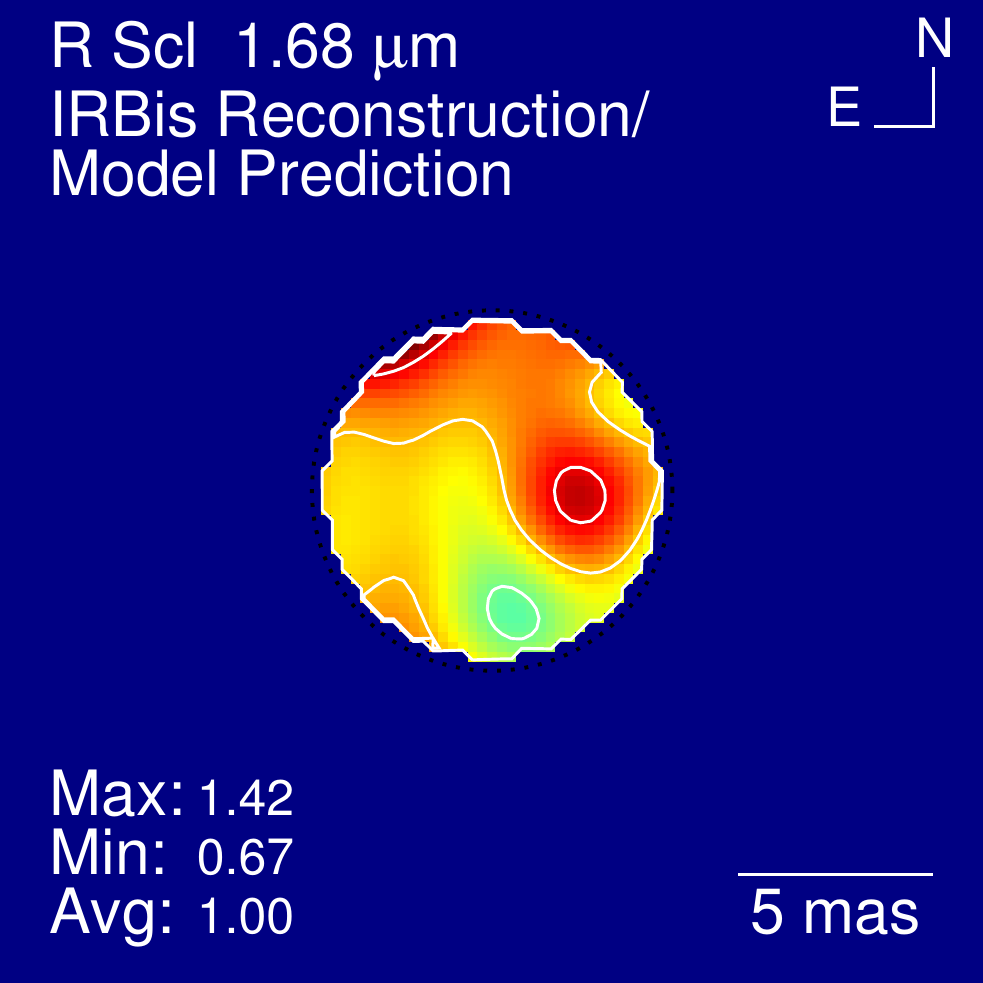}
  \includegraphics[width=0.33\hsize]{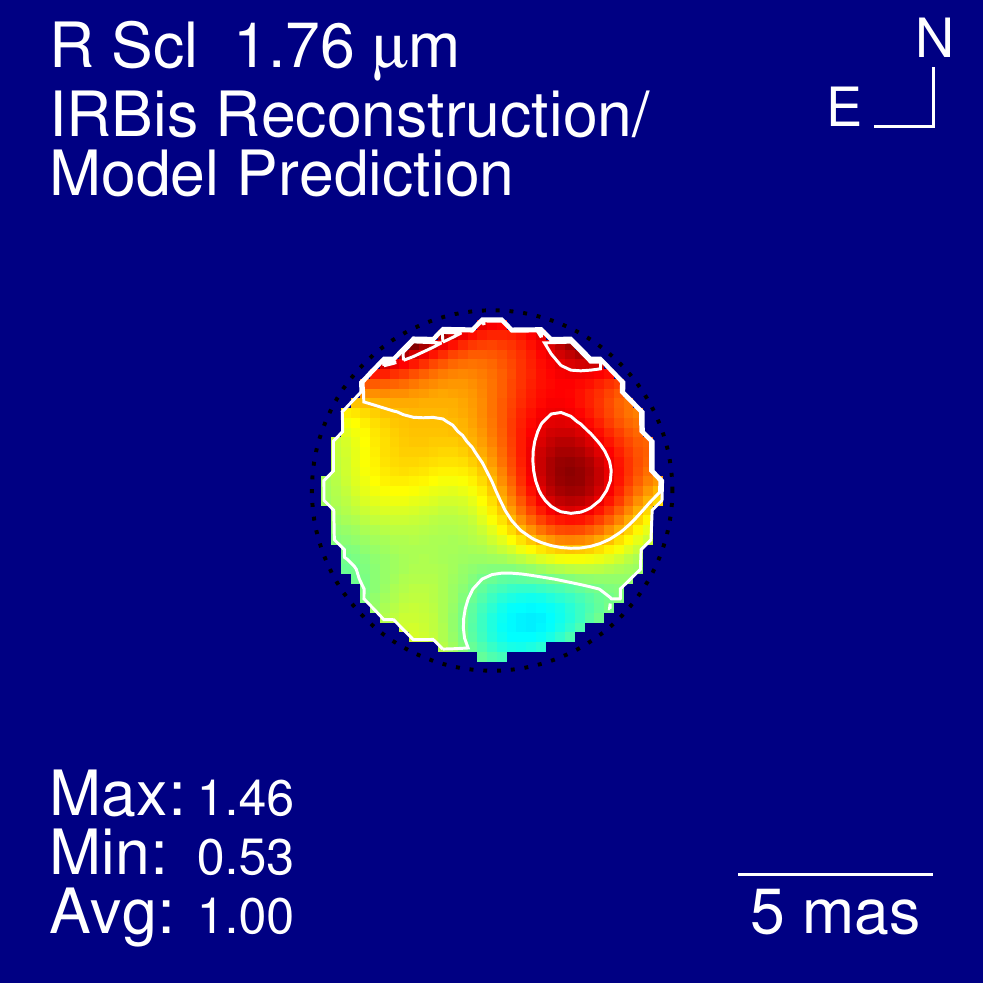}
\caption{Image reconstructions of R~Scl  from 
Fig.~\protect\ref{fig:pionier_image_IRB} divided by the model images
from Fig.~\protect\ref{fig:pionier_image_model} in order to remove the
global center-to-limb intensity variation and to highlight the
surface structure on top of this variation.
We use a cut-off radius at a model intensity level of 50\%, located 
slightly within the Rosseland radius, which is indicated by the 
black dashed circle. We calculated the maximum, minimum, and mean
intensities inside the cut-off radius to estimate the contrast
of the observed structure and printed them on the image.
These values are normalized to an average intensity of unity
for each spectral channel separately.
Contours are drawn at levels of 0.9, 0.7, 0.5.}
\label{fig:pionier_image_IRB_divided}
\end{figure*}
We used the {\tt IRBis} image reconstruction package by
\citet{Hofmann2014} to reconstruct images based on our
2014 PIONIER data at each of the three spectral channels.
This image reconstruction package currently
offers six different regularization functions. It offers 
the use of a start image. Most of the regularization functions
can be used with a prior image or with a flat prior.
For details on the image reconstruction algorithms, we 
refer to the original article by \citet{Hofmann2014}. We chose
to use the model images based on the best-fit 
model snapshot as start images. We used a flat prior, and 
smoothness 
as regularization function
\citep[regularization function 3 of][]{Hofmann2014}. 

In order to verify the imaging process, we studied the effects 
of using the different available
regularization methods of {\tt IRBis}.
We studied as well the effects of using other image 
reconstruction packages, namely {\tt MiRA} by \citet{Thiebaut2008} 
and {\tt BSMEM} by \citet{Buscher1994}. These studies also included
the use of different start images and different prior images.
The results of these imaging studies are detailed in 
appendix~\ref{sec:imagingstudies}. These studies show that
the reconstructed images are very stable with respect to 
using different image reconstruction packages and algorithms,
different regularization functions, different start images,
and different prior images.

We used start images based on the best-fit model snapshots
for the imaging with {\tt IRBis}.
Figure~\ref{fig:pionier_image_model} shows the model images
based on the intensity profiles of the best-fit dynamic model
snapshots. Indicated are the positions of the estimated
Rosseland radius as well as contours at levels of 
0.9, 0.7, 0.5, 0.3, and 0.1. The images show a center-to-limb
intensity variation with a steep decrease of the intensity 
profiles between intensity levels of 0.5 and 0.3, which is located
near the estimated Rosseland radius of 8.8\,mas. Beyond the
Rosseland radius, the images show a shallow extension reaching
a 10\% intensity level at an angular diameter of about 15.2\,mas
($\sim$\,1.7\,$\Theta_\mathrm{Ross}$) for the spectral channel
with central wavelength 1.59\,$\mu$m
and about 13.6\,mas ($\sim$\,1.5\,$\Theta_\mathrm{Ross}$)
for spectral channels with central wavelengths of 1.68\,$\mu$m and 
1.76\,$\mu$m. Likewise, the steep decrease near intensity levels
between 50\% and 30\% is slightly shallower for the spectral
channel at 1.59\,$\mu$m compared to the two redder spectral
channels. This difference between the 1.59\,$\mu$m channel
compared to the 1.68\,$\mu$m and 1.76\,$\mu$m channels was
confirmed by our measured visibility data in 
Sect.~\ref{sec:results_pionier} and may be an effect of extended
layers of C$_2$H$_2$ and HCN 
\citep[''1.53\,$\mu$m feature'',][]{Gautschy-Loidl2004}.
These model images are used as start images, but not as prior
images, for the following image reconstruction process. 
However, our image reconstructions using the {\tt MiRA} package
\citep{Thiebaut2008}, shown in appendix~\ref{sec:imagingstudies},
used a simple UD fit as a start image, and gave virtually identical
results, showing that the choice of the start image was not 
critical.

Figure~\ref{fig:pionier_image_IRB} shows our reconstructed images
for the three spectral channels. These images are convolved with
a theoretical point spread function (PSF) using a Gaussian
with FWHM $\lambda_\mathrm{central}/B_\mathrm{max}$.
We have added the squared
visibility amplitudes and closure phases based on the reconstructed
images to Fig.~\ref{fig:pionier2014_visspfr}, which shows the measured
values. The residuals between imaging values and measured
values are shown as well. The synthetic visibility
values based on the reconstructions are in very good agreement
with the measured values, except for high squared visibility 
amplitudes at small spatial frequencies. The latter was expected,
due to known
systematic calibration effects (Sect.~\ref{sec:results_pionier}
and appendix~\ref{sec:checkstar}). 
It is thus re-assuring that the
imaging processes kept these differences between reconstructed
images and measured visibility values, and did not correct for
these systematic calibration effects by artificial features
in the reconstructions.

As expected from the visibility 
results discussed in Sect.~\ref{sec:results_pionier}, the 
reconstructed images show a broadly circular stellar
disk with a very complex substructure. The global center-to-limb
intensity variation is broadly consistent with the model images
discussed above: our estimate of the Rosseland angular diameter
is on average located between the 50\% and 30\% intensity levels, 
as in the model images. The extension of the 10\% intensity level
is comparable to that of the model images as well. Furthermore,
the contours at 10\% and 30\% of the maximum intensity are roughly 
circular, while the higher intensity levels above about 50\%,
located inside the Rosseland angular diameter, that is, inside the 
stellar disk, show a very complex structure. Most strikingly, the 
images show one dominant bright spot located at the western part 
of the stellar disk, adjacent to a bright region toward the 
north-east of the bright spot.

The complex structure that we observed within the stellar disk
is superimposed
on the global center-to-limb intensity variation.
In order to remove the signature of the global center-to-limb 
intensity variation and to study the structure on top of this 
variation, we divided the reconstructed images shown in
Fig.~\ref{fig:pionier_image_IRB} by the model images shown
in Fig.~\ref{fig:pionier_image_model}. 
Figure~\ref{fig:pionier_image_IRB_divided} shows the divided
images. We used a cut-off radius at a model intensity level 
of 50\%, located slightly inside the Rosseland radius, outside
of which we set the divided images to zero.
We calculated the maximum, minimum, and mean intensities inside 
the cut-off radius to estimate the contrast of the 
observed structure.
The peak intensity of the dominant western
spot lies 58\%, 42\%, and 46\% above the average intensity (including the
spot itself) at the three spectral channels with central wavelengths
of 1.59\,$\mu$m, 1.68\,$\mu$m, and 1.76\,$\mu$m, respectively. 
The total contrast between the minimum and maximum
intensities within this part of the stellar disk are 1:2.3, 1:2.1,
and 1:2.8, respectively. We estimate the errors of the 
intensities to be about 5--10\%.
These divided images show more clearly than the original ones
that the dominant spot is slightly shifted from south-west
to west to north-west across the spectral channels at
1.59\,$\mu$m, 1.68\,$\mu$m, and 1.76\,$\mu$m. This may be related to 
the differences in the closure phases across these
spectral channels at $\sim$\,200\,cycles/arcsec 
(cf. Sect.~\ref{sec:results_pionier}, Fig.~6). However, we cannot currently
exclude an uncertainty of the position of the dominant spot
of this order in the imaging process. 
\subsection{Fundamental parameters of R~Scl}
\label{sec:fundpar}
\begin{table}
\centering
\caption{Fundamental properties of R~Scl used in this work. }
\label{tab:fundpar}
\begin{tabular}{llr}
\hline\hline
Parameter                               & Value            & Ref. \\\hline
Period(d)                               & 376              & 1 \\
Pulsation mode                          & Fundamental      & 1 \\
Distance (pc)                           & 370$^{+70}_{-100}$ & 1 \\
Interstellar extinction $A_V$           & 0.07             & 2 \\
Bolometric flux (10$^{-9}$ W/m$^2$)     & 1.28 $\pm$ 0.06  & 3 \\
Rosseland angular diameter (mas)        & 8.9 $\pm$ 0.3    & 4 \\
Rosseland radius ($R_\sun$)            & 355 $\pm$ 55     & 5 \\
Effective temperature (K)               & 2640 $\pm$ 80    & 6 \\
Luminosity ($\log L/L_\sun$))          & 3.74 $\pm$ 0.18  & 7 \\
Initial mass ($M_\sun$)                & 1.5  $\pm$ 0.5   & 8\\
Current mass ($M_\sun$)                & 1.3  $\pm$ 0.6   & 8\\
Surface gravity ($\log g$)              & -0.6 $\pm$ 0.4  & 9 \\\hline
\end{tabular}
\tablefoot{
1: Discussion in Sect.~\ref{sec:lightcurve};
2: \citet{Whitelock2008};
3: integrated photometry, see Sect. \ref{sec:lightcurve};
4: this work, average of our three epochs, see Sect.~\ref{sec:results};
5: calculated from 2 and 4;
6: calculated from 3 and 4;
7: calculated from 2 and 3;
8: 7 and 8 with evolutionary tracks by \citet{Lagarde2012};
9: 6 and 9.  
}
\end{table}
\begin{figure}
  \includegraphics[width=0.99\hsize]{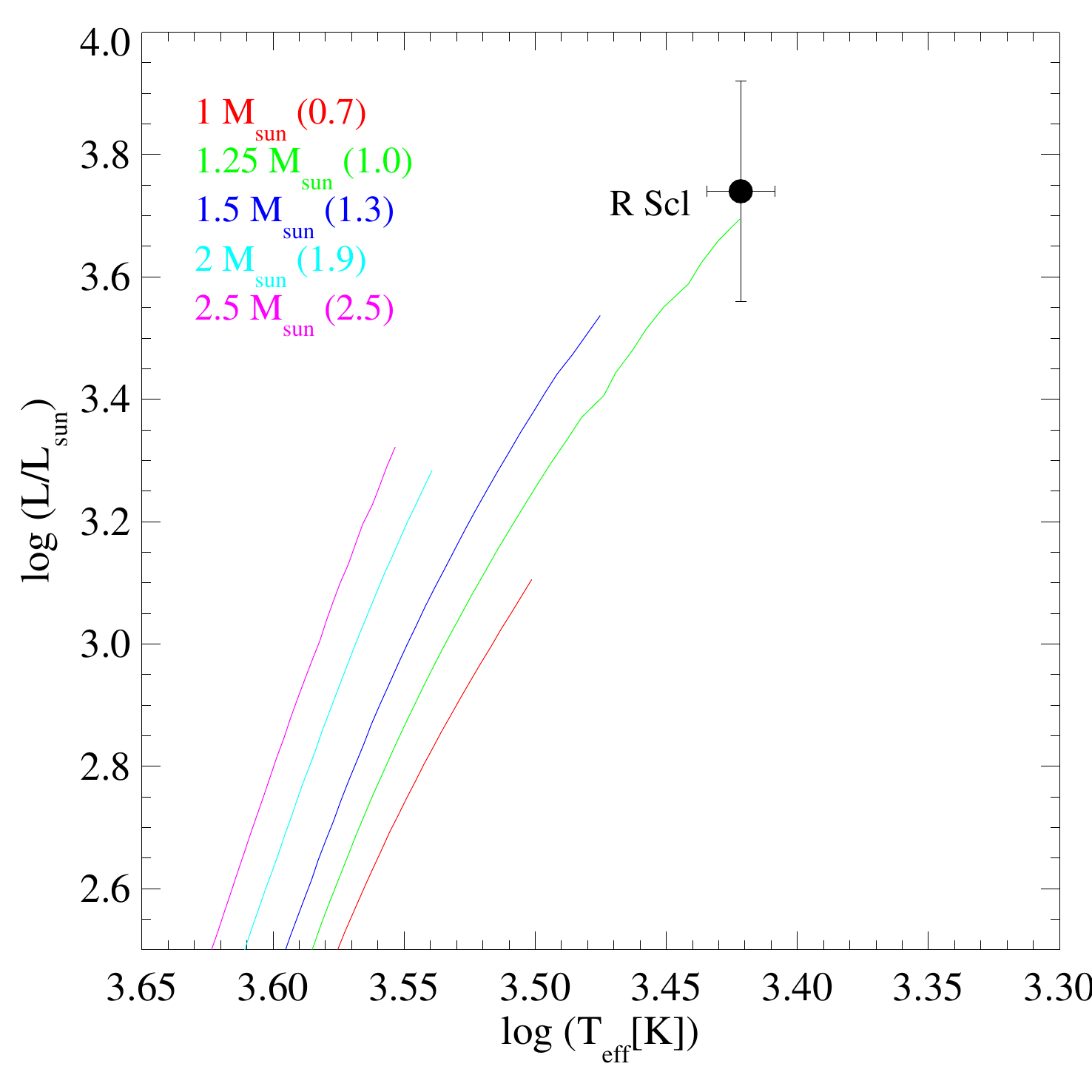}
\caption{Position of R~Scl in the HR diagram together with
evolutionary tracks by \citet{Lagarde2012}. The tracks are
those without rotation and reach up to the start of the 
TP-AGB. Masses are ZAMS masses; values in parenthesis are
masses at the start of the TP-AGB, that is, at the end of the shown
tracks.}
\label{fig:hrdiagram}
\end{figure}
Table~\ref{tab:fundpar} summarizes the fundamental parameters
of R~Scl that we used in this work, including our adopted values 
for the bolometric flux and distance, the measured Rosseland 
angular diameter, and the derived effective temperature and 
luminosity. We note that the distance to R~Scl is quite
uncertain, as discussed in Sect.~\ref{sec:lightcurve},
which is reflected in its large adopted error.
Figure~\ref{fig:hrdiagram} shows the position of R~Scl
in the Hertzsprung-Russell (HR) diagram based on the
effective temperature and luminosity.
This figure includes stellar evolutionary
tracks by \citet{Lagarde2012} for zero age main sequence 
(ZAMS; initial) masses between 
1\,$M_\sun$ and 2.5\,$M_\sun$. These tracks reach up to 
the start of the thermally pulsing (TP) AGB, at which point
the mass range is 0.7\,$M_\sun$ to 2.5\,$M_\sun$.
The position of R~Scl is consistent with an evolutionary 
phase shortly beyond the start of the TP-AGB and with
an initial mass interval between 1\,$M_\sun$ 
and 2\,$M_\sun$. We also compared the position of R~Scl
to the tracks of the TP-AGB by \citet{Marigo2013} as shown
by \citet{Klotz2013}. Here, R~Scl is also consistent
with masses at the start of the TP-AGB between about
1\,$M_\sun$ and 2\,$M_\sun$, that is, not significantly 
below 1\,$M_\sun$ and not significantly above 
2\,$M_\sun$.  These are evolutionary tracks 
for single stars, while a binary companion 
has been detected for R Scl. 
However, based on the estimates for the 
binary companion mass and separation \citep{Maercker2016}, 
it is unlikely that the companion has affected the 
evolution of R~Scl to the AGB. 
As a result, we estimate the initial mass 
of R~Scl to be 1.5\,$\pm$\,0.5\,$M_\sun$ and the current 
mass to be 1.3\,$\pm$\,0.6\,$M_\sun$. Together with our estimate
of the radius, these values correspond to a surface gravity
$\log g$ of -0.6\,$\pm$\,0.4.

The mass of R~Scl can also be estimated by the pulsation
properties. We used the formula 
\begin{equation*}
Q=P\times (M/R^2)^{0.5}
\end{equation*}
from \citet{Wood1990}, where $Q$ is the pulsation constant,
and $M$ and $R$ are the mass and radius in solar units.
We estimated a $Q$ value of 0.093$\pm$0.005 for the given 
ranges of masses and pulsation periods based on 
\citet{Fox1982} and assuming fundamental pulsation mode 
(cf. Sect.~\ref{sec:lightcurve}). We obtained a 
mass based on these pulsation properties and the measured radius
of 2.7\,$M_\sun$ and an uncertainty range between 
1.5$M_\sun$ and 4.7$M_\sun$. Within the errors, this mass
estimate is consistent with the estimate above based on the
position in the HR-diagram and the evolutionary models between
1.5$M_\sun$ and 1.9$M_\sun$. The mass based on the 
evolutionary tracks is mostly constrained by the effective temperature,
whose estimate is independent of the distance, while the pulsation
mass depends on the radius, and thus the distance. A lower distance
within the quoted errors would thus lead to a better agreement between 
these two mass estimates. For example, a distance of 300\,pc
at the lower end of our estimate would lead to a pulsation mass
of 1.4$M_\sun$.

We chose to use the mass estimate based on the evolutionary tracks
as our final mass estimate in Table~\ref{tab:fundpar}, because this
mass estimate is less affected by the large uncertainty of the distance
to R~Scl. However, the given errors do not include
errors inherent in choosing the particular set of evolutionary tracks.
\section{Discussion of the observed structure within the stellar disk}
\citet{Freytag2008} presented three-dimensional (3D) 
radiative hydrodynamical simulations
of the atmosphere and wind of an AGB star. They concluded that
large-scale convective flows, and the resulting
atmospheric shock waves, induce
certain non-radial structures in the region of the innermost dust shell. 
We interpret our observed structure on the surface of R~Scl within
this context.

In this scheme, our observed structure on the stellar disk of R~Scl
is mostly due to dust absorption and lack thereof, depending on how dense 
and close to the star the dust is with respect to the line of sight. 
In this view, the dominant
bright spot would be caused by a region where the 
stellar surface is less obscured by dust clouds. For example, Figure~2
of \citet{Freytag2008} shows arcs in two-dimensional (2D) slices of newly formed 
dust that are spread over typical angles of about 90 degrees.
These arcs of newly formed dust, if seen face-on on top of the
photosphere, would be qualitatively consistent with our 
observed structure in terms of the angular scales. Dust clumps of 
amorphous carbon (amC) located near the condensation radius
can efficiently block the light at near-infrared 
wavelengths so that this scheme is also qualitatively consistent 
with our observations in terms of the high observed intensity contrasts.
These scales are representative of the density structure at
2-3 stellar radii, where the more small-scale photospheric shocks
corresponding to the convective flows have merged, forming these 
partial near-spherical shells that cover a wide angle, and where
new dust forms.
Molecules at these radii should reflect these density structures 
corresponding to the large-scale shocks as well, but may not lead to the
observed high contrasts. We expect the actual convective 
surface patterns at photospheric layers to be obscured in the near-IR
by molecular (cf. Fig.~\ref{fig:mol_contrib}) and dusty layers located
above the photosphere.

\citet{Freytag2008} already noted that such dust structures, which
developed in their 3D simulations, should be detectable by 
high spatial resolution techniques. Indeed, \citet{Ohnaka2016} 
and \citet{Khouri2016} observed clumpy dust clouds around
the oxygen-rich AGB stars W~Hya and R~Dor using VLT/SPHERE, which are
seen in scattered visual light around the stellar photosphere.
The sizes of these clumps are qualitatively consistent with
those predicted by \citet{Freytag2008}. In particular, the
dust clouds shown by \citet[][Fig. 8]{Ohnaka2016} are interpreted as
atmospheric structures with a 
half opening angle of 45 degrees, that is, 90 degrees total, and a 
density enhancement of four compared to the surrounding medium.
Equivalently, we interpret 
our detected structure of the carbon-rich AGB R~Scl as being caused 
by absorption by amC dust in front of the star instead of by scattering next
to the star. Moreover, amC dust has a higher absorption coefficient
than oxygen-rich dust, and can thus lead to higher intensity 
contrasts.


The 3D simulations do not necessarily predict one single region
that vastly dominates our view onto the stellar disk as observed
in our reconstructed images. The simulations show a variability 
on a timescale of a few months, with features that may vary from
one to a few months. In our picture, we would have observed the star when
the view was dominated by one feature. Unfortunately,
our second epoch observations did not result in images so we 
could not
probe the time variability of the observed structure. 
Yet, such time-linked 
observations are of great importance to further constrain
and verify the model predictions. With our single epoch of 
reconstructed images we cannot exclude whether the dominant bright
spot might be more stable than predicted by simulations based on
convection, or whether there may be additional physical mechanisms 
such as magnetic effects that enhance or stabilize the 
structure. 

Compared to other carbon AGB stars, our observation of 
a strong clumpiness of the CSE of R~Scl
is consistent with earlier detections of asymmetries or 
inhomogeneities toward carbon AGB stars 
\citep{Paladini2012,vanBelle2013}. Compared to the extreme
clumpiness of the circumstellar environment of IRC\,+10\ 216
\citep{Weigelt1998,Osterbart2000,Weigelt2002,Kim2015,Stewart2016},
our observations of R~Scl differ in two important aspects:
(1) In our reconstructed near-infrared $H$-band images of R~Scl, 
the stellar disk is still visible and only partially obscured by an 
optically thin CSE, while the stellar disk 
of IRC\,+10\ 216 is completely obscured at near-infrared wavelengths
to the point that there has been a decade-long debate on the 
position of the star with respect to the clumps of the 
circumstellar environment; (2) In our interpretation of R~Scl,
the near-infrared observations
show newly formed dust clumps near the photosphere at 
2-3 stellar radii, while the near-infrared observations of 
the dust clumps of IRC\,+10\ 216 cover a volume of roughly 
10-20 stellar radii. In both interpretations, the bright spots
are interpreted as cavities in a dense circumstellar shell.

The binary companion of R~Scl at an estimated separation of 60\,AU
\citep{Maercker2012,Maercker2016} is not expected to have a direct
influence on the surface and close circumstellar environment
of the primary AGB star. Only if the observed structure and the
dominant bright spot turn out 
to be more stable than predicted by the convection simulations
on timescales of several years, localized mass loss together with
the binary motion may induce shaping close to the surface of R~Scl.
Again, time series of observations are needed to probe the timescales of the detected structure in order to constrain the 
convection simulations and to study whether further physical mechanisms
may have to be taken into account.
\section{Summary and conclusions}
We observed the atmosphere and the close CSE 
of the carbon-rich AGB star R~Scl employing 
near-infrared interferometric techniques. We obtained 
near-infrared $K$-band interferometry with a spectral resolution
of 1500 using the AMBER instrument in 2012, at a 
maximum visual phase of R~Scl of 1.0. We also obtained near-infrared
$H$-band interferometry with spectral resolutions of about 
20 and 40 in 2014 and 2015 at pre-maximum and maximum phases 
of R~Scl of 0.8 and 1.0. The PIONIER observations obtained in 2014
had a sufficient coverage of the $uv$-plane so that we were successful
in reconstructing
images of R~Scl at the three spectral channels with
central wavelengths of 1.59\,$\mu$m, 1.68\,$\mu$m, and 1.76\,$\mu$m,
and bandwidths of about 0.09\,$\mu$m.

The visibility data in the first lobe indicate a broadly circular 
resolved stellar disk with a uniform disk (UD) diameter of 
about 10.9$\pm$0.5\,mas, which is consistent with previous estimates.
The AMBER data indicate more extended layers of CN and $^{12}$CO lying 
above the pseudo-continuum. Lines of $^{13}$CO are visible in the flux spectrum
but are less prominent in the visibility spectra, possibly indicating that
$^{13}$CO is less extended.
Likewise, the PIONIER data indicate a larger
extension at the spectral channel at 1.59\,$\mu$m compared to the 
other spectral channel, which is likely caused by the 
''1.53\,$\mu$m feature'' of inhomogeneous extended layers of 
C$_2$H$_2$ and HCN.
The visibility data in the second lobe and the closure phase data
indicate a complex substructure within the resolved stellar disk.

We compared our visibility data to a recent grid of one-dimensional (1D) dynamic atmosphere 
and wind
models for carbon-rich AGB stars. We obtained a best fit to all data 
for a model with
averaged effective temperature 2800\,K, luminosity $\log L/L_\sun$ 3.85,
mass 1\,$M_\sun$, and period 390\,d, which does not produce 
a mass loss. The latter is unexpected, as R~Scl is surrounded by
a large circumstellar environment and estimates of the present day
outflow velocity suggest a value of about 10\,km/s. However, we do
not know the current mass-loss rate, which may be smaller than
10$^{-6}$\,$M_\sun$/year. Considered models with mass loss generally
show a CSE that is too extended compared to our
observations.

Based on the model comparisons, we estimate a Rosseland angular
diameter of R~Scl of 8.9$\pm$0.3\,mas. Together with the estimated
bolometric flux and distance of R~Scl, this value corresponds to a
Rosseland radius of 355$\pm$55\,$R_\sun$, an effective temperature
of 2640$\pm$80\,K, and a luminosity of $\log L\sim$3.74$\pm$0.18.
Compared to evolutionary tracks, these parameters are consistent
with initial masses of 1.5$\pm$0.5\,$M_\sun$ and current masses 
of 1.3$\pm$0.7\,$M_\sun$.

The reconstructed images of R~Scl based on the 2014 PIONIER data
show a broadly circular stellar disk with a complex, non-spherical
substructure, as expected based on a visual inspection of the visibility
data. The global center-to-limb intensity variation (CLV) of the 
reconstructed images is consistent with that of the best-fit models.
Inside the stellar disk, the reconstructed images show a very
complex structure. Most strikingly, they exhibit one dominant bright 
spot located at the western part of the stellar disk, adjacent to a 
bright region toward the north-east of the bright spot.
We corrected the reconstructed images for the overall 
CLV, and estimated a peak intensity 
of the dominant western spot between about of 45\% and 60\% on top
of the average CLV-corrected intensity. 
The total contrast between the minimum and maximum CLV-corrected
intensities of the stellar disk are between about 1:2 and 1:3.

We interpreted this complex structure within the stellar disk
of R~Scl as caused by the effects of large-scale convection cells that
lead to almost spherical shock fronts, but which develop certain
non-radial structures in the innermost dust shell as predicted
by 3D radiative hydrodynamical simulations. Within this interpretation,
our observed structure on the stellar disk of R Scl is mostly due to 
dust absorption and lack thereof, depending on how dense and close 
to the star the dust is with respect to the line of sight. The dominant 
bright spot would be caused by a region where the 
stellar surface is less obscured by dust clouds. The scales and 
contrasts of the
observed structure roughly agree with the model predictions as far
as they can be probed by our single epoch of broad-band PIONIER images.
Certainly, time-resolved images at higher spectral resolution are needed
to confirm and constrain this interpretation.
\bibliographystyle{aa}
\bibliography{RSCL}
\begin{acknowledgements}
CP is supported by the Belgian Fund for Scientific Research F.R.S.- FNRS.
We acknowledge with thanks the variable star observations from the AAVSO 
International Database contributed by observers worldwide and used in this 
research.
This research has made use of the SIMBAD and AFOEV databases, 
operated at CDS, France.
\end{acknowledgements}
\begin{appendix}
\section{PIONIER observations of the resolved 
K5/M0\,III star $\upsilon$\,Cet}
\label{sec:checkstar}
\begin{figure*}
  \includegraphics[width=0.49\hsize]{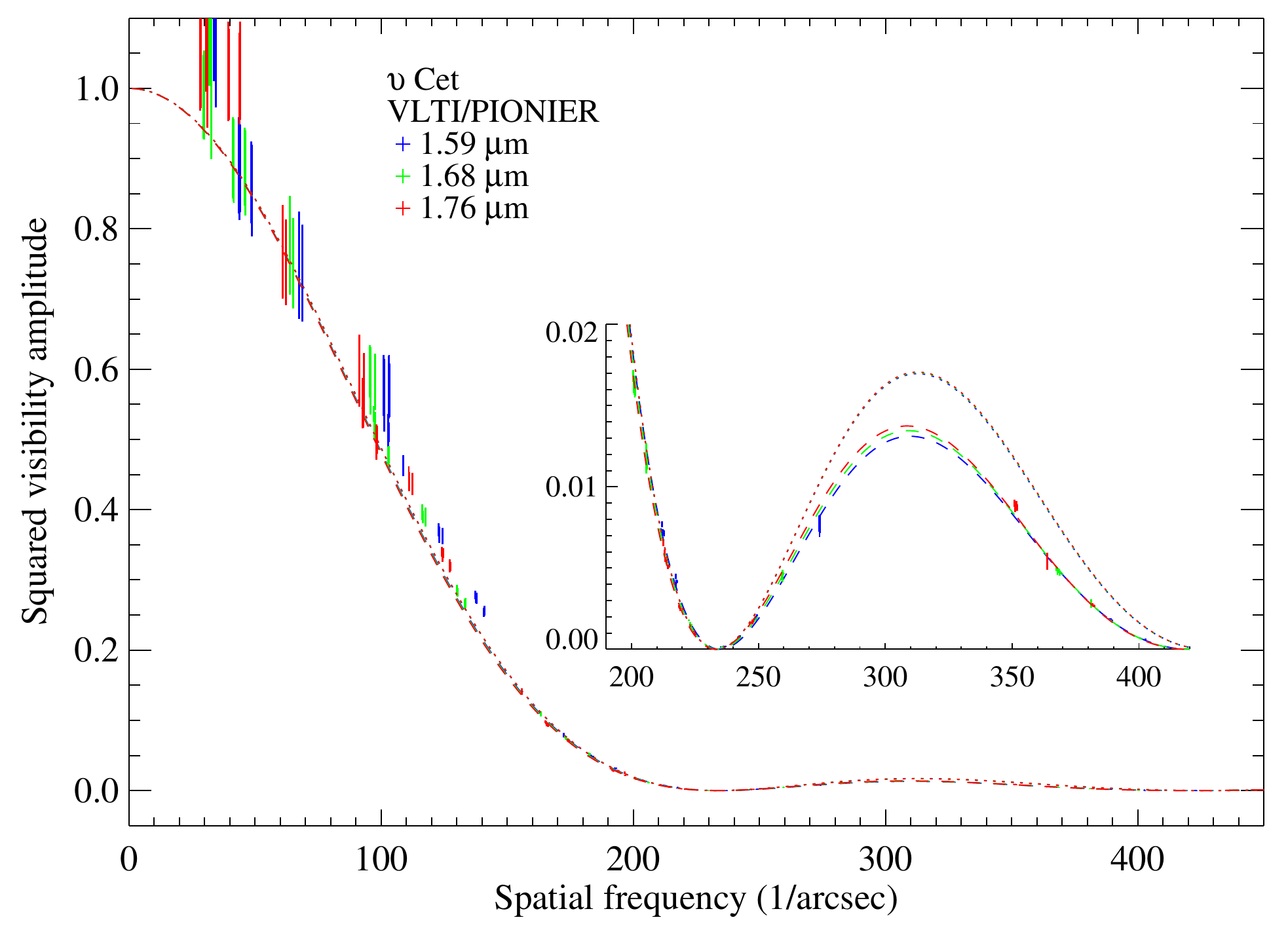}
  \includegraphics[width=0.49\hsize]{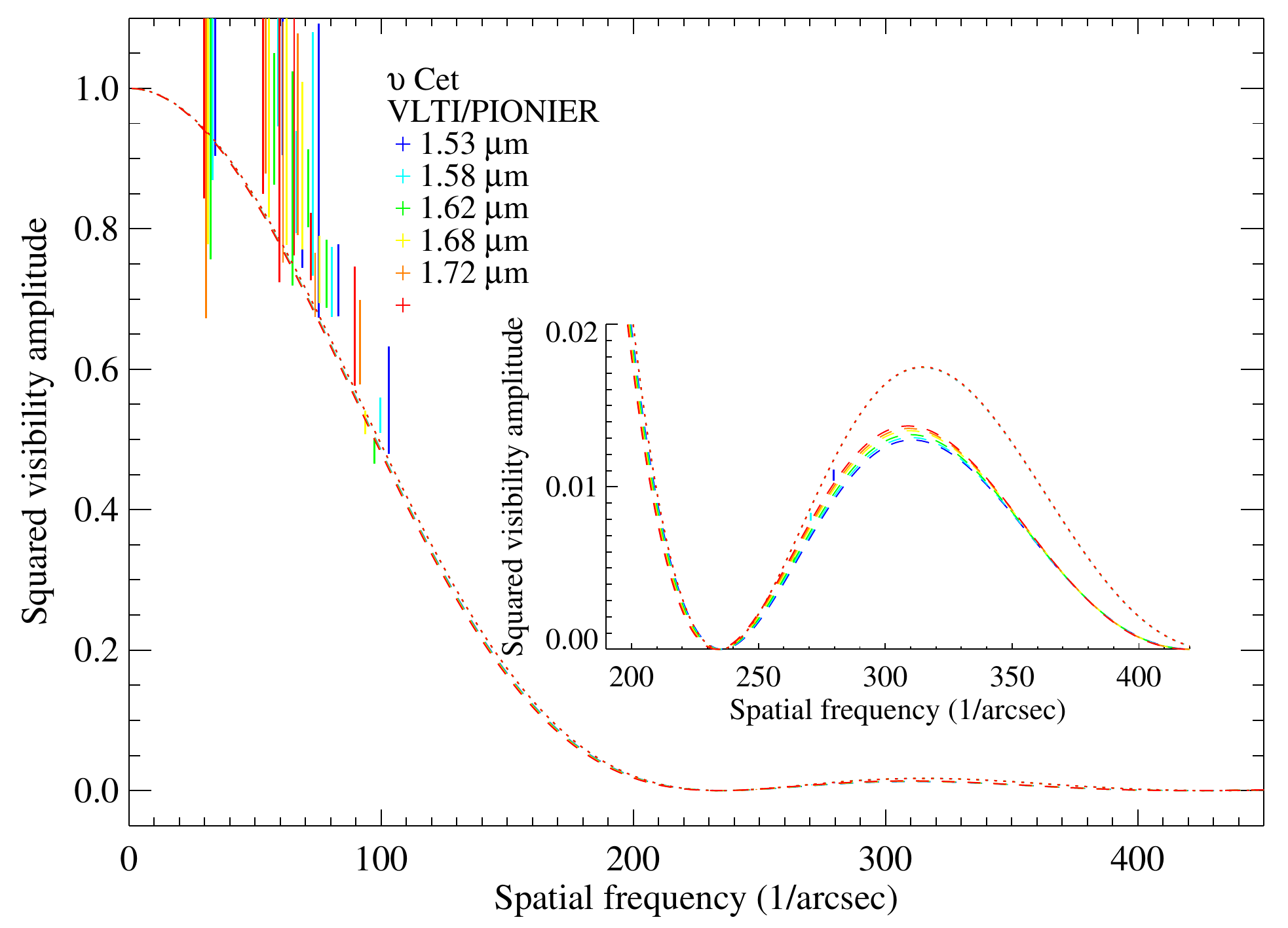}

  \includegraphics[width=0.49\hsize]{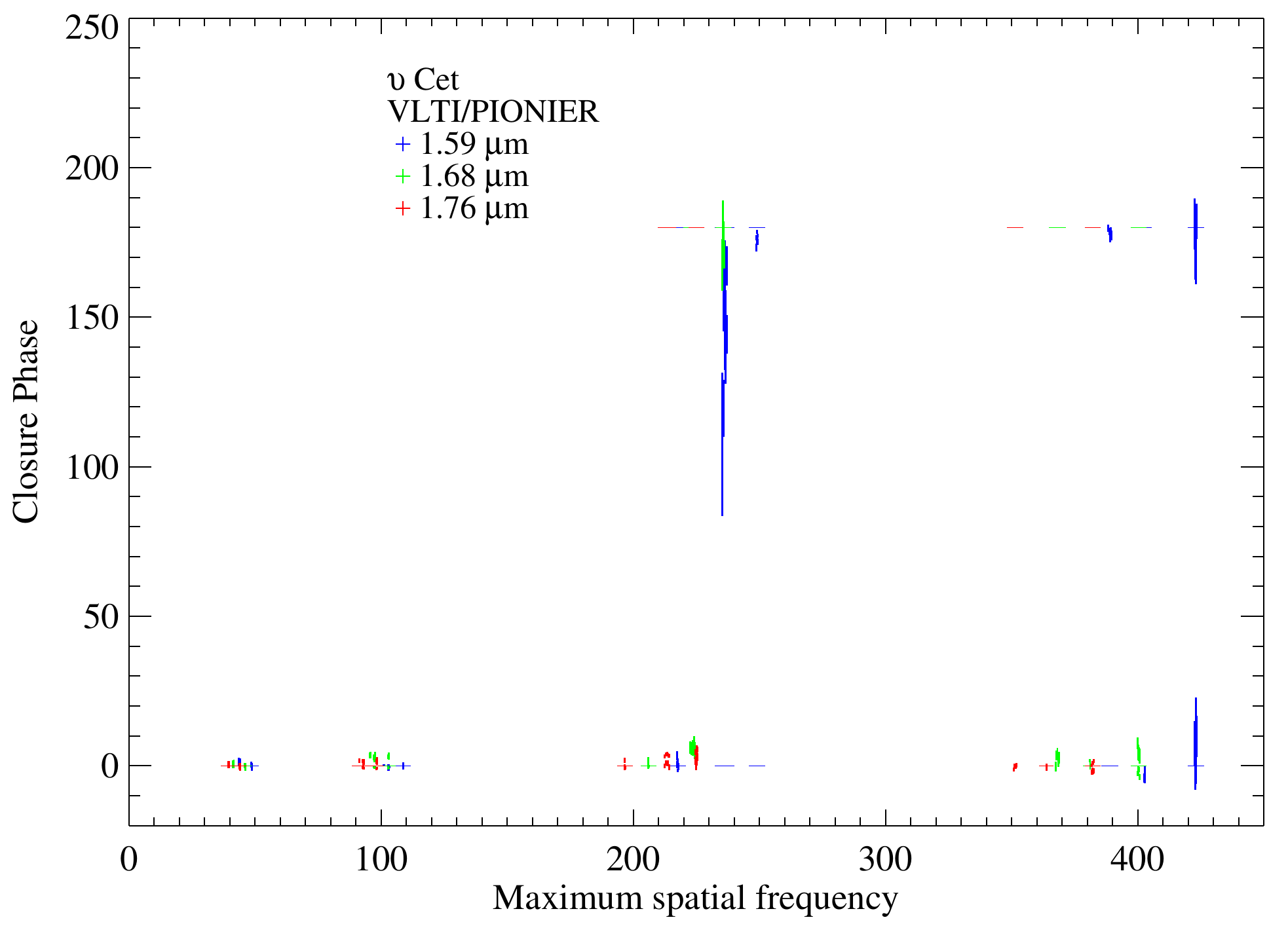}
  \includegraphics[width=0.49\hsize]{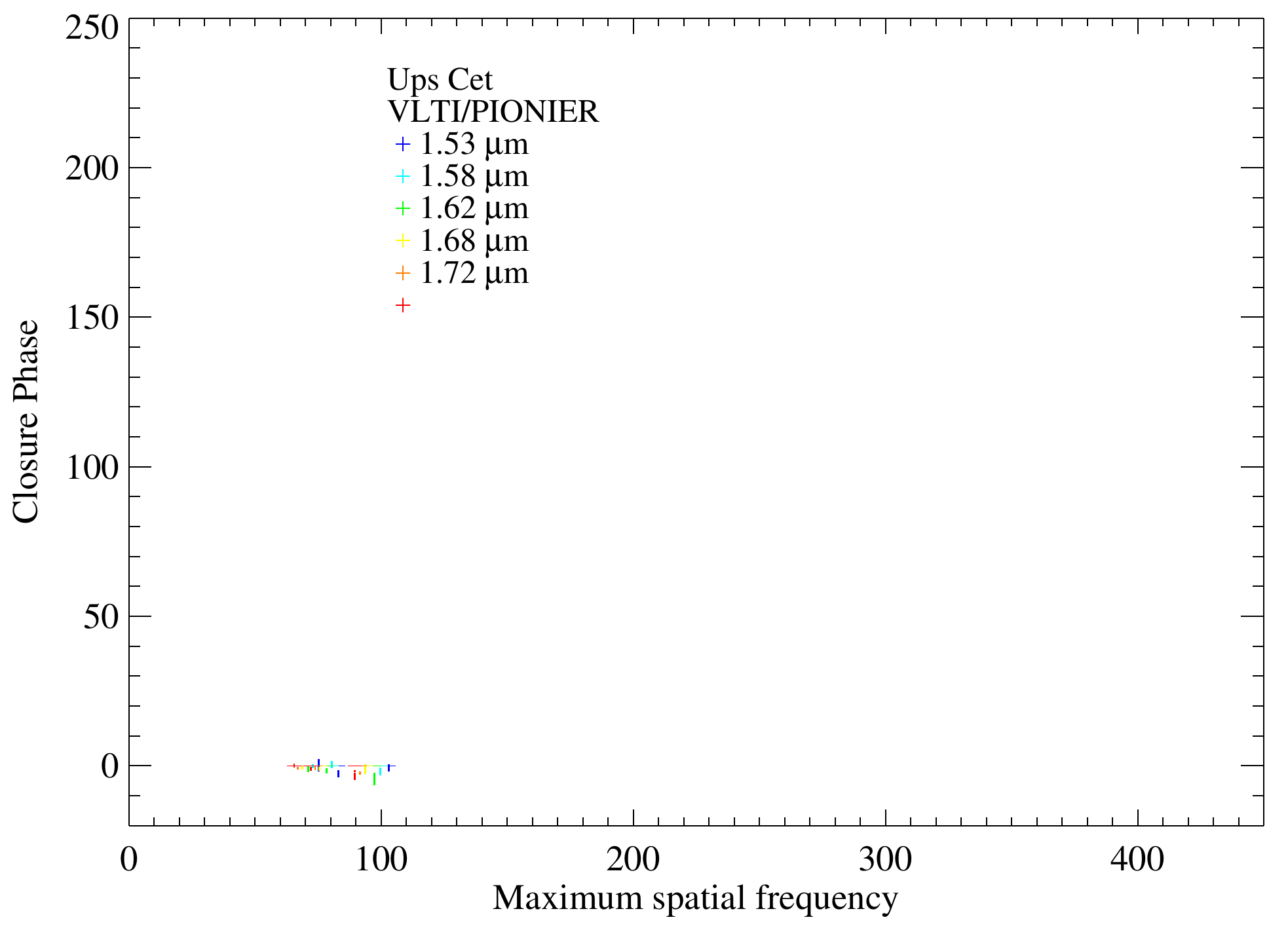}
\caption{PIONIER results for the check star 
$\upsilon$~Cet from 2014 (left) and 2015 (right).}
\label{fig:pionier20142015_upscet_visspfr}
\end{figure*}
In addition to the PIONIER observations of R~Scl, we secured a few 
data points on the nearby resolved
K5/M0 giant $\upsilon$~Cet  as a ``check star'',
using every baseline configuration that was used for R~Scl.
The aim of these observations was to test the validity of the 
visibility and closure phase data based on a star that can be 
well described a priori. To describe the star, we used the grid 
of PHOENIX model atmospheres as introduced by \citet{Arroyo2013} 
and chose an effective temperature of 4000\,K and a surface
gravity of $\log g=1$ based on the spectral type. We can expect
that a regular K or early M giant can be well described by
this model atmosphere.
We took data on $\upsilon$~Cet on 2014-08-25
(1 observation), 2014-09-02 (2 obs.), 2014-09-06 (1 obs.),
2014-09-07 (1 obs.), 2015-11-28 (1 obs.), and 2015-12-02 (1 obs.).
We tested the consistency of measured and modeled 
visibility values and closure phases, and investigated the
stability over the nights and between the 2014 and 2015 epochs.
The data reduction was performed in the same way as for R~Scl, 
and we used the same calibrator observations for $\upsilon$~Cet 
as for R~Scl.
We obtained best-fit Rosseland angular diameters of 
5.89\,$\pm$\,0.05\,mas and 5.88\,$\pm$\,0.05\,mas for the data 
obtained in 2014 and 2015, respectively. This shows that the 
two epochs are consistent, in particular in view of the change
of the detector of PIONIER between these epochs.
Figure~\ref{fig:pionier20142015_upscet_visspfr} shows the observed
and modeled squared visibility values and closure phases for 2014
and 2015. The low visibility values close to the visibility null 
and beyond are confirmed to be well consistent with the model
atmosphere and to be very accurate. In particular, the visibility
values in the second lobe are consistent with the model-predicted
strength of the limb-darkening and are significantly different to 
a uniform disk (UD) function as expected. The measured visibility 
values in the first
lobe with values larger than about 0.2-0.3 lie systematically above 
the model prediction for both epochs. We interpret this as due to 
known systematic calibration effects, caused by different
magnitudes or airmass between science and calibrator 
measurements.\footnote{http://www.eso.org/sci/facilities/paranal/instruments/pionier/manuals.html.}
This effect is thus also expected for high visibilities of R~Scl, 
and we needed to consider it for our R~Scl data as well (see main text). 
The closure phases show nicely the flip between 
0\,$\deg$ and 180\,$\deg$ at the position of the visibility minimum. 
The intermediate values for the spectral
channel at 1.59\,$\mu$m close to a maximum spatial frequency of
230\,cycles/arcsec are most likely caused by a flip within the width
of the spectral channel. Most closure phase positions are consistent 
between observed and modeled data. The few deviations may indicate
either a diameter that is slightly different to the closure
phase data or a variation of the strength of limb-darkening as a 
function of wavelength that is slightly different between 
observations and model.
\section{Imaging studies}
\label{sec:imagingstudies}
\begin{figure*}
  \includegraphics[width=0.18\hsize]{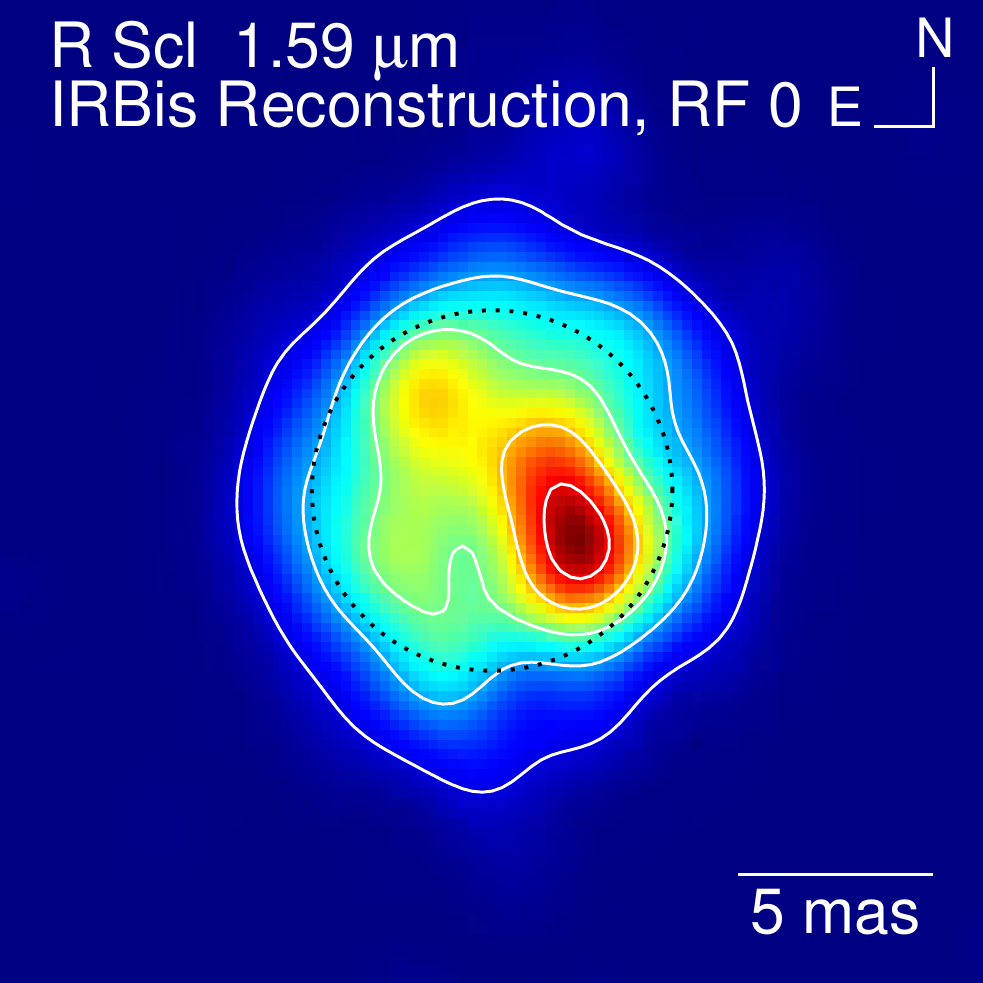}
  \includegraphics[width=0.18\hsize]{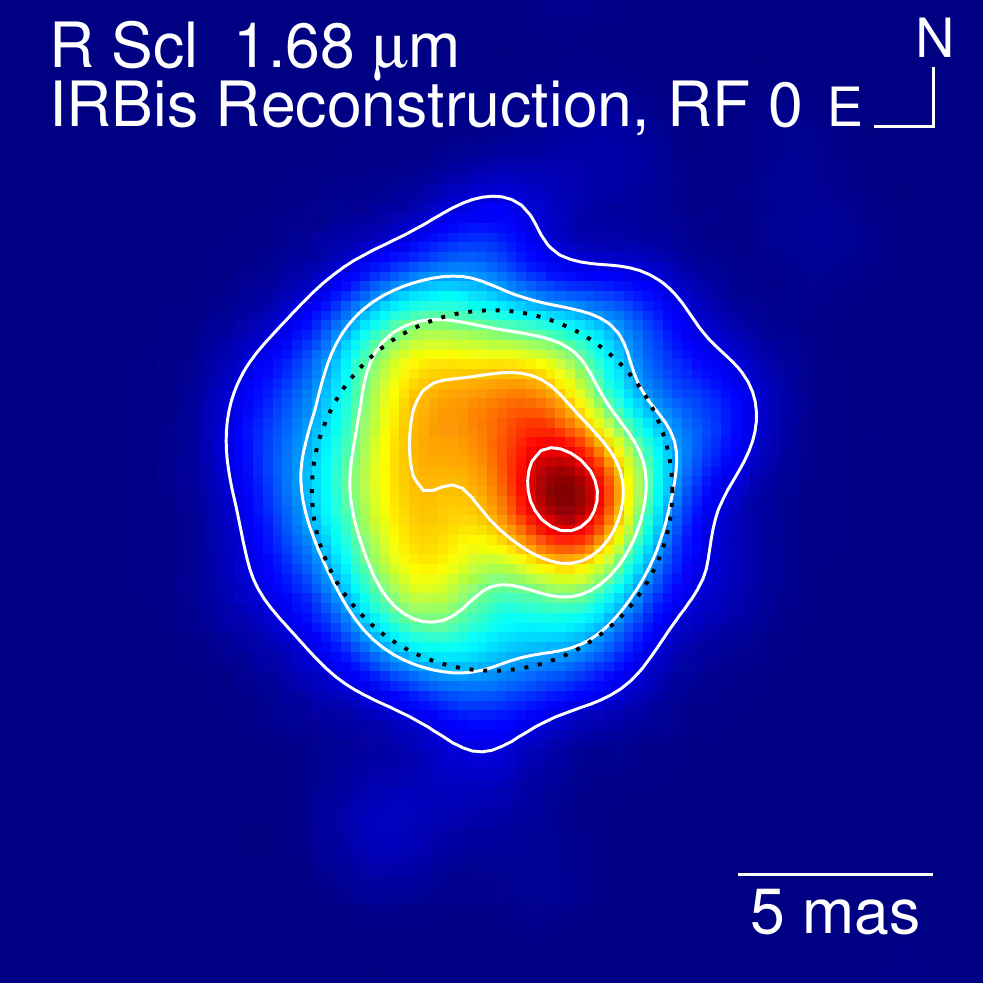}
  \includegraphics[width=0.18\hsize]{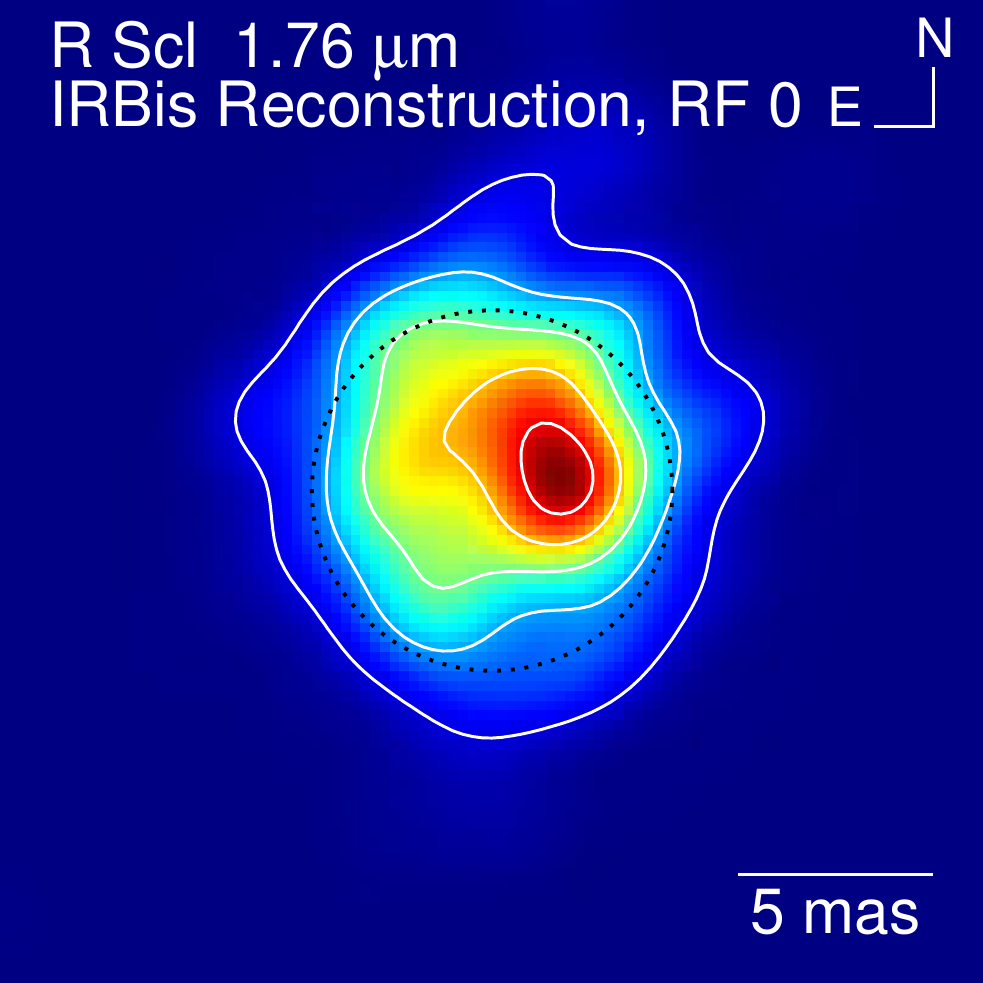}

  \includegraphics[width=0.18\hsize]{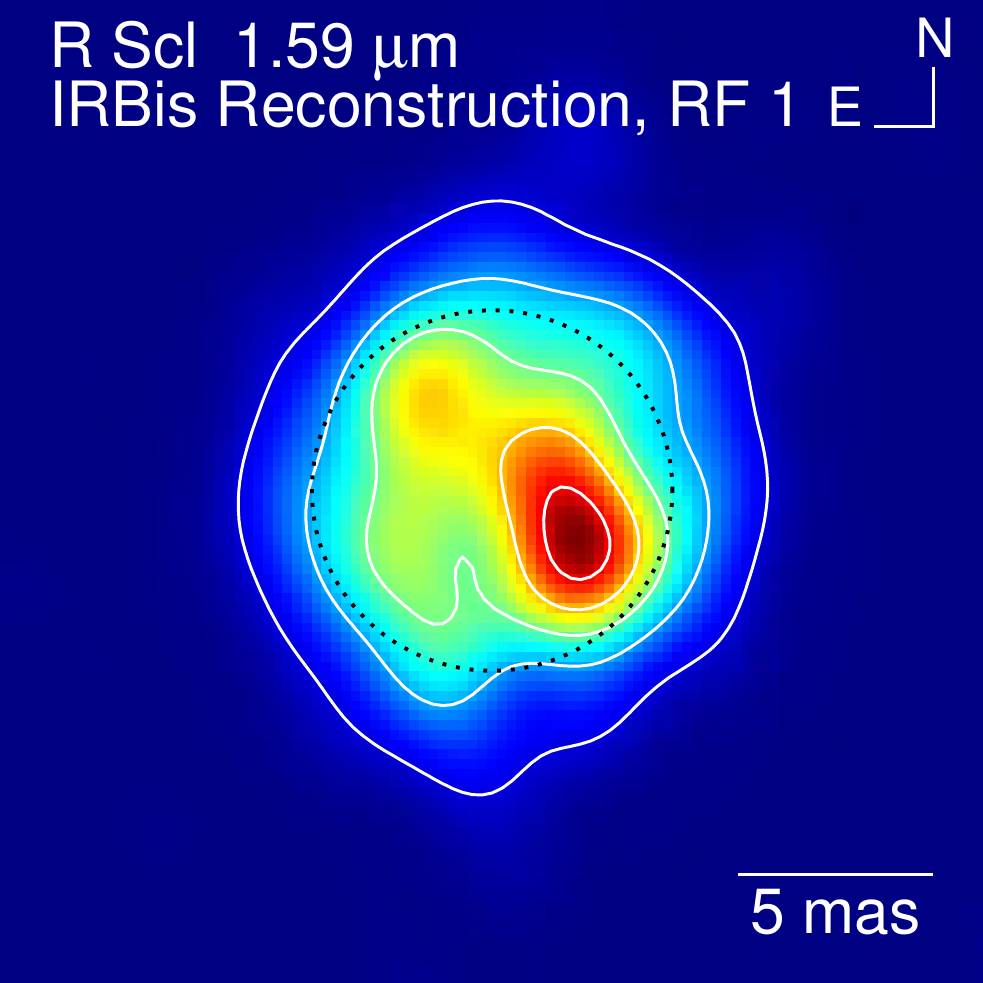}
  \includegraphics[width=0.18\hsize]{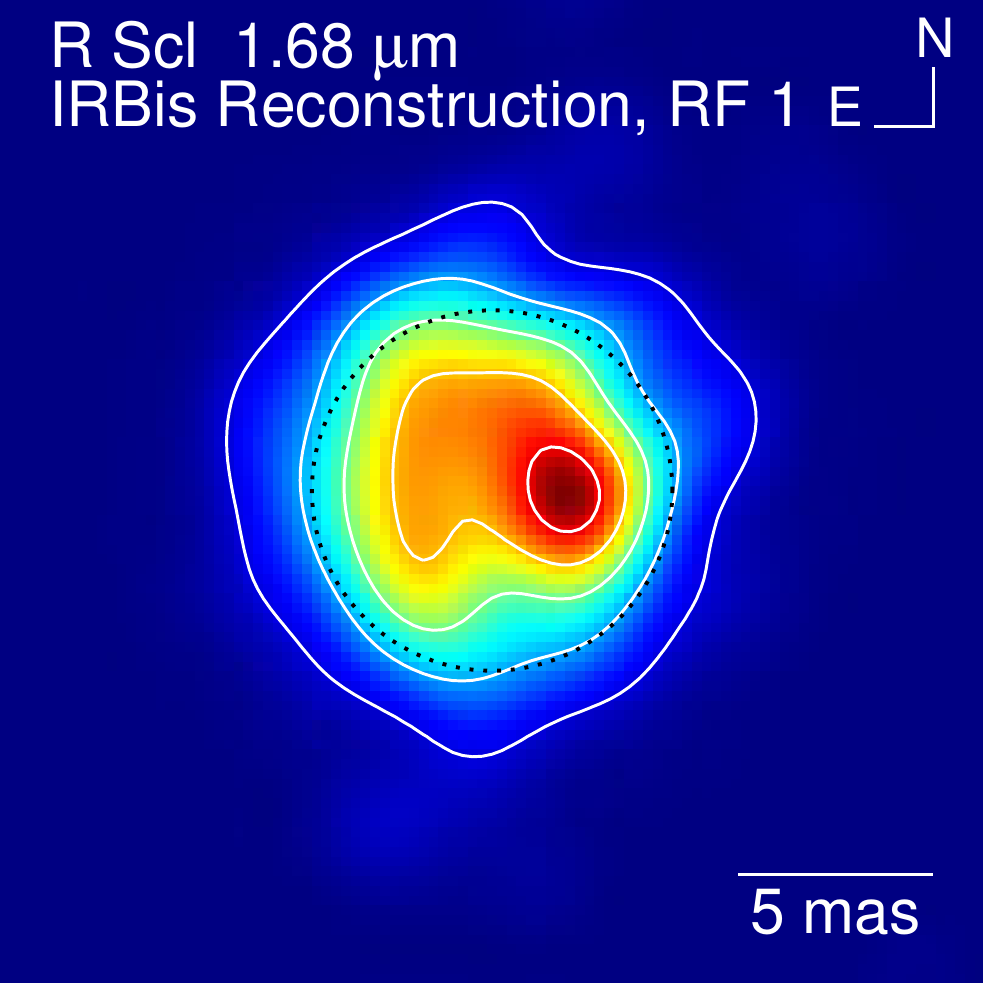}
  \includegraphics[width=0.18\hsize]{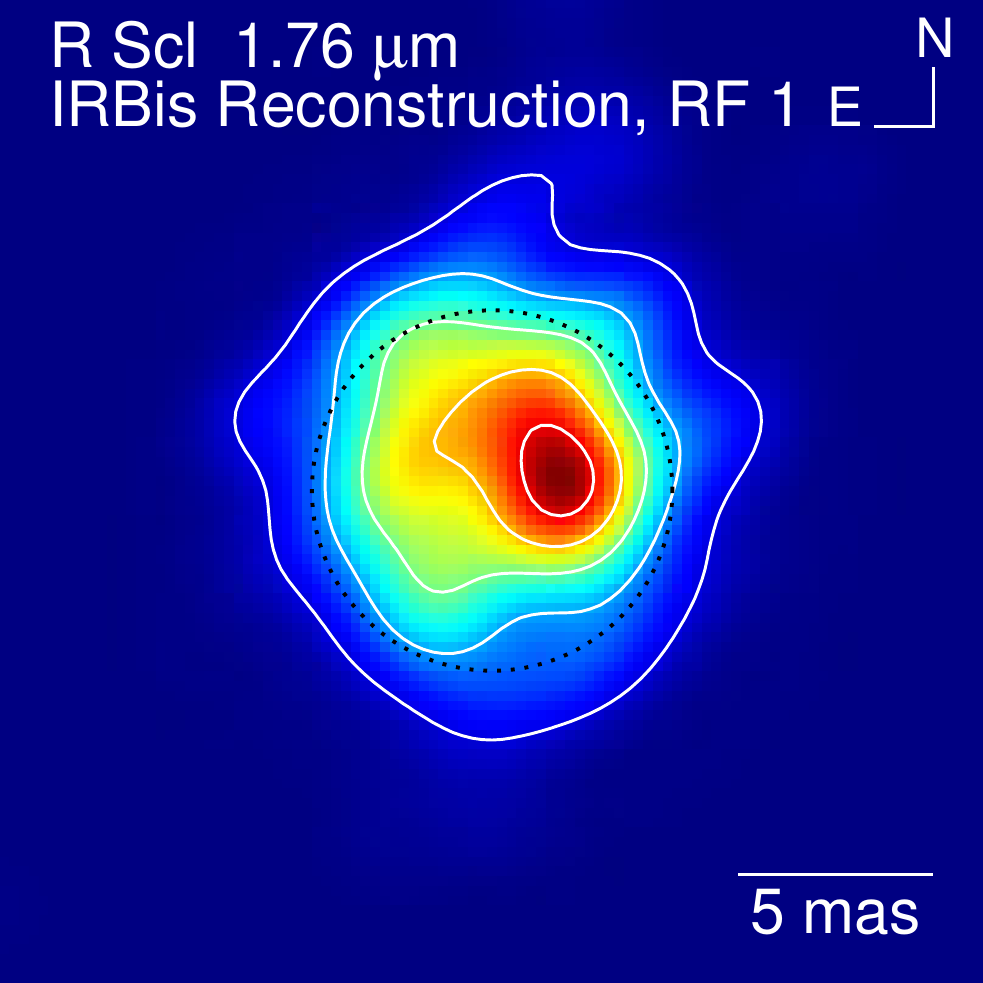}

  \includegraphics[width=0.18\hsize]{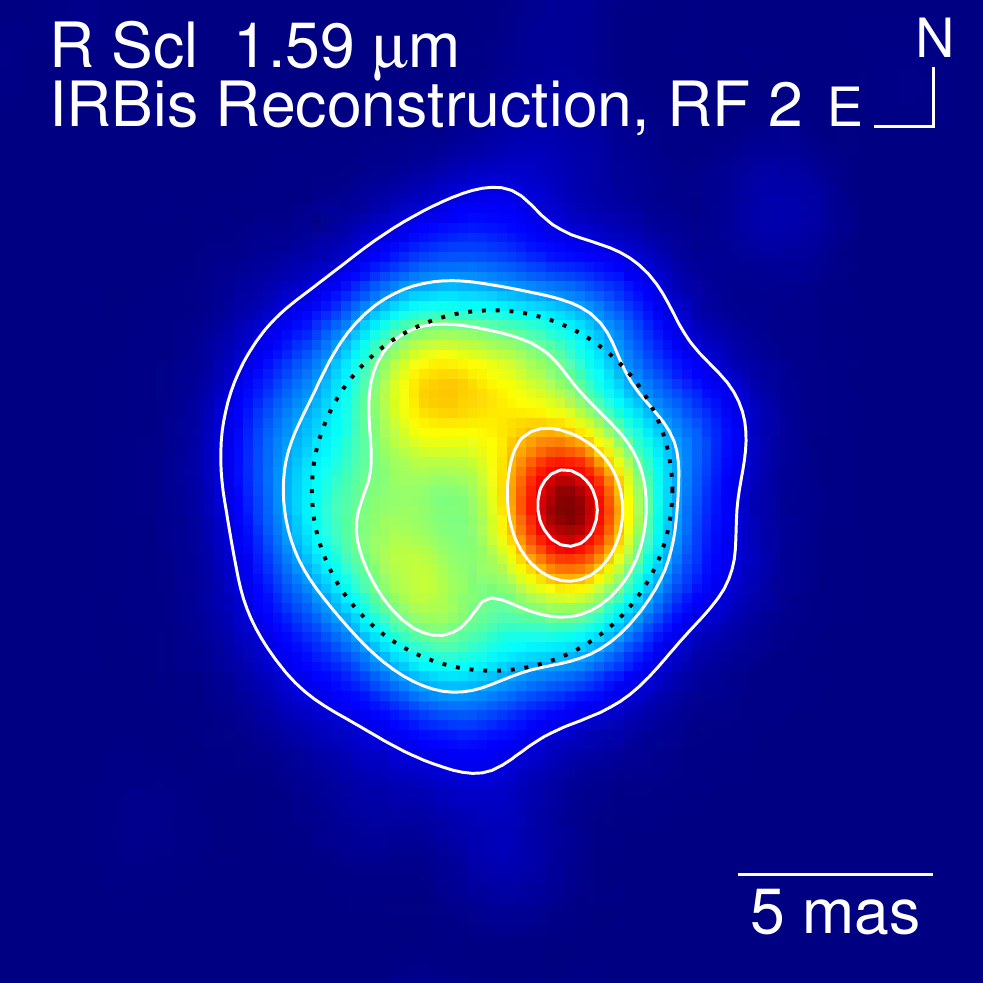}
  \includegraphics[width=0.18\hsize]{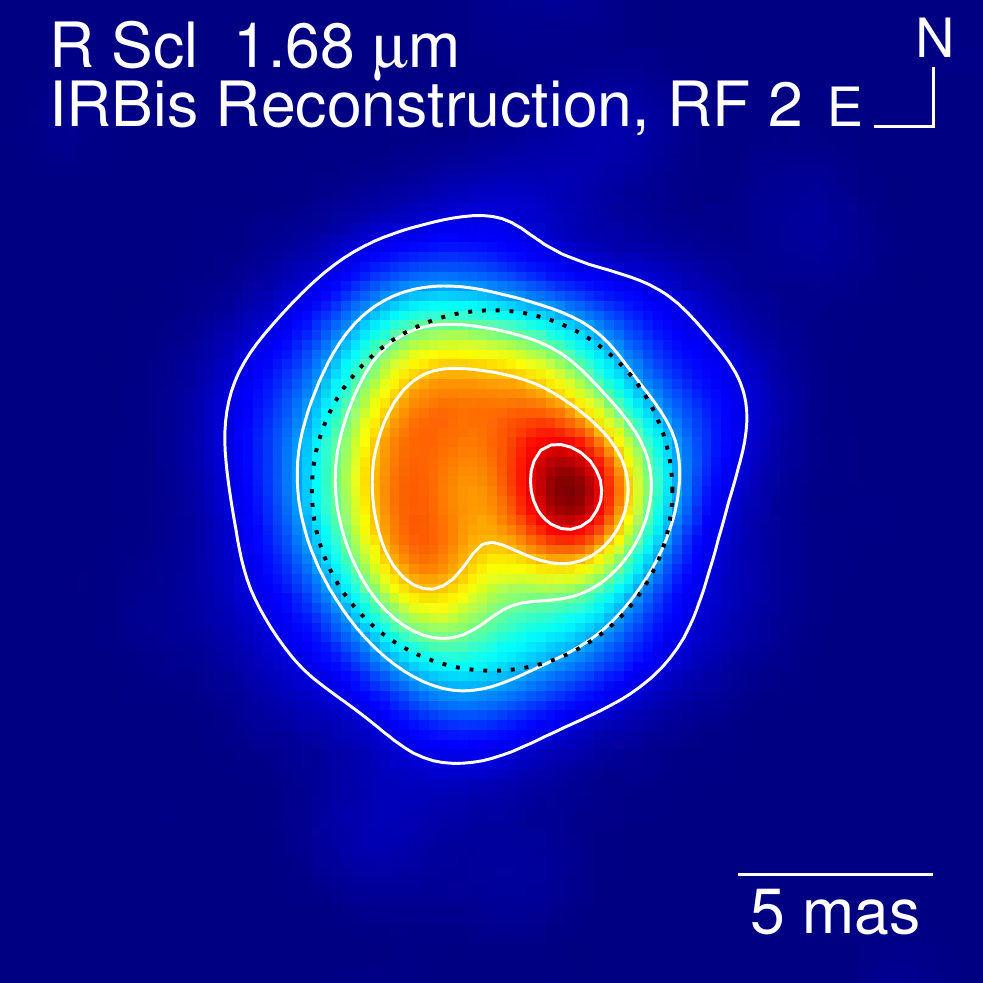}
  \includegraphics[width=0.18\hsize]{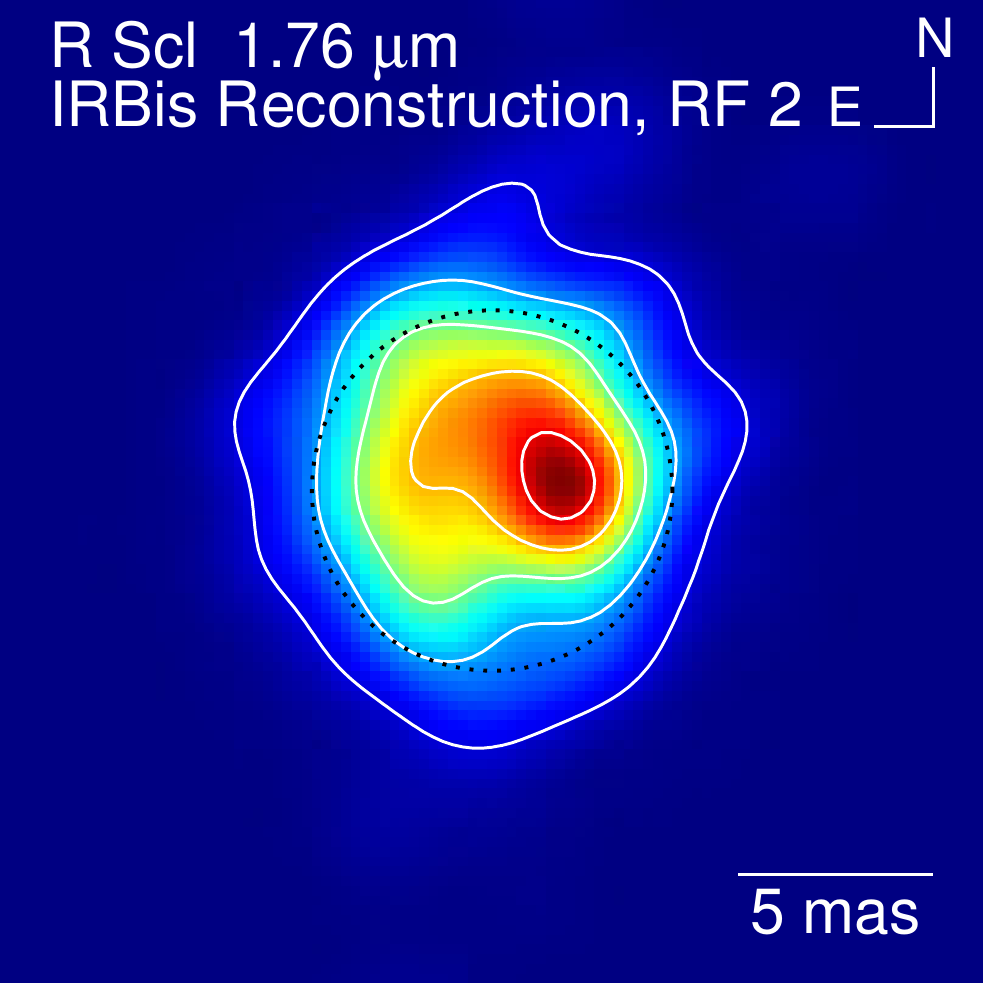}

  \includegraphics[width=0.18\hsize]{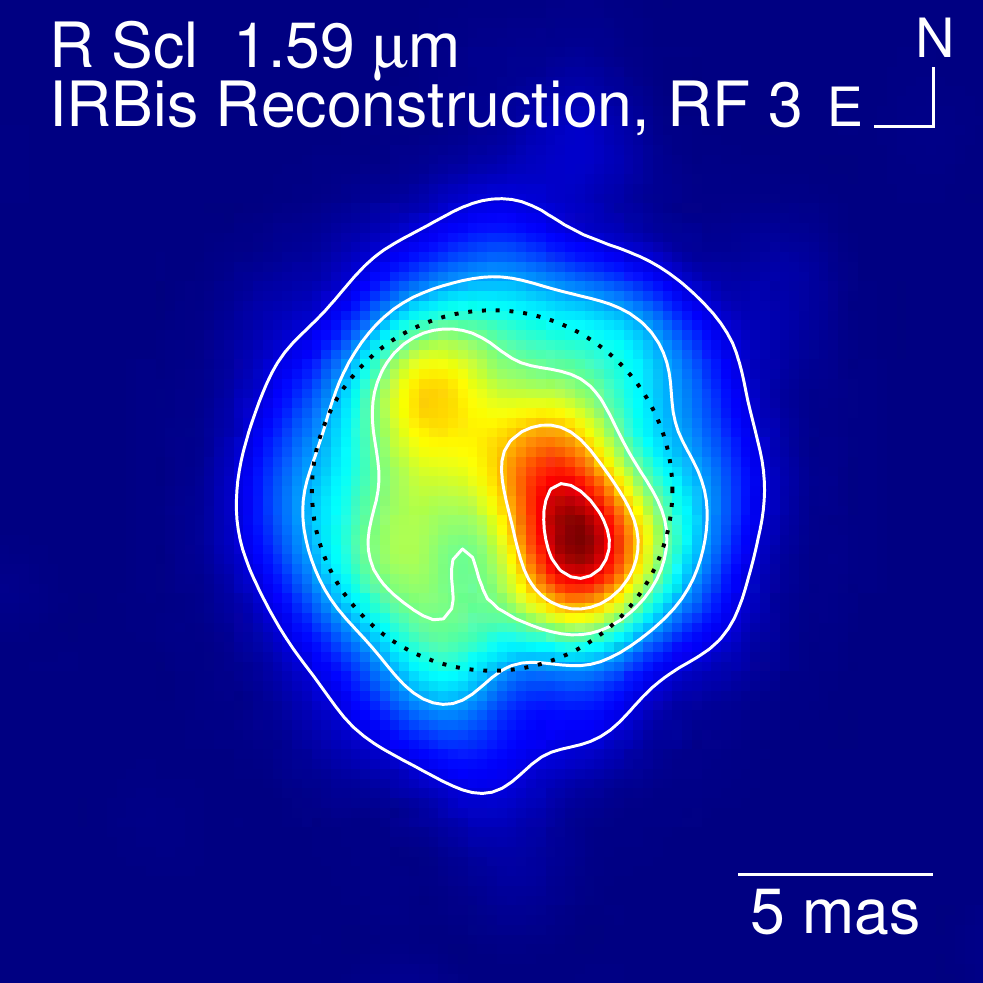}
  \includegraphics[width=0.18\hsize]{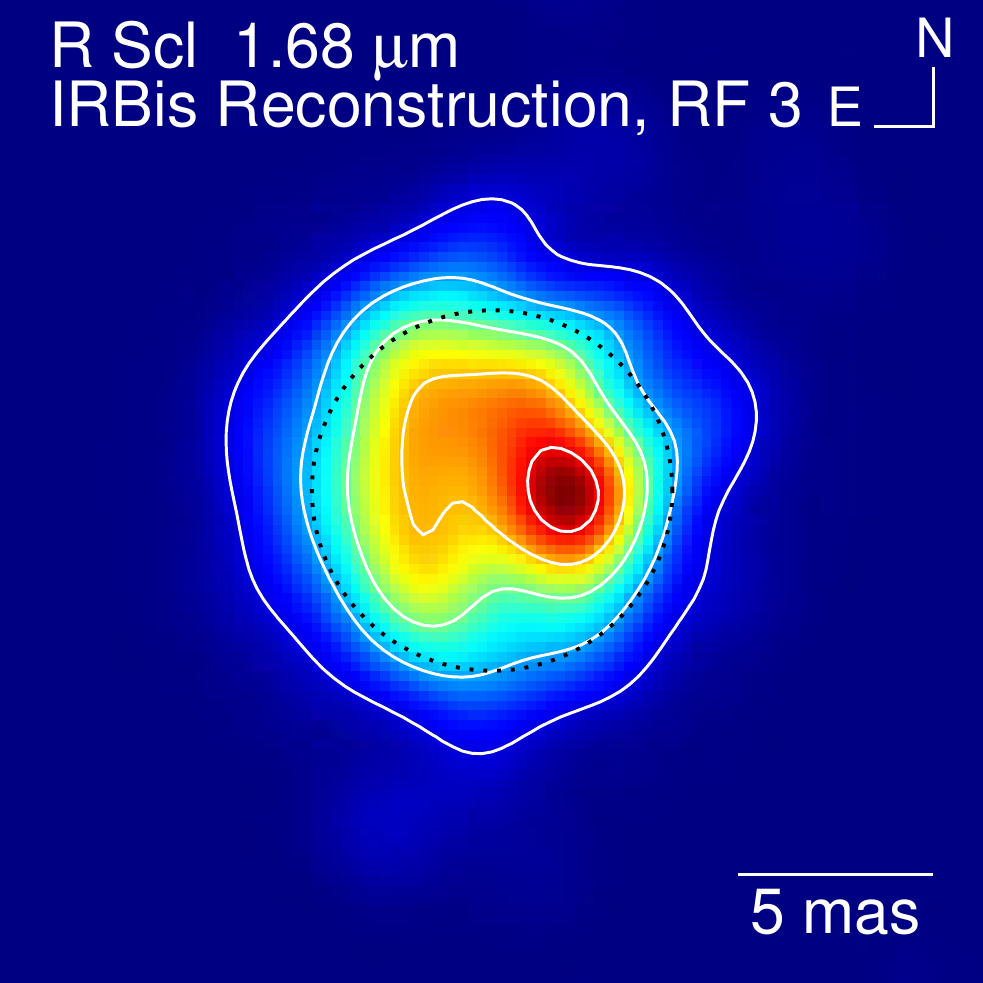}
  \includegraphics[width=0.18\hsize]{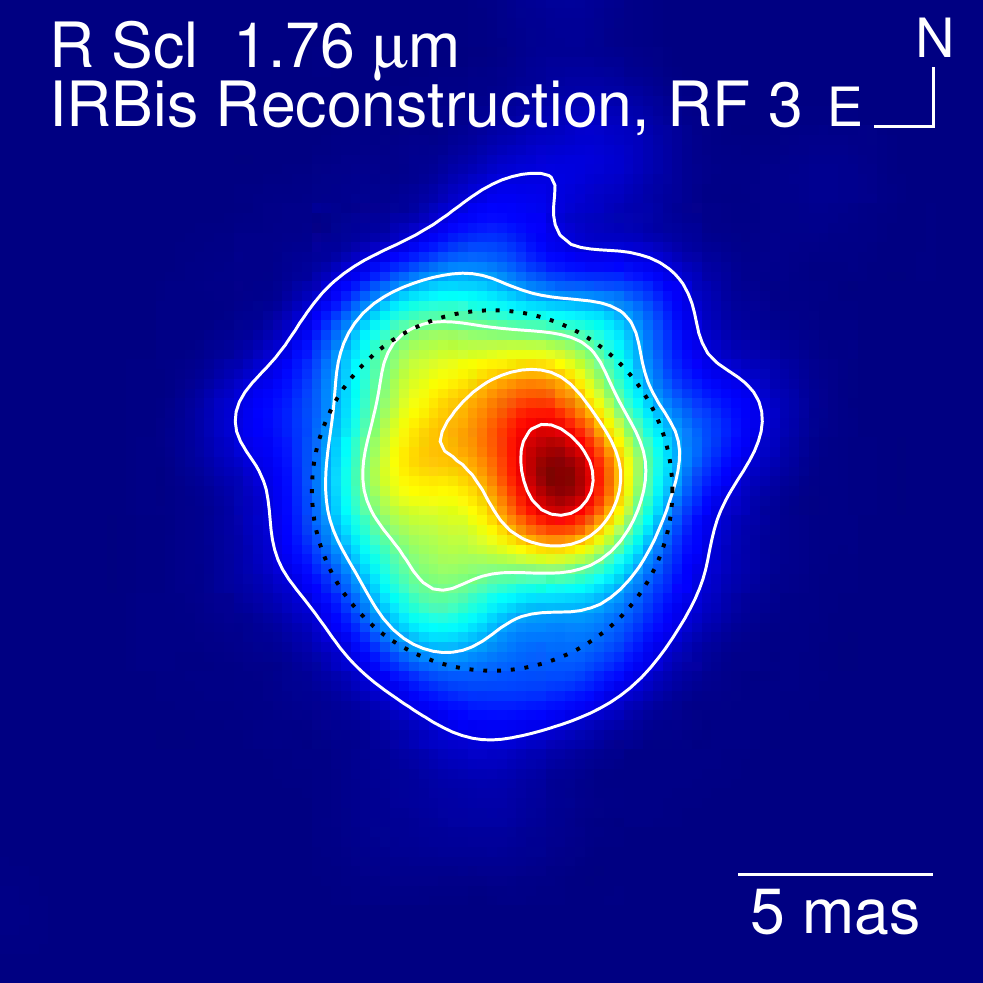}

  \includegraphics[width=0.18\hsize]{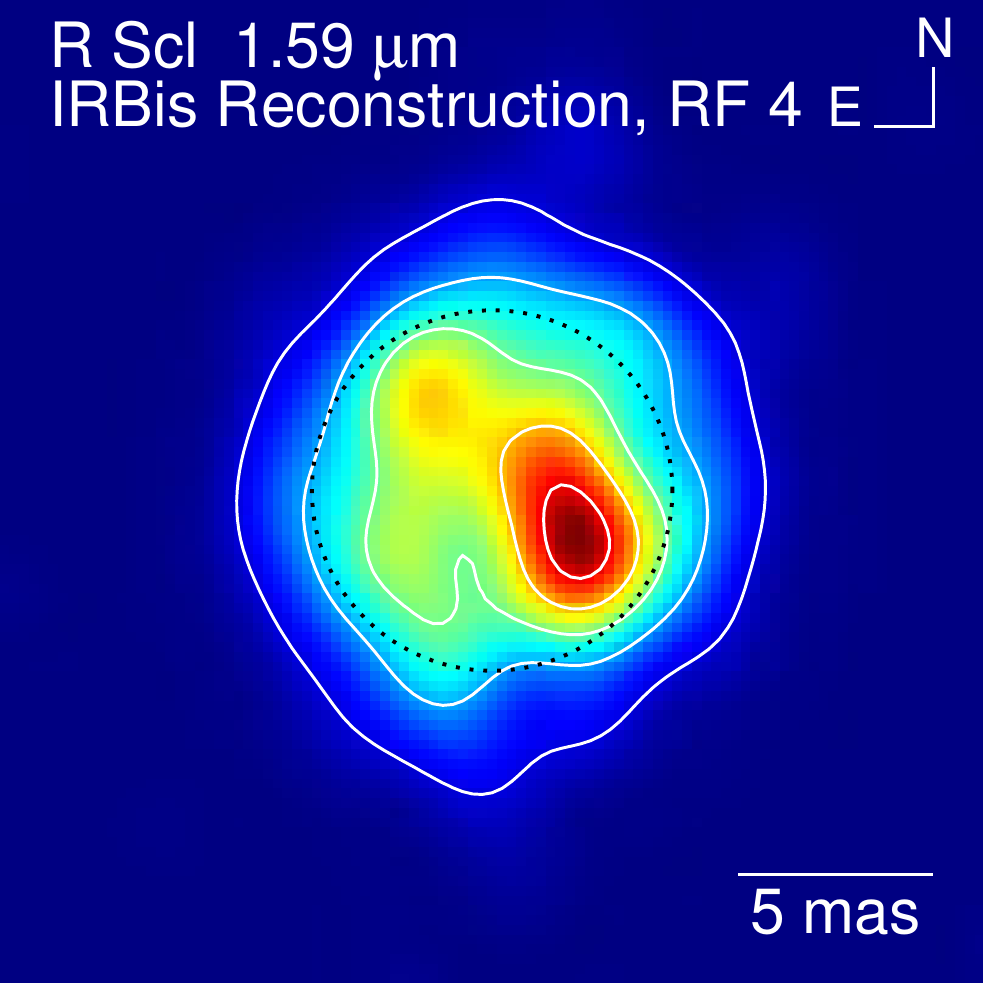}
  \includegraphics[width=0.18\hsize]{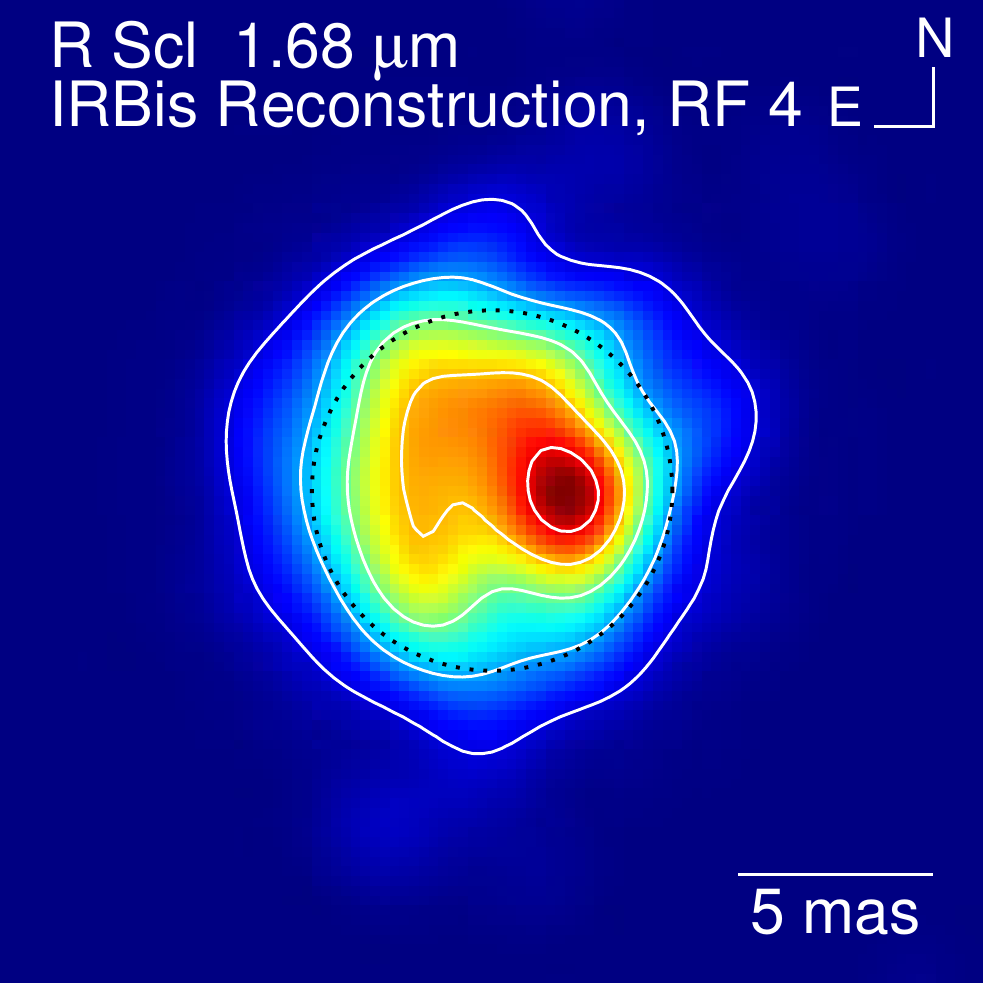}
  \includegraphics[width=0.18\hsize]{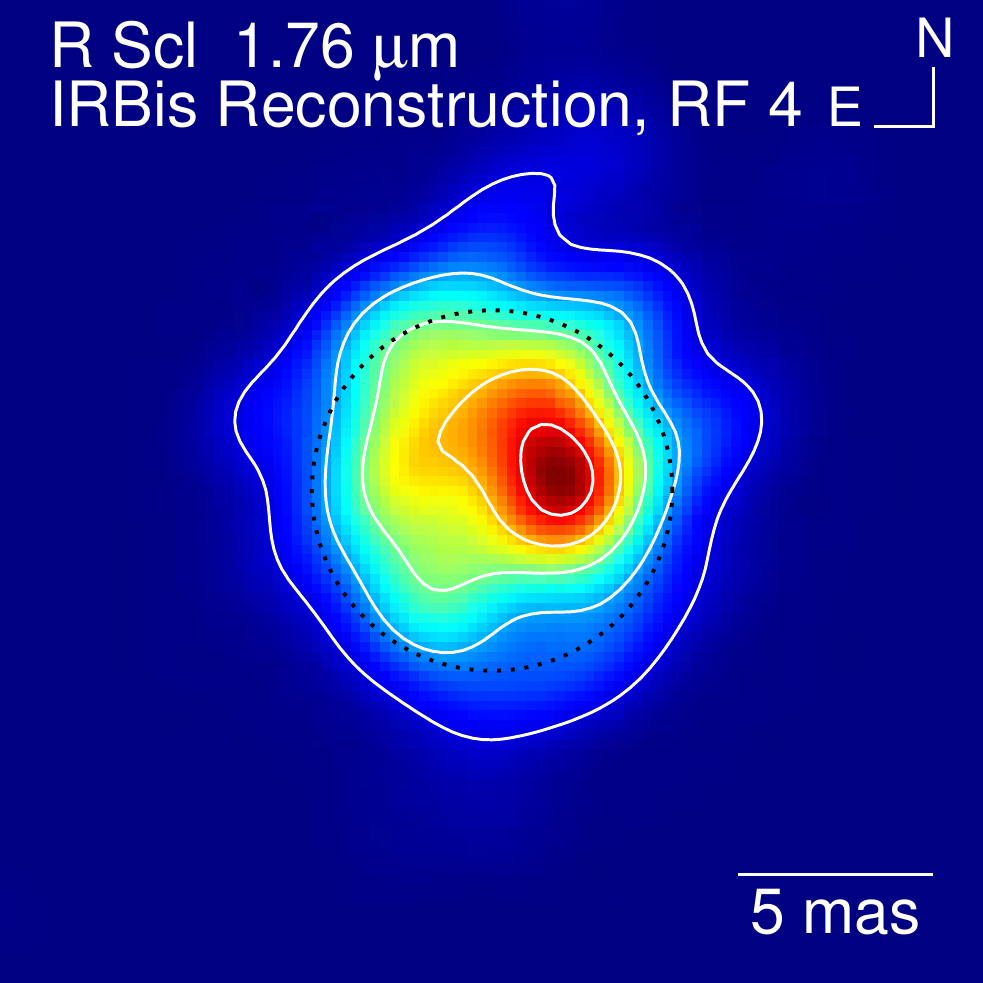}

  \includegraphics[width=0.18\hsize]{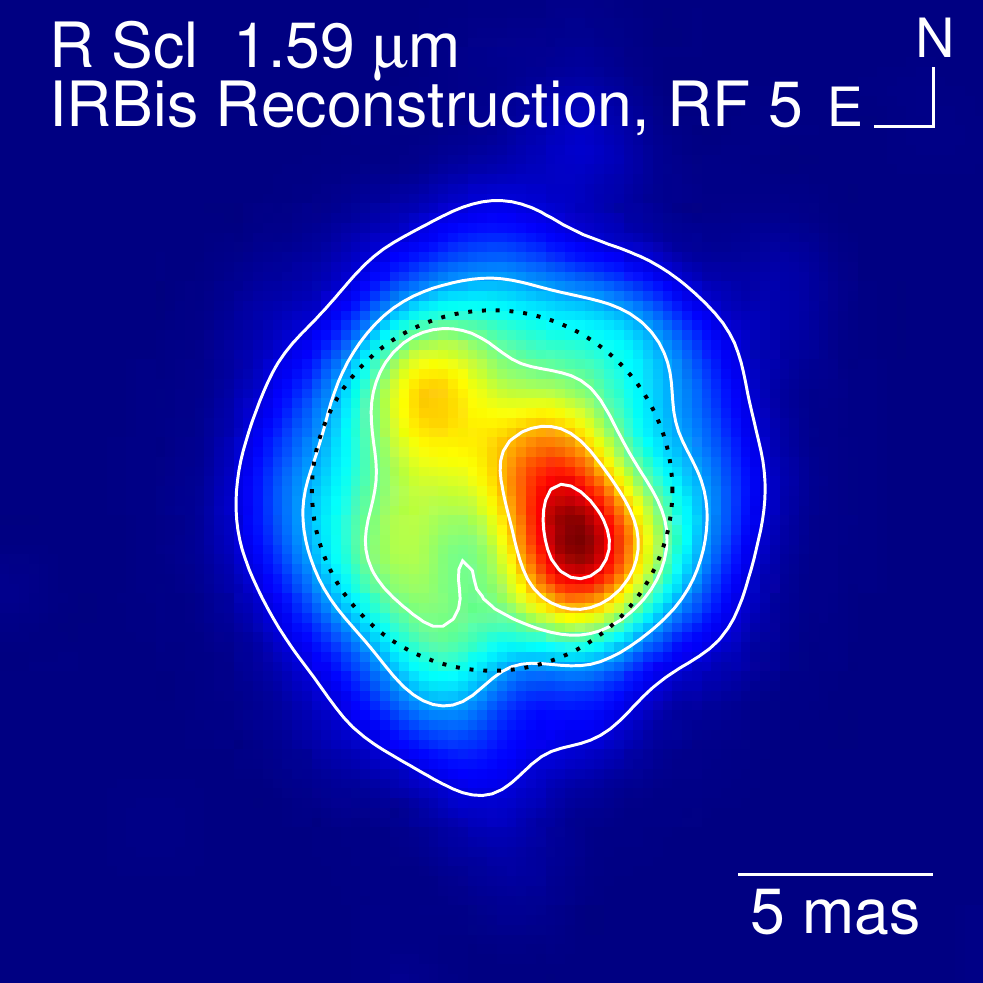}
  \includegraphics[width=0.18\hsize]{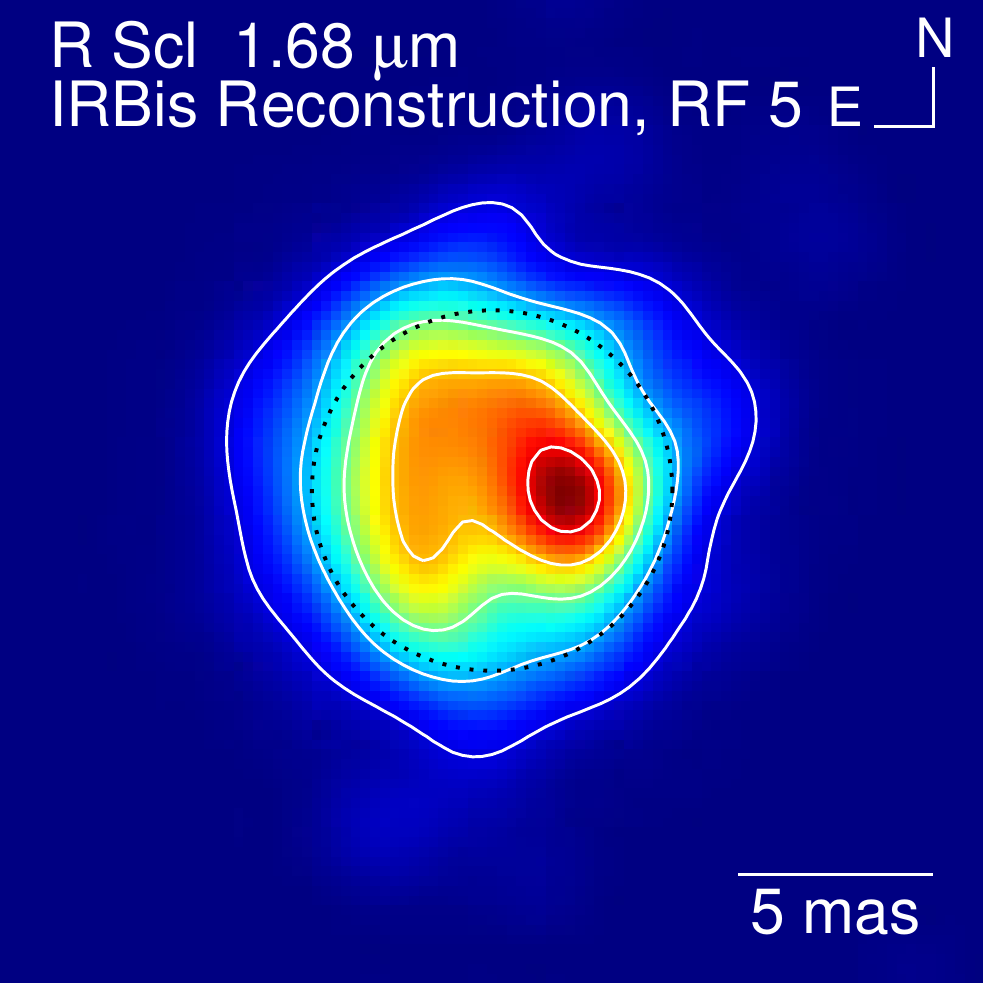}
  \includegraphics[width=0.18\hsize]{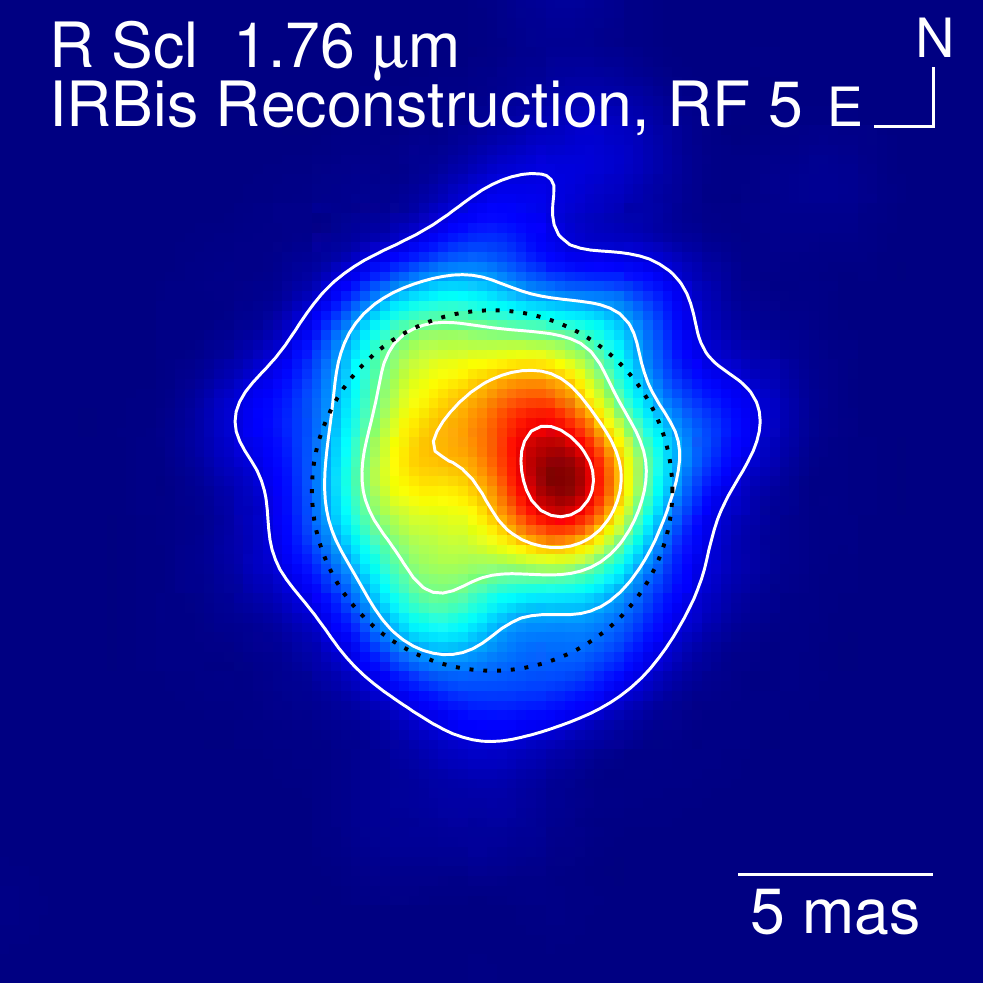}

  \includegraphics[width=0.18\hsize]{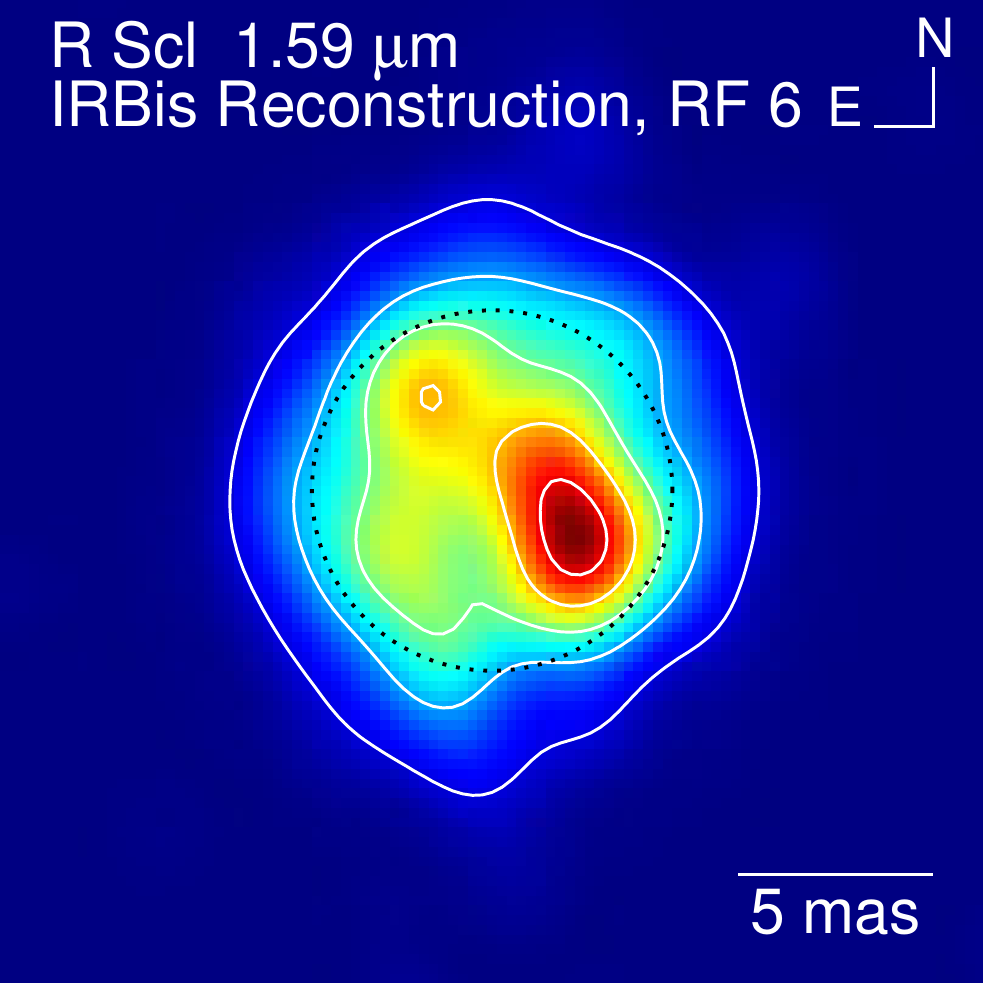}
  \includegraphics[width=0.18\hsize]{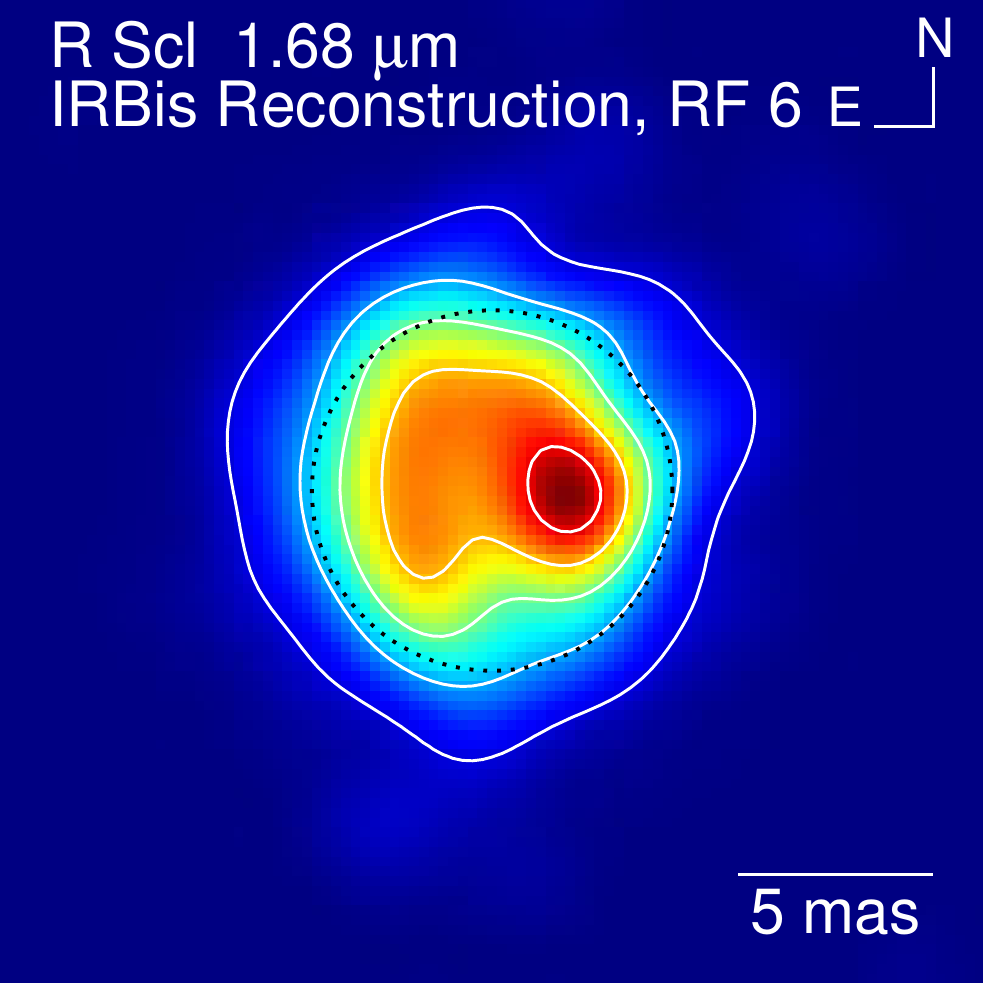}
  \includegraphics[width=0.18\hsize]{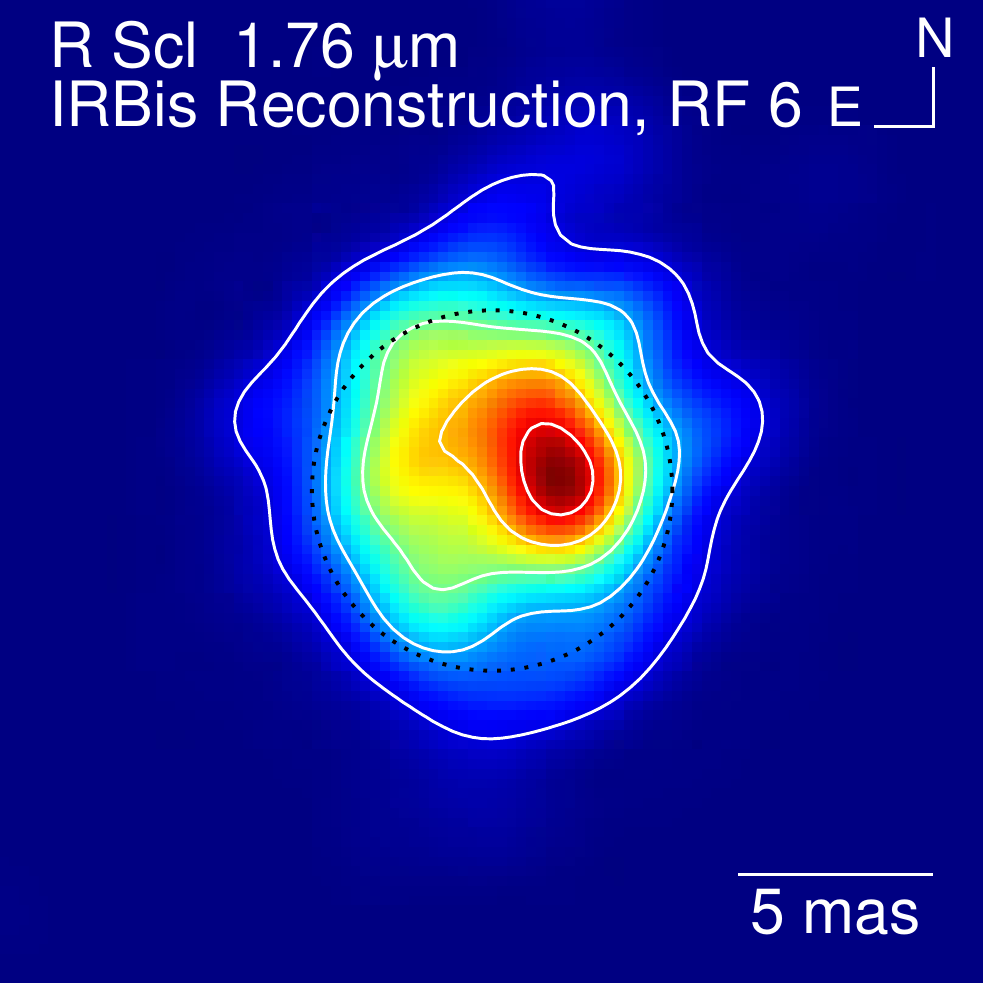}
\caption{R~Scl image reconstruction based on {\tt IRBis} with different 
regularization functions. The image reconstructions are based
on (from top to bottom) no regularization function, followed
by regularization functions no. 1--6 from 
\protect\citet{Hofmann2014}. 
}
\label{fig:pionier_image_IRB_differentreg}
\end{figure*}

We studied the effects on the reconstructed R~Scl images
using different regularization functions, different
imaging reconstruction packages with different algorithms,
and different start and prior images.
In particular, we used all the seven different regularization
methods that are available in the {\tt IRBis}
package \citep{Hofmann2014}, including no regularization at all
(regularization function 0).
For most of the reconstructions, we used a flat prior
(regularization functions 1, 3, 4, 5). The remaining
regularization functions (2 and 6) require us to use the
start image also as a prior image.
The regularization functions of {\tt IRBis} are described 
in \citet{Hofmann2014}, except for regularization functions 5
and 6, which were added later.
Regularization function 5 is called ''smoothness'' and is defined as
\begin{equation*}
H5[o_k(x,y)]  := \sum_{dx,dy=-1,1} \sum_{x,y}\frac{{|o_k(x,y)-o_k(x+dx,y+dy)|^2}}{prior(x,y)}.
\end{equation*}
Regularization function 6 is called ''quadratic Tikhonov'' and
is defined as:
\begin{equation*}
H6[o_k(x,y)]  := \sum_{x,y} |o_k(x,y)-prior(x,y)|^2.
\end{equation*}
Figure~\ref{fig:pionier_image_IRB_differentreg} shows the 
reconstructed images based on the seven regularization functions
(0: no regularization; 1--6: regularization functions 1--6).
Table~\ref{tab:image_qc} lists the quality parameters
for each of the reconstructions as defined by \citet{Hofmann2014}.
Here, $\chi^2_{V^2}$ is the reduced $\chi^2$ between the measured and
synthetic squared visibility values, $\chi^2_{CP}$ is the reduced 
$\chi^2$ between the measured and synthetic closure phase values, and
$\rho\rho$ is the ratio between the sum of all positive residuals and
the sum of all negative residuals. The quantity $q_\mathrm{rec}$ adds
up the differences from unity of the $\chi^2$ and $\rho\rho$
values, providing one final quality parameter. All reconstructed
images as well as the quality parameters are very similar among the
different regularization functions. The $\chi^2_{V^2}$ values
are larger than unity because of the known systematic calibration 
effect described in the main text and in appendix~\ref{sec:checkstar},
which increases across the three spectral channels at 1.59\,$\mu$m,
1.68\,$\mu$m, and 1.76\,$\mu$m.
We chose to use the reconstruction based on regularization function 3
as the final result.

\begin{figure*}
  \includegraphics[width=0.33\hsize]{figures/RSCL-IRB-159-3f.pdf}
  \includegraphics[width=0.33\hsize]{figures/RSCL-IRB-168-3f.pdf}
  \includegraphics[width=0.33\hsize]{figures/RSCL-IRB-176-3f.pdf}

  \includegraphics[width=0.33\hsize]{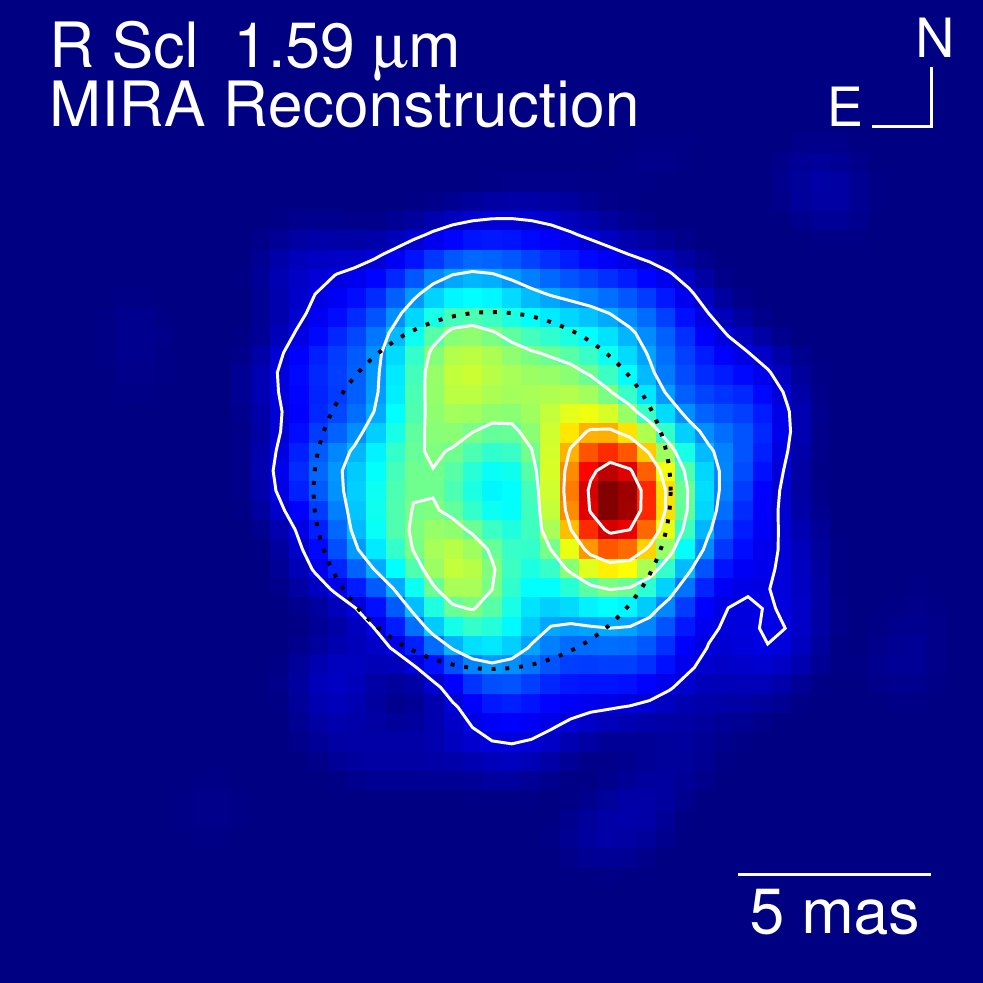}
  \includegraphics[width=0.33\hsize]{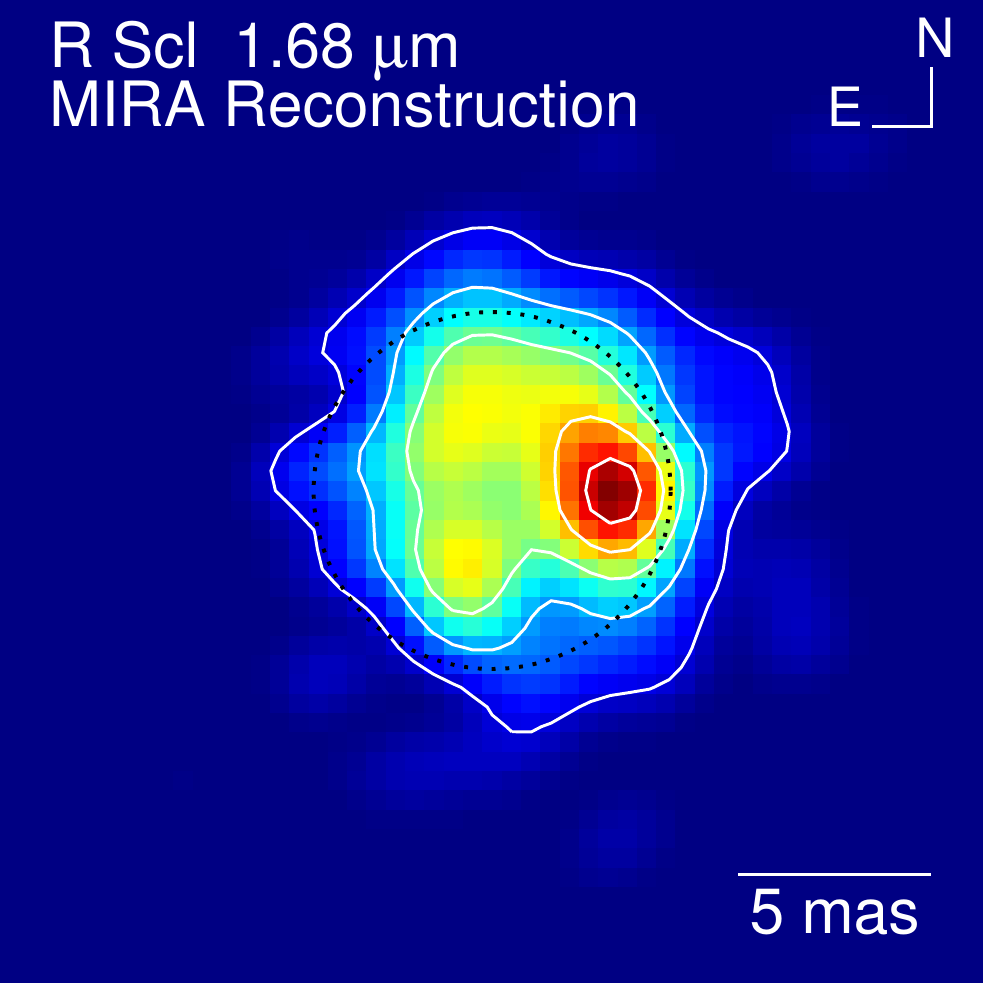}
  \includegraphics[width=0.33\hsize]{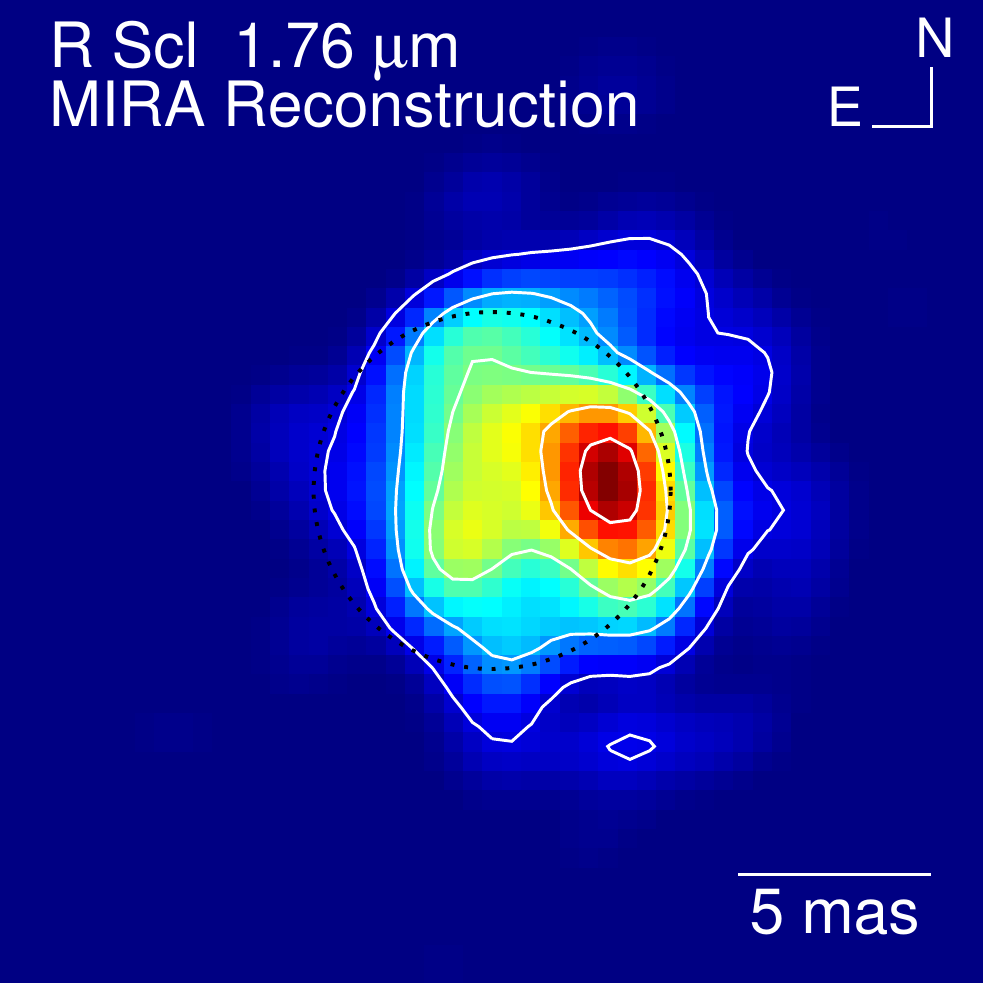}

  \includegraphics[width=0.33\hsize]{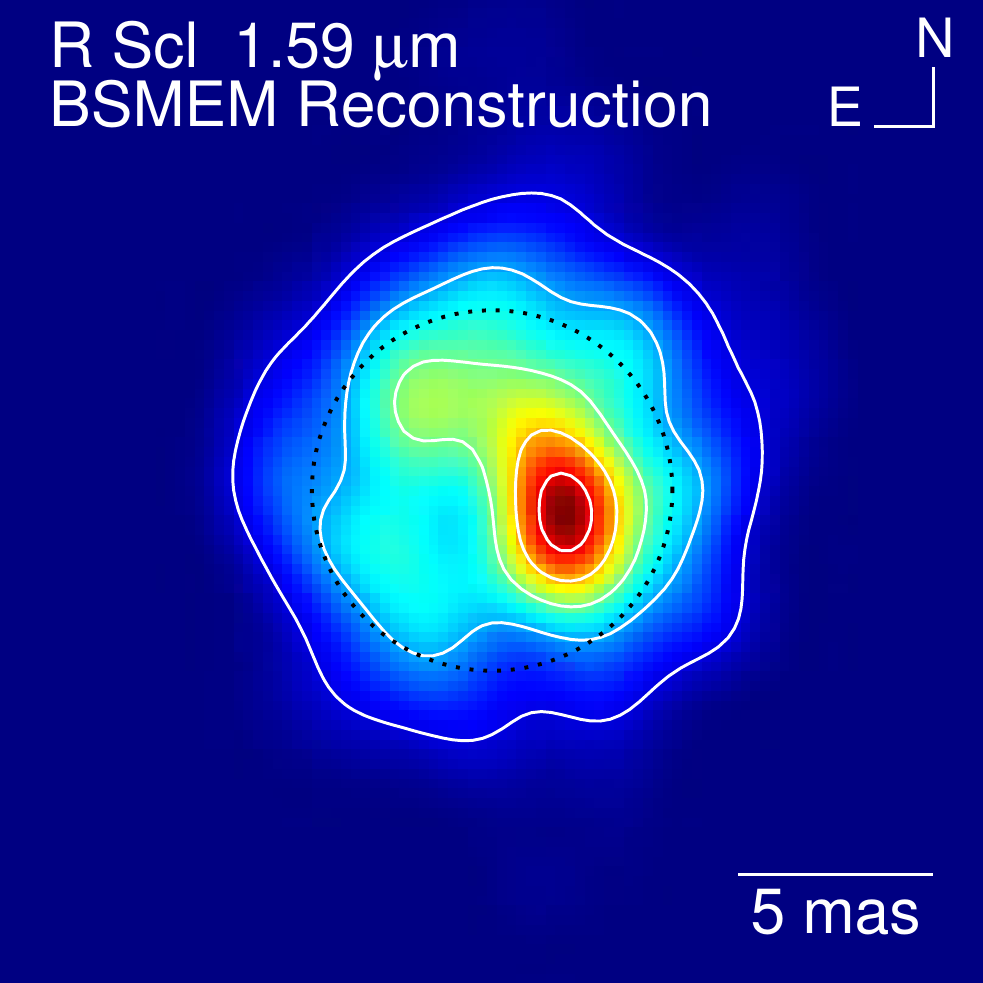}
  \includegraphics[width=0.33\hsize]{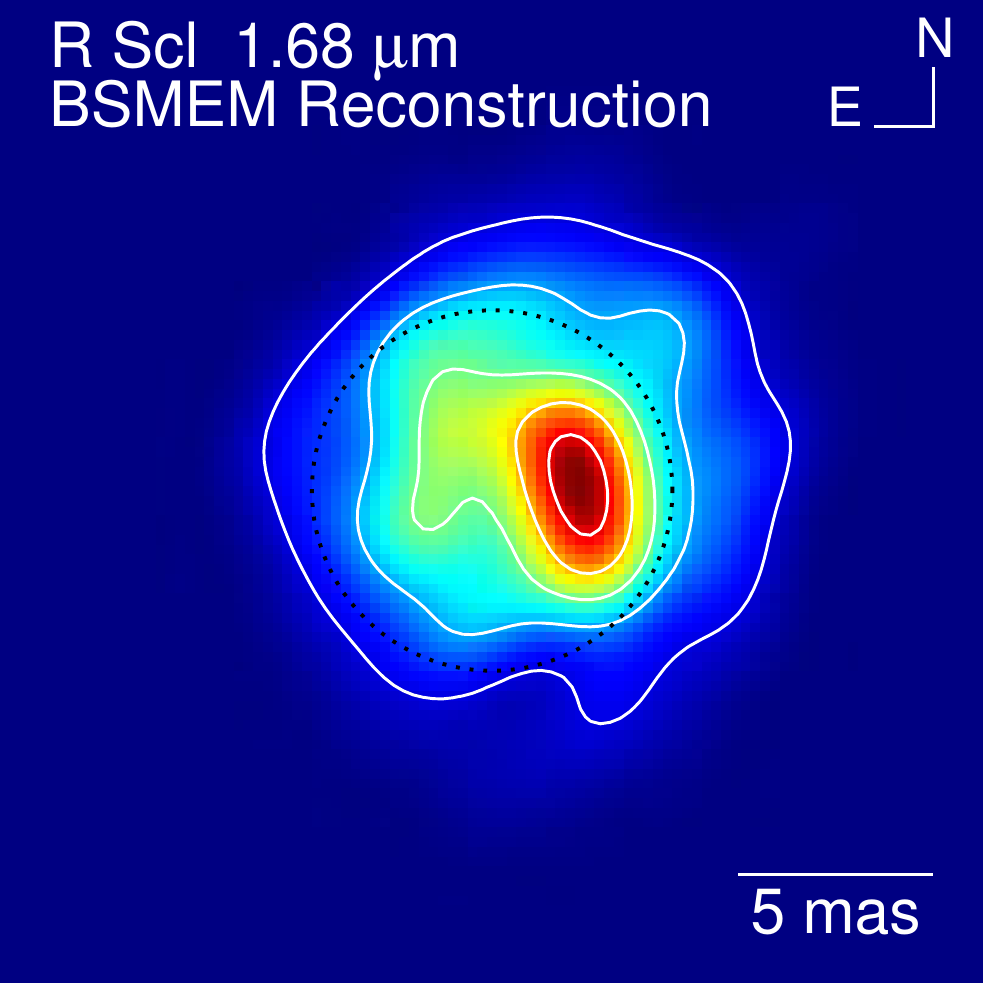}
  \includegraphics[width=0.33\hsize]{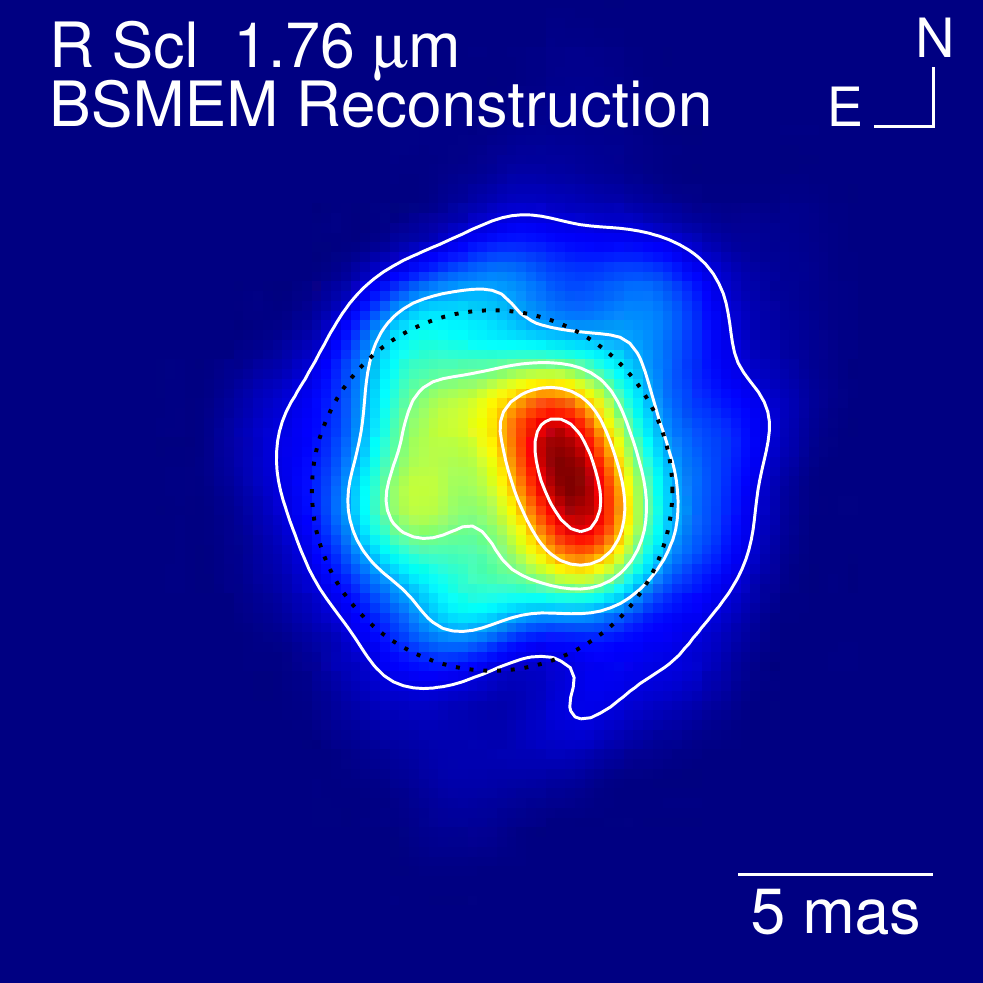}
\caption{R~Scl image reconstructions based on different image
reconstruction packages. For comparison, the top row shows the 
reconstructions based on {\tt IRBis} \protect\citep{Hofmann2014}
that we adopted as the final result
(as shown in Fig.~\protect\ref{fig:pionier_image_IRB}), followed
by reconstructions based on (middle row) {\tt MiRA} 
\protect\citep{Thiebaut2008}
and (bottom row) {\tt BSMEM} \protect\citep{Buscher1994}.
The {\tt MiRA} reconstruction uses a UD fit as a start image,
smoothness as regularization function, and does not use a prior. 
The {\tt BSMEM} reconstruction uses the model atmosphere images
as first start images and priors and is based on maximum entropy 
as regularization function. The MiRA reconstructions use double
the pixel size compared to the other two reconstructions.}
\label{fig:pionier_image_diffpack}
\end{figure*}

Furthermore, we reconstructed images using two other image 
reconstruction packages, namely {\tt MiRA} \citep{Thiebaut2008} 
and {\tt BSMEM} \citep{Buscher1994}. For the {\tt MiRA} 
reconstructions, we used smoothness as regularization function, 
a simple UD fit as a start image, and we used a flat prior image. 
For the {\tt BSMEM} reconstruction, we used a two-step approach. 
We first used the model images as 
start images to reconstruct a first image that contains only
the shorter baselines up to spatial frequencies of
about 200\,cycles/arcsec. In a second step, we used a smoothed version
of the first reconstructions as start images for a final
reconstruction based on all data.
Figure~\ref{fig:pionier_image_diffpack} shows the resulting
reconstructions compared to the final result from {\tt IRBis}.
Again, the resulting images are very similar.
Our {\tt MiRA} reconstruction used double the pixel
size compared to our {\tt IRBis} and {\tt BSMEM} reconstructions.

In summary, we show that our reconstructed images are
robust in view of these variations, which cover in total
three different reconstruction packages based on different algorithms,
seven different regularization functions including no regularization,
different start images, and different priors.
%

\begin{table}
\centering
\caption{Quality parameters of the {\tt IRBis} image reconstructions based on different regularization functions.}
\label{tab:image_qc}
\begin{tabular}{lrrrrr}
\hline\hline
Reg. function & $q_\mathrm{rec}$ & $\chi^2_{V^2}$ & $\rho\rho_{V^2}$ & $\chi^2_\mathrm{CP}$ & $\rho\rho_\mathrm{CP}$ \\
\hline
\multicolumn{6}{c}{Spectral channel 1.59\,$\mu$m:}\\
0: no reg.         &  0.310 & 1.617 & 1.401 & 1.163 & 1.057\\
1: compactness     &  0.312 & 1.630 & 1.398 & 1.168 & 1.053\\
2: max. entropy    &  0.299 & 1.631 & 1.318 & 1.177 & 1.071\\
3: smoothness      &  0.310 & 1.622 & 1.404 & 1.168 & 1.047\\
4: edge pres.      &  0.313 & 1.626 & 1.405 & 1.168 & 1.054\\
5: smoothness      &  0.313 & 1.627 & 1.403 & 1.168 & 1.055\\
6: qu. Tikhonov    &  0.317 & 1.629 & 1.412 & 1.170 & 1.055\\
\multicolumn{6}{c}{Spectral channel 1.68\,$\mu$m:}\\
0: no reg.         &  0.765 & 3.727 & 1.227 & 1.006 & 1.102\\
1: compactness     &  0.768 & 3.785 & 1.187 & 1.008 & 1.092\\
2: max. entropy    &  0.772 & 3.994 & 1.004 & 0.998 & 1.086\\
3: smoothness      &  0.767 & 3.752 & 1.209 & 1.006 & 1.101\\
4: edge pres.      &  0.765 & 3.750 & 1.205 & 1.006 & 1.099\\
5: smoothness      &  0.769 & 3.786 & 1.184 & 1.008 & 1.098\\
6: qu. Tikhonov    &  0.766 & 3.799 & 1.161 & 1.010 & 1.093\\
\multicolumn{6}{c}{Spectral channel 1.76\,$\mu$m:}\\
0: no reg.         &  2.294 & 5.739 & 3.623 & 2.779 & 1.033\\
1: compactness     &  2.314 & 5.789 & 3.648 & 2.775 & 1.044\\
2: max. entropy    &  2.413 & 6.031 & 3.716 & 2.783 & 1.122\\
3: smoothness      &  2.299 & 5.755 & 3.618 & 2.781 & 1.042\\
4: edge pres.      &  2.305 & 5.773 & 3.636 & 2.783 & 1.028\\
5: smoothness      &  2.323 & 5.814 & 3.650 & 2.782 & 1.044\\
6: qu. Tikhonov    &  2.300 & 5.760 & 3.626 & 2.779 & 1.036\\
\hline
\end{tabular}
\end{table}
%
\end{appendix}
\end{document}